\documentclass[apj]{emulateapj}
\usepackage{lscape}
\usepackage[hyperfootnotes=false]{hyperref}

\newcommand{\cxo}{\textit{CXO}}
\newcommand{\rosat}{\textit{ROSAT}}

\newcommand{\xmm}{\textit{XMM-Newton}}
\newcommand{\chandra}{\textit{Chandra}}
\newcommand{\nh}{\ensuremath{N_{\rm H}}}

\newcommand{\hi}{{H\,{\sc i}}}
\newcommand{\lsim}{\lesssim}
\newcommand{\gsim}{\gtrsim}
\newcommand{\expnt}[2]{\ensuremath{#1 \times 10^{#2}}}   
\def\mc {\multicolumn}

\newcommand{\Ga}{G65.3+5.7}
\newcommand{\Gb}{G74.0$-$8.5}
\newcommand{\Gc}{G156.2+5.7}
\newcommand{\Gd}{G160.9+2.6}
\newcommand{\Ge}{G205.5+0.5}
\newcommand{\Gf}{G330.0+15.0}
\newcommand{\snra}{SNR~\Ga}
\newcommand{\snrb}{SNR~\Gb}
\newcommand{\snrc}{SNR~\Gc}
\newcommand{\snrd}{SNR~\Gd}
\newcommand{\snre}{SNR~\Ge}
\newcommand{\snrf}{SNR~\Gf}
\newcommand{\rxsa}{1RXS~J193458.1+335301}
\newcommand{\rxsb}{1RXS~J205042.9+284643}
\newcommand{\rxsc}{1RXS~J150139.6$-$403815}
\newcommand{\rxsd}{1RXS~J205812.8+292037}
\newcommand{\lnsa}{1E~0627.4+0537}
\newcommand{\lnsb}{1E~0630.9+0611}
\newcommand{\lnsc}{1E~0636.8+0517}

\newcommand{\hr}{\ensuremath{^{\rm h}}}
\newcommand{\mn}{\ensuremath{^{\rm m}}}

\defcitealias{kfg+04}{Paper~I}

\slugcomment{Accepted for publication in ApJS}

\shorttitle{Search for Compact Central Sources in Large SNRs}
\shortauthors{Kaplan et al.}

\begin{document}

\title{An X-ray Search for Compact Central Sources in Supernova
  Remnants II: Six Large Diameter SNRs}

\author{D.~L.~Kaplan\altaffilmark{1},
  B.~M.~Gaensler\altaffilmark{2,3}, S.~R.~Kulkarni\altaffilmark{4},
  \&\ P.~O.~Slane\altaffilmark{2}} 
\altaffiltext{1}{Pappalardo
  Fellow; MIT Kavli Institute for Astrophysics and Space Research,
  Massachusetts Institute of Technology, 77 Massachusetts Avenue, Room
  37-664H, Cambridge, MA, 02139; \texttt{dlk@space.mit.edu}}
\altaffiltext{2}{Harvard-Smithsonian Center for Astrophysics, 60
  Garden Street, MS-6, Cambridge, MA 02138;
  \texttt{bgaensler,slane@cfa.harvard.edu}.} 
\altaffiltext{3}{Alfred P.\ Sloan Research Fellow}
\altaffiltext{4}{Department of Astronomy, 105-24 California Institute
  of Technology, Pasadena, CA 91125; \texttt{srk@astro.caltech.edu} }

\begin{abstract}
We present the second in a series of results in which we have searched
for undiscovered neutron stars in supernova remnants (SNRs).  This
paper deals with the largest six SNRs in our sample, too large for
\chandra\ or \xmm\ to cover in a single pointing.  These SNRs are
nearby, with typical distances of $<1$~kpc.  We therefore used the
\rosat\ Bright Source Catalog and past observations in the literature
to identify X-ray point sources in and near the SNRs.  Out of 54
sources, we were immediately able to identify optical/IR counterparts
to 41 from existing data.  We obtained \chandra\ snap-shot images of
the remaining 13 sources.  Of these, 10 were point sources with
readily identified counterparts, two were extended, and one was not
detected in the \chandra\ observation but is likely a flare star.  One
of the extended sources may be a pulsar wind nebula, but if so it is
probably not associated with the nearby SNR.  We are then left with no
identified neutron stars in these six SNRs down to luminosity limits
of $\sim 10^{32}\mbox{ ergs s}^{-1}$.  These limits are generally less
than the luminosities of typical neutron stars of the same ages, but
are compatible with some lower-luminosity sources such as the neutron
stars in the SNRs CTA~1 and IC~443.
\end{abstract}

\keywords{ISM: individual (\snra, \snrb, \snrc, \snrd, \snre, \snrf)
  --- pulsars: general --- stars: neutron --- supernova remnants
  --- X-rays: stars}

\section{Introduction}
The connection between core collapse supernovae and neutron stars
\citep{bz34} has had a solid observational footing for almost forty
years, due largely to the discovery of young radio pulsars in
supernova remnants (SNRs) like Vela \citep*{lvm68} and in the Crab Nebula
\citep{sr68}.  Energetic young pulsars like these are strong radio and
X-ray sources, and often power synchrotron nebulae called pulsar wind
nebulae or PWNe \citep{gs06} that are indirect markers of pulsars
\citep[e.g.,][]{camilo03}.

The idea that young neutron stars resemble the Crab pulsar came to
dominate the search for the products of supernovae
\citep[e.g.,][]{kmj+96}.  Recently, though, young neutron stars have
been revealed in a wide variety of manifestations, from Anomalous
X-ray pulsars (AXPs) and soft $\gamma$-ray repeaters (SGRs), to nearby
thermal and radio quiet neutron stars, to long period radio pulsars
with high inferred magnetic fields.  As exemplified by the identification
of the central compact object (CCO) in the \object[SNR 111.7-02.1]{Cas~A} SNR \citep{t99}, much
of this diversity has come from X-ray observations.

While this diversity is clearly demonstrated observationally, theory
and simulation cannot yet constrain the fundamental birth properties
of neutron stars \citep[e.g.,][]{bom04,che05}.  Models still have
difficulties achieving explosions, much less following the activity in
the post-collapse object in any detail.

\citet[][hereafter \citetalias{kfg+04}]{kfg+04} have attempted to
address our lack of understanding of stellar death and neutron star
cooling by defining a volume-limited ($d < 5$~kpc) sample of supernova
remnants (SNRs), examining the neutron stars that they contain, and
outlining a survey designed to detect or significantly constrain
neutron stars in the remaining remnants.  The primary subsample discussed
in \citetalias{kfg+04} is one where the SNR diameter is $<45\arcmin$,
so that the \textit{Chandra X-ray Observatory} can observe a
significant fraction of the SNR interior with its ACIS-I detector and
hence cover the area where neutron stars would be with a reasonable
range of velocities ($v_{\perp} < 700\mbox{ km s}^{-1}$, where
$v_{\perp}$ is the velocity perpendicular to the line of sight).
\citetalias{kfg+04} also discuss two other subsamples of SNRs: one with
diameters $45\arcmin < \theta < 90\arcmin$ for which \xmm\ is suitable
(and which we will present in a forthcoming paper), and one with
$\theta > 90\arcmin$.  It is this subsample of the six largest SNRs from
\citetalias{kfg+04} that we consider here.

The organization of the paper is as follows. In
\S~\ref{sec:snrs} we give brief summaries of the six SNRs discussed
here.  In \S~\ref{sec:bsc} we describe our identification of
candidate X-ray sources in and around the SNRs.  In \S~\ref{sec:cpt}
we detail the initial identification of optical/IR counterparts to the
X-ray sources using available sky surveys: as discussed in
\citetalias{kfg+04}, optical/IR observations are a powerful way to
reject X-ray sources that are not neutron stars (see also e.g.,
\citealt{rfbm03}).  With the sky surveys we were able to identify most
of the X-ray sources with high confidence: those for which we were not
certain were selected for additional \chandra\ observations and
optical/IR observations (\S~\ref{sec:cxo}).  Finally, we give our
discussion and conclusions in \S~\ref{sec:disc}.  All coordinates are
J2000.0.

\begin{deluxetable*}{l l c c l c c}
\tablewidth{0pt}
\tablecaption{Large SNRs\label{tab:snrs}}
\tablehead{
\colhead{SNR} & \colhead{Other} & \colhead{Size} & \colhead{D} &
\colhead{Distance} & \colhead{$N_{\rm H}/10^{21}$} &
\colhead{$L_{\rm X}/10^{31}$} \\
 & \colhead{Name} & \colhead{(arcmin)} & \colhead{(kpc)} &
\colhead{Method} & \colhead{(cm$^{-2}$)\tablenotemark{a}} & \colhead{($\mbox{ergs s}^{-1}$)\tablenotemark{b}} \\}
\startdata
\object[SNR 065.3+05.7]{\Ga} & G65.2+5.7   & $310 \times 240$ & 0.8 & optical velocity & 1.4 & 6.7 \\
\object[SNR 074.0-08.5]{\Gb} & Cygnus Loop & $230 \times 160$ & 0.44 & optical proper motion &
0.8 & 1.7\\
\object[SNR 156.2+05.7]{\Gc} & & 110 & 1.3 & NEI fits &3.5 & 29\\
\object[SNR 160.9+02.6]{\Gd} & HB9 & $140 \times 120$ & $1.5$ & \hi, optical velocity & 1 & $21$\\
\object[SNR 205.5+00.5]{\Ge} & Monoceros & 220 & 1.2 & optical velocity& 0.8 & 13\\
\object[SNR 330.0+15.0]{\Gf} &Lupus Loop & 180 & 1.2 & NEI fits & 0.5 & 12.0\\
\enddata
\tablenotetext{a}{Hydrogen column density to SNR.  Derived from
 previous observations (if
 available), otherwise determined from measured \hi\ 
 absorption or using {\tt COLDEN}\  integrated over
 velocity range appropriate for the SNR distance.}
\tablenotetext{b}{Unabsorbed X-ray luminosity (0.3--8.0~keV) of a
nominal $0.05\mbox{ s}^{-1}$ \rosat\
   PSPC source at the distance and absorption of the SNR, assuming a
   blackbody spectrum with $kT=0.25$~keV (this allows for easy
   conversion of count-rates to luminosities, assuming that the sources
 are associated with the SNRs.}
\tablecomments{See \S~\ref{sec:snrs} for a general discussion about
  the quality of the remnant distances and for detailed discussions
  about each remnant.}
\end{deluxetable*}

\ \\

\section{Supernova Remnants}
\label{sec:snrs}
We list the SNRs for this paper, along with relevant parameters, in
Table~\ref{tab:snrs}.  Each SNR has a distance determined from a more
reliable method than the $\Sigma$-$D$ method
\citep[e.g.,][]{ht85,cb98}, but they are not all of the same quality.
In the best cases, the distances are from kinematic observations of
optical or radio lines.  In the worst cases, the distances are from
fitting shock models to the X-ray data \citep[e.g.,][]{khvdw94}.
These distances involve many uncertainties beyond the kinematic
distances, including the assumption of a Sedov-phase remnant, the
state of equilibration in the system, the non-sphericity of the
explosion, and the unknown total explosion energy.  For kinematic
distances, the uncertainties are probably $\lsim 30$\%, but for
distances from X-ray fitting they could exceed 50\% \citep{khvdw94}.
The ages tend to be derived from X-ray fits for all sources, although
having an independently determined distance for some sources makes for
better constraints.  Below we discuss each SNR in more detail.

\begin{figure}
\plotone{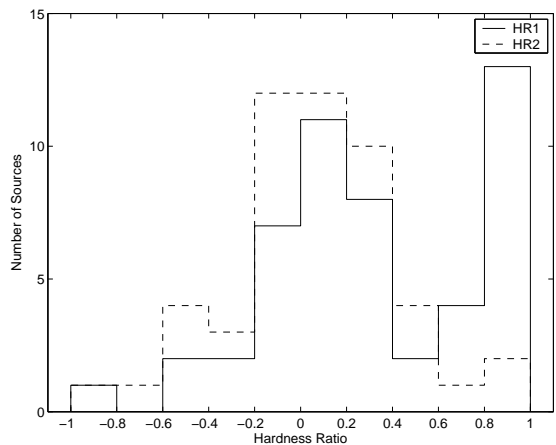}
\caption{Hardness ratios of BSC sources from Table~\ref{tab:srcs}.
  HR1 is the solid line, while HR2 is the dashed line.  HR$1\equiv
  (B-A)/(B+A)$ and HR$2\equiv(D-C)/(D+C)$, where $A$, $B$, $C$, and
  $D$ are the count-rates in the PHA ranges 11--41, 52--201, 52--90,
  and 90--201 respectively, and the PHA values correspond roughly to
  the energies in eV \citep{rbs2}.  As discussed in
  \citet{rbs2}, we note that HR2 is constructed only from counts in
  the $B$ range, so it is not a contradiction to have (for example)
  HR$1=-1$ and HR$2=1$. }
\label{fig:hard}
\end{figure}

\subsection{\snra}
\snra\ (also known as G65.2+5.7) was identified as a SNR by
\citet*{gkp77} by its filamentary line emission.  It has major axes of
$310\arcmin \times 240\arcmin$.  According to \citet{mbpv02}, the age
is 20--25~kyr, and the distance is $\approx 0.8$--1.0~kpc (a kinematic
distance derived from the velocity of optical emission lines; also see
\citealt{loz81}).  \citet{mbpv02} show data from pointed \rosat\
observations \citep{sbla02} but do not discuss point sources; the
\rosat\ data detect emission from much of the interior at
$>\expnt{2.5}{-4}\mbox{ cts s}^{-1}\mbox{ pixel}^{-1}$, with
$45\arcsec$ pixels.  \citet*{skp04} do mention an extended ($\gsim
6\arcmin$ radius) soft source at $19^{\rm h}36^{\rm m}46^{\rm s}$
$+30^{\degr}40^{\arcmin}07\arcsec$ but come to no conclusion as to its
identity.  The inner $3\arcmin$ were searched for radio pulsars by
\citet{gra+96} at 1410~MHz down to a limit of 0.1~mJy, but given the
size of \snra\ a transverse velocity of only $v_{\perp}=20\mbox{ km s}^{-1}$
would have moved a neutron star outside the search region so the lack
of detection was not very constraining,

The 0.58-s radio pulsar \object[PSR J1931+30]{PSR~J1931+30} lies
$45\arcmin$ from the center of the remnant.  However, with no estimate
of the spin-down rate (and hence no spin-down age), the period seems
rather large to be associated with an SNR (it would require an unusual
but not unheard of magnetic field of $\sim 10^{13}$~G), and no
definite claim of an association can be made \citep{sbla02}.

\begin{figure*}
\epsscale{.9}\plotone{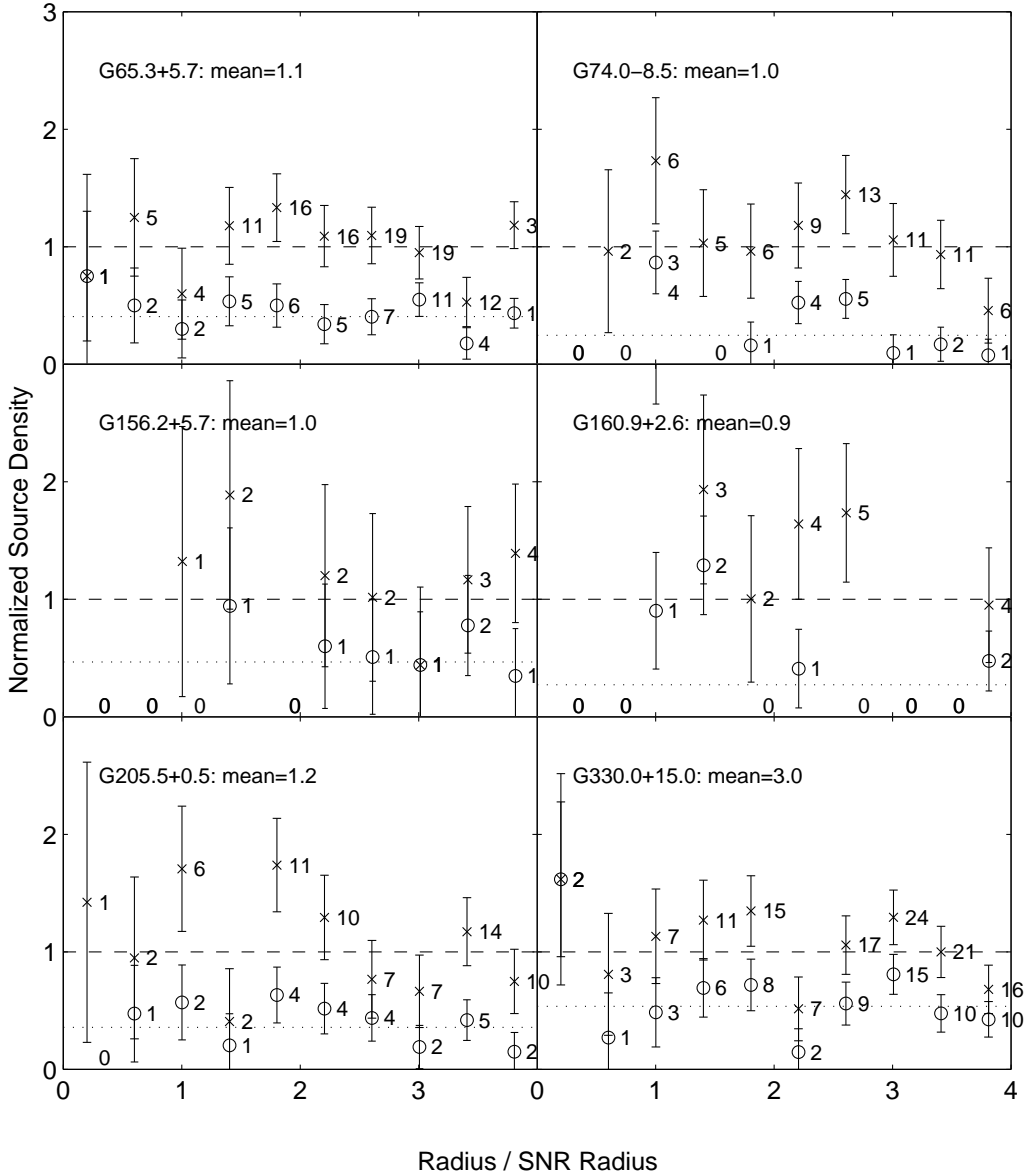}\epsscale{1}
\caption{Normalized density of BSC sources in each of the SNRs from
  Table~\ref{tab:snrs}.  The number of sources per square arcminute
  divided by the mean density is plotted against radius (in units of
  the SNR radius).  All sources are shown as the x's, while only the
  unresolved sources are shown as the circles.  The means of the
  different source densities are shown as the dashed and dotted lines,
  respectively.  The means of the total source densities (in units of
  $10^{-4} \mbox{ arcmin}^{-2}$) are given next to the SNR names.  At
  the position of each bin is printed the number of sources
  contributing to that bin.  For bins with no sources plotted, these
  deficits are in all cases consistent with the small number counts
  (i.e.\ we expect $\lsim 1$ source in each of those bins) except for
  the third and fourth bins of \snrd, where 3.3 and 3.8 sources are
  expected, respectively. However, even in these bins there is no
  significant deficit of point sources.
\label{fig:dens}}
\end{figure*}

\subsection{\snrb}
\snrb, also known as the Cygnus Loop, is a $230\arcmin \times
160\arcmin$ radio and X-ray shell.  The distance, estimated from
measurements of the shock velocity and proper motion is 0.44~kpc
\citep{bsrl99} and the age is 8~kyr \citep*{lgw+02}. \citet{mtk+98}
reported $\approx 8.8\mbox{ count s}^{-1}$ in the interior over the
$22\arcmin$ field of the SIS for their \textit{ASCA} observation, and
used these data to conclude that \snrb\ was likely the result of a
Type~II supernova on the basis of elemental abundances.

\citet{mtt+98} searched \snrb\ for promising X-ray point sources that
might be compact objects, and identified two, one of which they later
concluded was an active galactic nucleus (AGN) on the basis of its
long-term variability, X-ray spectrum, and radio counterpart
\citep{mot+01}, and the other of which they conclude may be a neutron
star.  The inner $10\arcmin$ were searched for radio pulsars by
\citet{gra+96} down to a 430~MHz flux limit of 0.3~mJy, but given the
size of \snrb\ the lack of detection was not very constraining
($v_{\perp} \leq 80\mbox{ km s}^{-1}$).  The
inner $30\arcmin$ were also searched by \citet{bl96} for pulsars down
to a 400-MHz flux of 3~mJy.  Assuming an average radio spectrum for
radio pulsars of $S_{\nu}\propto \nu^{-1.5}$, this
translates to a 1400-MHz luminosity limit of $0.01\mbox{ mJy kpc}^2$,
which is considerably fainter than the very low luminosity
\object[PSR J0205+6449]{PSR~J0205+6449} \citep[$0.5\mbox{ mJy kpc}^2$;][]{csl+02}.

\subsection{\snrc}
\snrc\ was discovered in the \rosat\ All-Sky Survey by \citet*{pap91}.
It has a faint $110\arcmin$ shell in both X-rays and radio, and
non-equilibrium fits to the X-ray data place it at a distance of
$\approx 1.3$~kpc with an age of 15~kyr \citep{ykt+99}.
\citet*{llc98} searched \snrb\ for radio pulsars, tiling seven
pointings of the 76-m Lovell telescope at Jodrell Bank, each of which
covered $\approx 0.5\degr$.  The search did not find any pulsars, down
to a flux limit of $0.7$~mJy at 606~MHz, or a 1400-MHz luminosity
limit of $0.3\mbox{ mJy kpc}^2$.

\subsection{\snrd}
\snrd, also known as HB~9, is a $140\arcmin\times 120\arcmin$ radio
shell with bright X-rays in the interior.  \citet{la95} use X-ray
fitting to estimate a distance of 1.5~kpc and an age of 8--20~kyr.
This distance is consistent with the upper limit of 4~kpc derived from
other measurements \citep{loz81,lr91}.

The inner $30\arcmin$ were also searched by \citet{bl96} for pulsars
down to a 610-MHz flux of 15~mJy.  \citet*{dth78} discovered an old
radio pulsar (\object[PSR B0458+46]{PSR~B0458+46}) in the interior of
the SNR, although the association between the pulsar and the SNR is
generally considered to be false \citep[e.g.,][]{kh02} based on the
large spin-down age and low spin-down energy loss rate for the pulsar
($10^6$~yr and $10^{33}\mbox{ ergs s}^{-1}$, respectively).

\subsection{\snre}
\snre, also known as the Monoceros nebula, is a $220\arcmin$ radio
shell.  The systemic velocity of  optical line emission puts the
SNR at a distance of 0.8~kpc \citep{loz81}, although distances up to
1.6~kpc are preferred by low-frequency radio data that show the SNR
within the Mon~OB2 association \citep{ode86}.  The age is likely $\sim
30$~kyr, as inferred from fits to X-ray data \citep*{lns86}.

\subsection{\snrf}
\snrf, the Lupus Loop, is a low surface brightness radio shell
approximately $180\arcmin$ in diameter.  Non-equilibrium fits to the
X-ray data and comparison with the column density of the nearby
remnant of SN~1006 suggest a distance of 1.0--1.2~kpc and an age of
50~kyr \citep*{lnh91}.

\begin{figure*}
\plotone{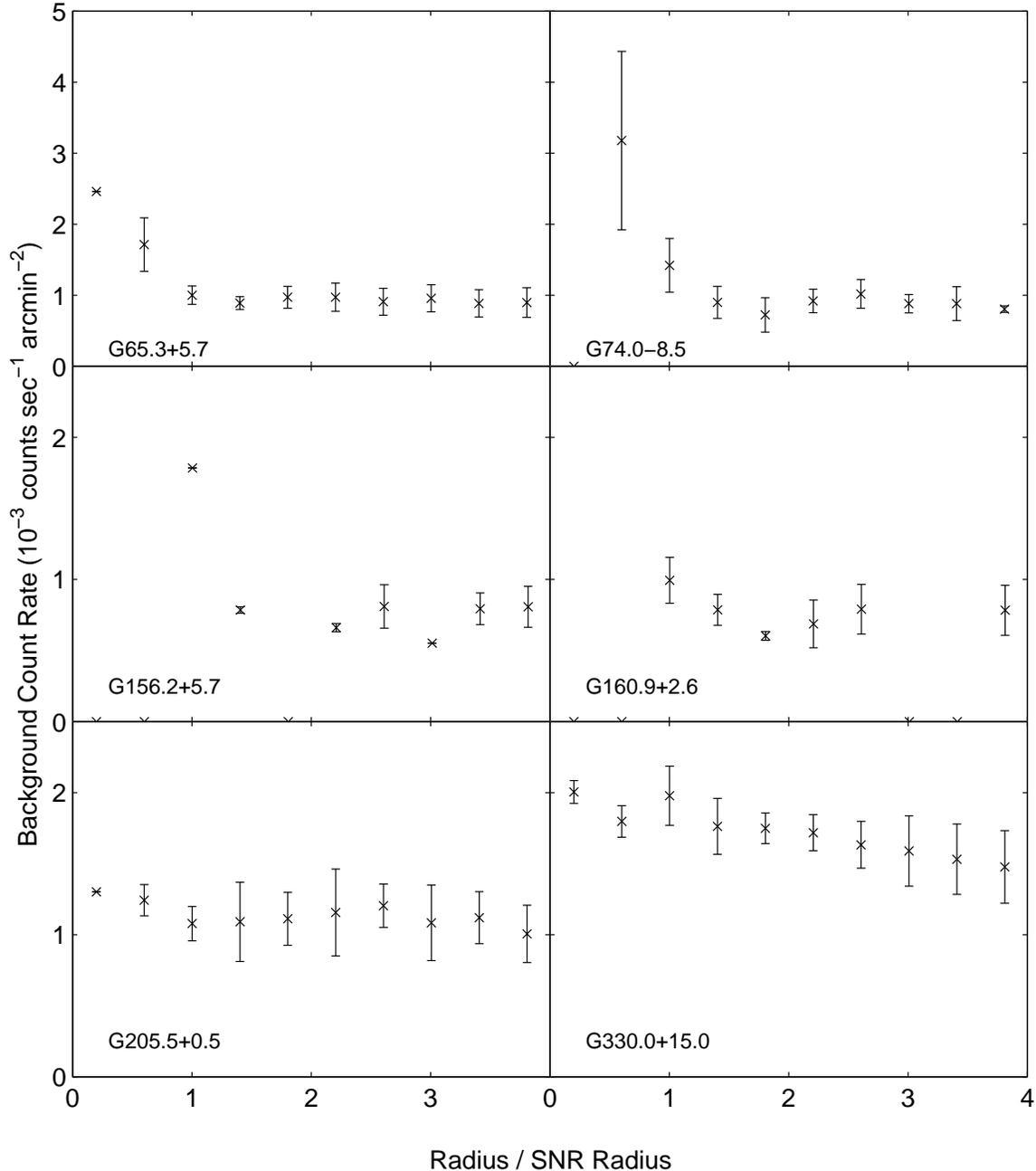}
\caption{Background count-rate vs.\ radius for the BSC sources in each of the SNRs from
  Table~\ref{tab:snrs}.  The average background count-rate
  ($10^{-3}\mbox{ counts s}^{-1}\mbox{ arcmin}^{-2}$) in each of
  10 radial bins between 0 and 4 times the SNR radius is shown, along
  with uncertainties showing the standard deviation in each bin.  SNRs
  \Ga\ and \Gb\ do show a factor of 2--3 increase in background rate
  inside the SNRs.  For SNRs \Gc\ and \Gd, the situation is not as
  clear because there are very few sources inside (see
  Fig.~\ref{fig:dens}).  For SNRs \Ge\ and \Gf, there do not appear
  to be a significant rises toward the interiors.
\label{fig:bg}}
\end{figure*}

\section{Source Selection}
\label{sec:bsc}
\chandra\ or \xmm\ imaging of the entire fields of the large-diameter
SNRs listed in Table~\ref{tab:snrs} is impractical because of their
sizes.  Their proximities ($d\lsim 1$~kpc), though, means that we do not
need the high sensitivities of \chandra\ or \xmm\ to achieve the same
luminosity limit as in \citetalias{kfg+04}.  We therefore used
the \rosat\ All-Sky Survey Bright Source Catalog (hereafter BSC;
\citealt{rbs2}) for our source selection.  This was a survey of the
entire sky with the Position-Sensitive Proportional Counter (PSPC)
aboard \rosat.

The positional accuracy of the PSPC does not approach that of
\chandra\ or even \xmm\ (typical uncertainties are $10\arcsec$), and
the observations are not as deep as the pointed \chandra\ and \xmm\
observations used for the other SNRs.  Nonetheless, the BSC is useful.
As seen in Table~\ref{tab:snrs}, its limit of $0.05\mbox{ count
s}^{-1}$ in the PSPC is actually of roughly comparable depth to our
\chandra\ observations in \citetalias{kfg+04} ---
$\expnt{(1-10)}{31}\mbox{ ergs s}^{-1}$ --- when the smaller distances
and column densities of the SNRs in this paper are taken into account.
While the X-ray positions do not in all cases allow unambiguous
optical identifications, the relative brightness and softness of the
X-ray sources compared to those in \citetalias{kfg+04} means that very
often stars from the Digitized Sky Survey (DSS) or 2MASS \citep{2mass} can
be identified as counterparts.

We selected the BSC sources within twice the nominal radii (for
elliptical sources, we took the semi-major axes) of the SNRs in
Table~\ref{tab:snrs} (as defined by their positions and sizes given by
\citealt{g00}) that had $\geq 0.05\mbox{ count s}^{-1}$ and were
listed as unextended (a value of 0 in the \texttt{ext} column of the
BSC catalog).  Searching outside the remnants allowed us to find
neutron stars that have overtaken the SNR shocks --- not an uncommon
occurrence \citep*{vdsdk04} in SNRs of the ages considered here
(10--30~kyr).  This gave us all of the X-ray sources listed in
Table~\ref{tab:srcs}.  For the sake of comparison between sources, we
plot the distribution of hardness ratios in Figure~\ref{fig:hard}.

\subsection{Extended Sources}
\label{sec:ext}
In our analysis, we rejected those BSC sources that were identified as
extended.  This was for several reasons: we eliminated peaks in
diffuse background emission that may have been identified as discrete
sources, and we eliminated large extended objects such as galaxy
clusters.  Practically, point sources offer much better astrometry and
are better suited to counterpart identification.

However, in some sense our selection was less than ideal.  We would
have eliminated any bright PWNe, although these might have been
identified by previous searches.  Also, source confusion makes our
resulting luminosity limits less constraining than they might
otherwise be, as two nearby point-sources could have been identified
as a single extended source and hence been rejected.  Given the
relatively low space density of BSC sources (Fig.~\ref{fig:dens}) this
should not be a major effect, but it should still be noted.  In
contrast, our \chandra\ observations do not suffer from any confusion
limitations.

One might ask if the diffuse emission from the SNRs themselves will
limit the depth of the BSC in the SNR interiors.  We have found in
general that this is not the case.  Figure~\ref{fig:dens} shows the
density of BSC sources (both point-like and of all sizes) within
different radii from the SNR centers.  While the inner reaches of the
SNRs have few sources and therefore poor statistics, in no case is
there a statistically significant deficit of point sources inside the
SNR.  There might be a slight deficit inside \snrc\ or \snrd, but
these are also the smallest of the SNRs and therefore have the fewest
total sources.  Similarly, in Figure~\ref{fig:bg} we show the average
background count-rates determined when extracting the sources plotted
in Figure~\ref{fig:dens}, with the same binning.  Two of the SNRs
(\Ga\ and \Gb) do show background increases in the interiors, two do
not (\Ge\ and \Gf), and two are uncertain due to few counts (\Gc\ and
\Gd), but even an increase of a factor of three above the mean
background rate ($\approx 10^{-3}\mbox{ counts s}^{-1}\mbox{
arcmin}^{-2}$) would give only $\approx 0.005\mbox{ counts s}^{-1}$
within the 90\% confidence radius of a PSPC source (for
0.3~keV\footnote{\url{http://heasarc.gsfc.nasa.gov/docs/heasarc/caldb/docs/rosat/cal\_ros\_92\_001/cal\_ros\_92\_001.html.}})
which is a factor of 10 less than the minimum source count-rate for
the BSC.  Therefore the diffuse SNR emission should not have
significantly affected the BSC source detection, and it is unlikely
that there were any point sources that were missed.

\subsection{Additional Sources}
\label{sec:addl}
Besides the BSC, we took advantage of X-ray observations in the
literature to identify additional sources for \chandra\ followup.
These were: for \snrb, AX~J2049.6+2939 \citep[from
\textit{ASCA};][]{mtt+98,mot+01}; and for \snre: \textit{Einstein} sources 1,
3, and 6 from \citet{lns86}, known as \lnsa, \lnsb, and \lnsc,
respectively.

\section{Counterpart Identification}
\label{sec:cpt}
\begin{figure*}
\plottwo{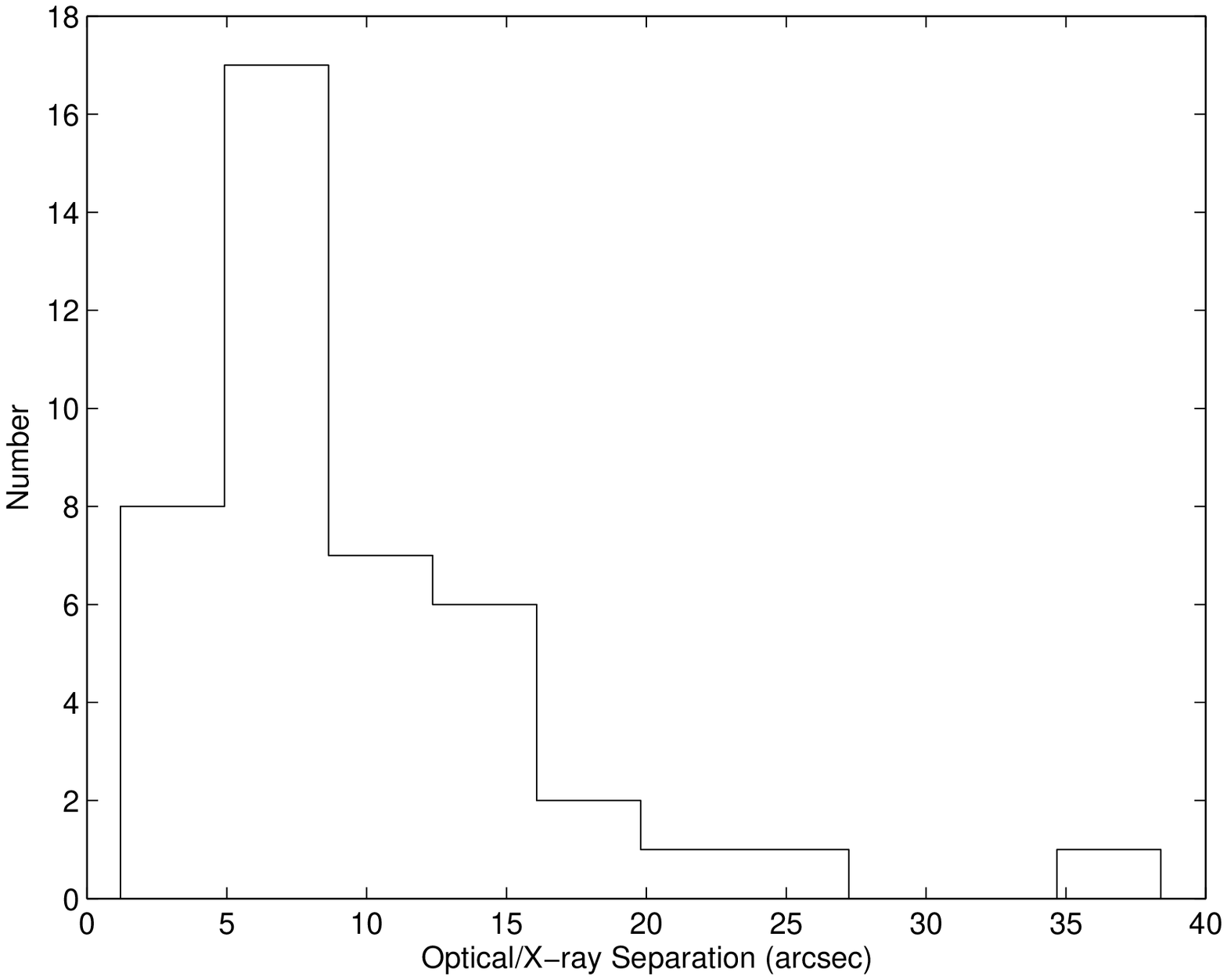}{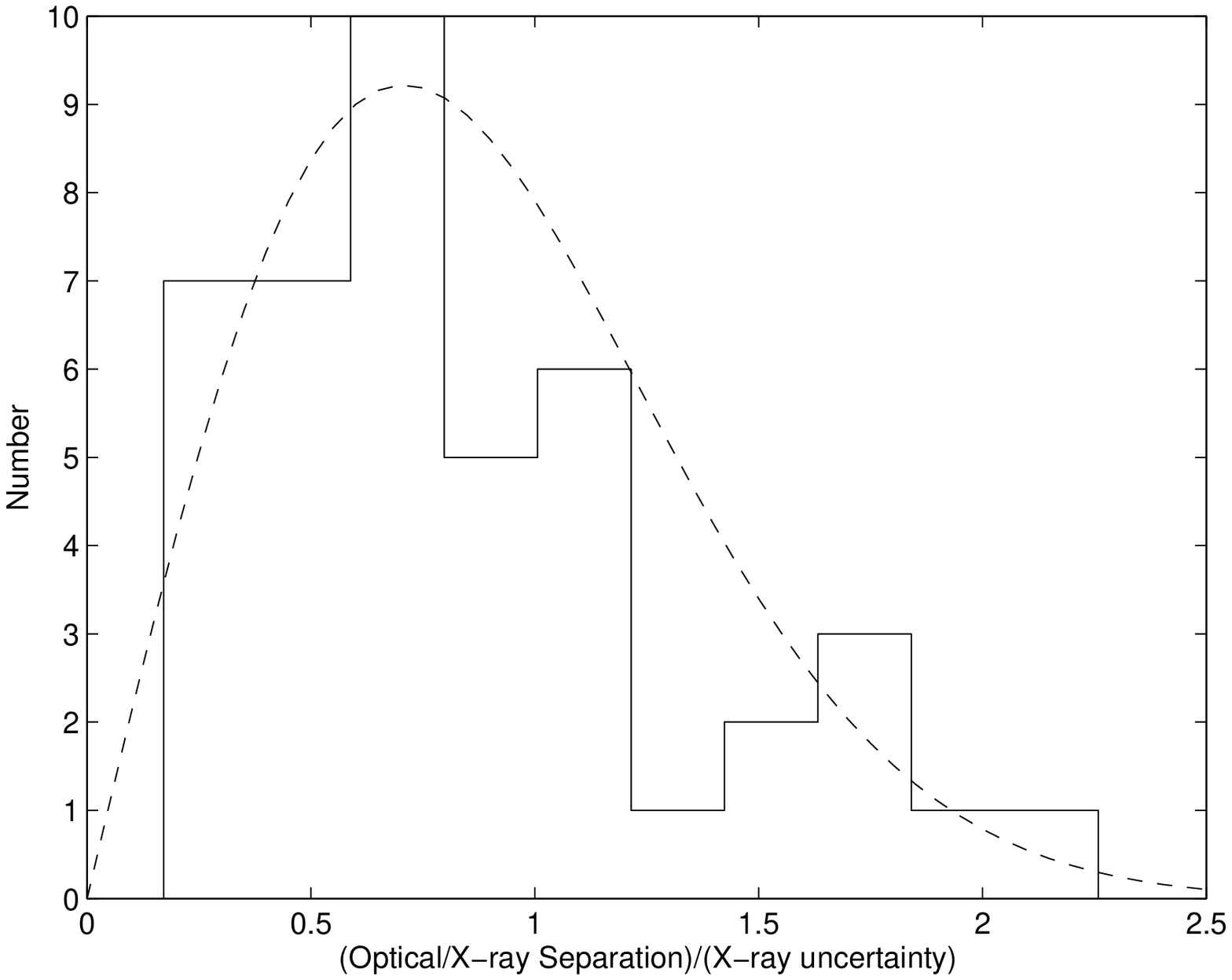}
\caption{Distribution of the separation between the \rosat\ positions
  and the positions of their optical counterparts.  Left: absolute
  separation in arcsec; right: separation normalized to the X-ray
  position uncertainty, with the expected distribution for true
  associations ($f_{r} \propto r \exp (-r^2)$, where $r$ is the
  normalized separation; dashed line) also plotted.  This shows that we
  have largely identified the correct counterparts for the X-ray
  sources, and that the position uncertainties are
  reasonable.\label{fig:offset}}
\end{figure*}

Once we had assembled the list of X-ray sources, we then examined the
publicly available surveys (DSS, 2MASS\footnote{When we were doing the
initial source selection, the final 2MASS data had not been released,
so there were cases where we made decisions based only on DSS data.},
and NRAO VLA Sky Survey\footnote{For all SNRs but \snrf, which is
below the $\delta=-40\degr$ limit of the NVSS.} [NVSS;
\citealt{ccg+98}]), as well as examination of SIMBAD and the relevant
literature.  With these sources of information, we were able to
identify likely counterparts to 41 of the 50 sources in
Table~\ref{tab:srcs}.  We list the relevant data (X-ray and optical)
of the identifications in Table~\ref{tab:srcs}, with a summary of all
identifications and additional notes in Table~\ref{tab:id}.  The
separations between the nominal X-ray and optical positions were
consistent with the predicted X-ray position uncertainties
(Fig.~\ref{fig:offset}).

There are a number of cases where there were multiple stars within the
X-ray error circles, some of which were known to be physically
associated with each other (as noted in SIMBAD).  In these cases we
list multiple possible counterparts in Table~\ref{tab:srcs}.  The true
source of the X-ray emission may be any one of the stars, or may in
fact come from the interactions between them.

\begin{figure}
\plotone{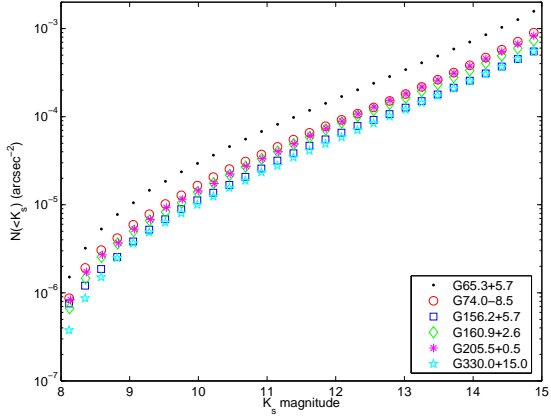}
\caption{Cumulative number density of IR sources.  The number of 2MASS
  sources per square arcsecond brighter than a given $K_s$ magnitude
  is plotted against $K_s$ magnitude for SNRs \Ga\ (black points),
  \Gb\ (red circles), \Gc\ (blue squares), \Gd\ (green diamonds), \Ge\
  (maroon asterisks), and \Gf\ (cyan stars).  Typical PSPC error
  circles have radii of $\approx 10\arcsec$ (Tab.~\ref{tab:srcs}).
\label{fig:stars}
}
\end{figure}

While the identifications were made only on the basis of positional
coincidence with bright stars, in many cases we can be additionally
confident.  This is because the stars are so bright ($V < 5$~mag) that
the chances of a false association are negligible or the stars are of
types known to have X-ray emission (e.g., T Tauri stars).  To aid in
the evaluation of our identifications, we plot the cumulative number
density of 2MASS sources for each SNR in Figure~\ref{fig:stars}.  For
a typical position uncertainty of $10\arcsec$, we expect that all
identifications with 2MASS sources brighter than $K_s\approx11$~mag
will be real (chance probability $<1$\%), and even for sources
brighter than $K_s\approx13$~mag the identifications will be probable
(chance rates $<10$\%).  We note, though, that while \rxsd\ is
consistent with having an association with one of the identified 2MASS
sources, the association is not secure.  A \chandra\ followup
observation likely would have been definitive as it was for the
majority of the ambiguous sources discussed in Section~\ref{sec:cxo},
but due to an oversight on our part this source was not selected for
followup \chandra\ observations.  We
therefore discuss \rxsd\ in extra detail in Section~\ref{sec:rxsd}.

\begin{figure}
\plotone{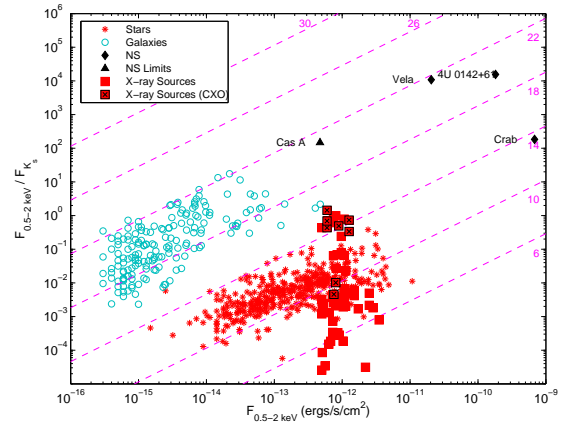}
\caption{X-ray-to-$K_s$ flux ratio vs.\ X-ray flux for sources from
  Table~\ref{tab:srcs}, with sources from the CDF/Orion studies and
  selected neutron stars.  Stars from CDF/Orion are blue asterisks,
  galaxies are green circles.  Selected neutron stars are black
  diamonds/limits, and are labeled.  The X-ray sources from
  Table~\ref{tab:srcs} (including detections from
  \S~\ref{sec:notescxo}) are the red squares (those with \chandra\
  followup, plus \rxsd, have  x's in their squares).  The diagonal
  lines represent constant magnitude, and are labeled by that
  magnitude.  For the X-ray sources from Table~\ref{tab:srcs}, the
  PSPC were converted to a flux by $F_{0.5-2.0\mbox{ \scriptsize
  keV}}={\rm PSPC} \times \expnt{1}{-11}\mbox{ ergs s}^{-1}\mbox{
  cm}^{-2}$, appropriate for a blackbody with $kT_\infty=0.25$~keV and
  $\nh=\expnt{5}{20}\mbox{ cm}^{-2}$.  The X-ray source are largely
  consistent with foreground stars, especially considering the
  possible range of temperatures and column densities, but a number
  may also be active galaxies such as those found by \citet{rfbm03}.
\label{fig:fxk}
}
\end{figure}

\begin{figure*}
\centering
\includegraphics[width=0.5\textwidth]{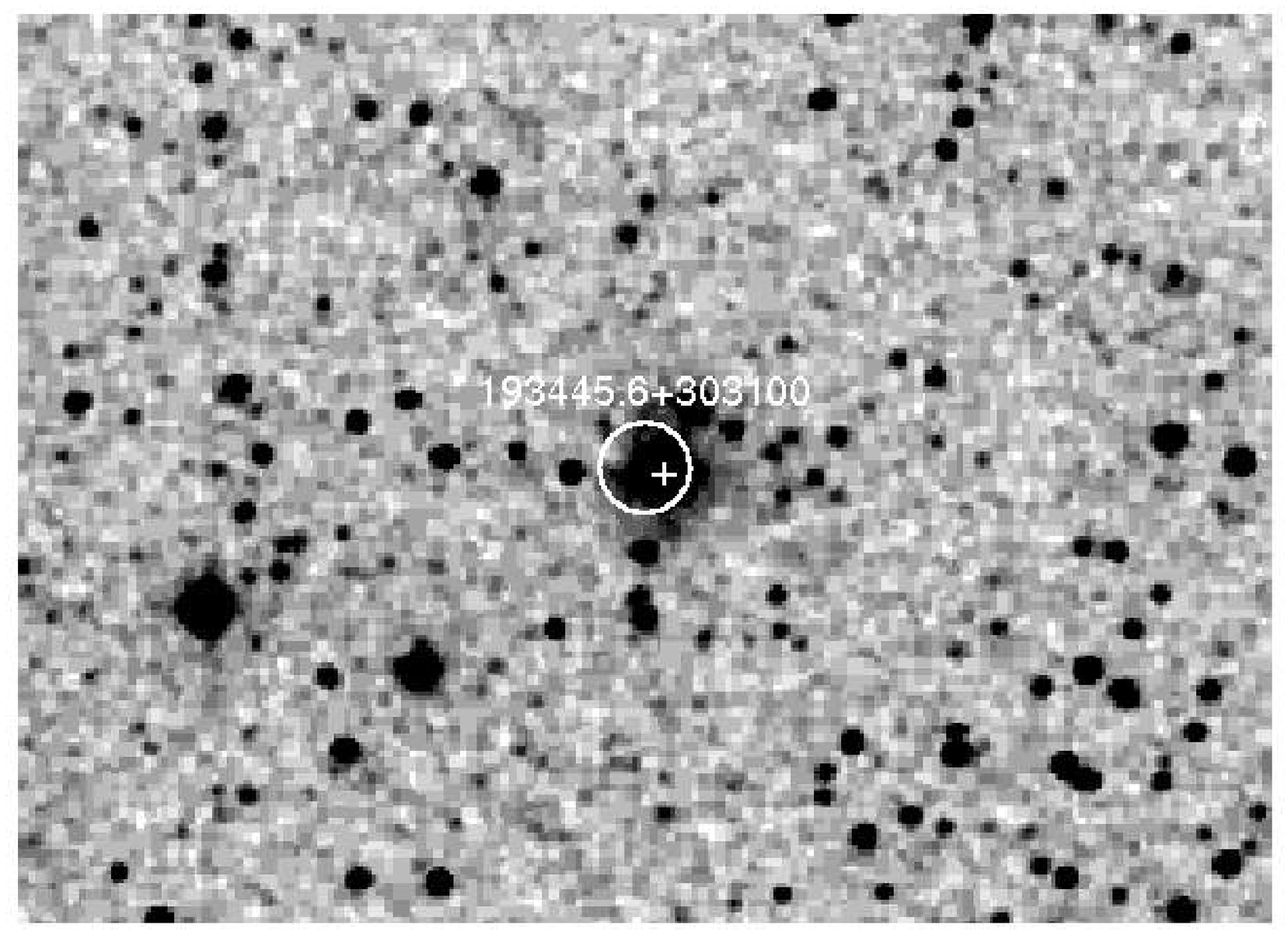}\includegraphics[width=0.5\textwidth]{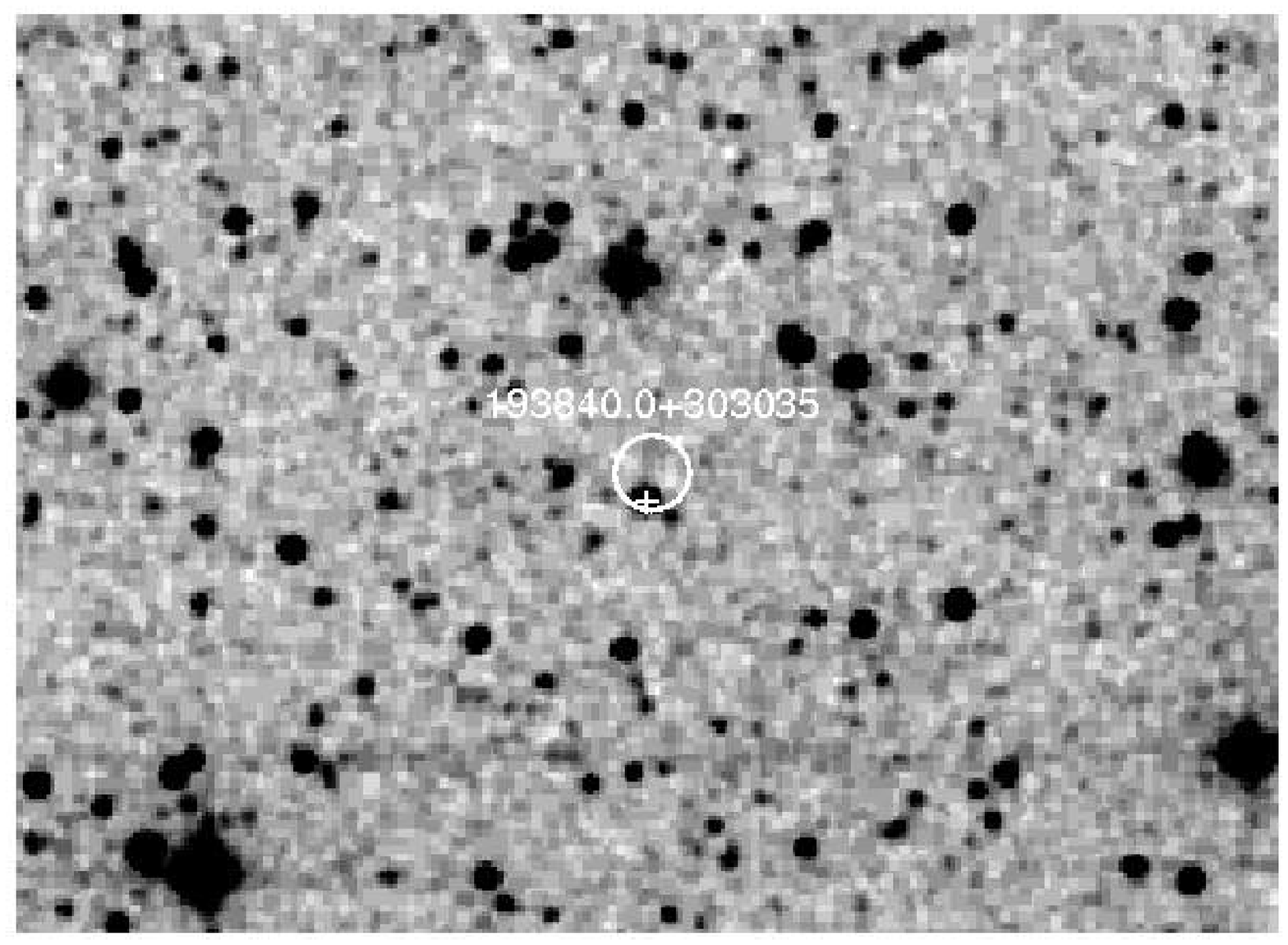}
\includegraphics[width=0.5\textwidth]{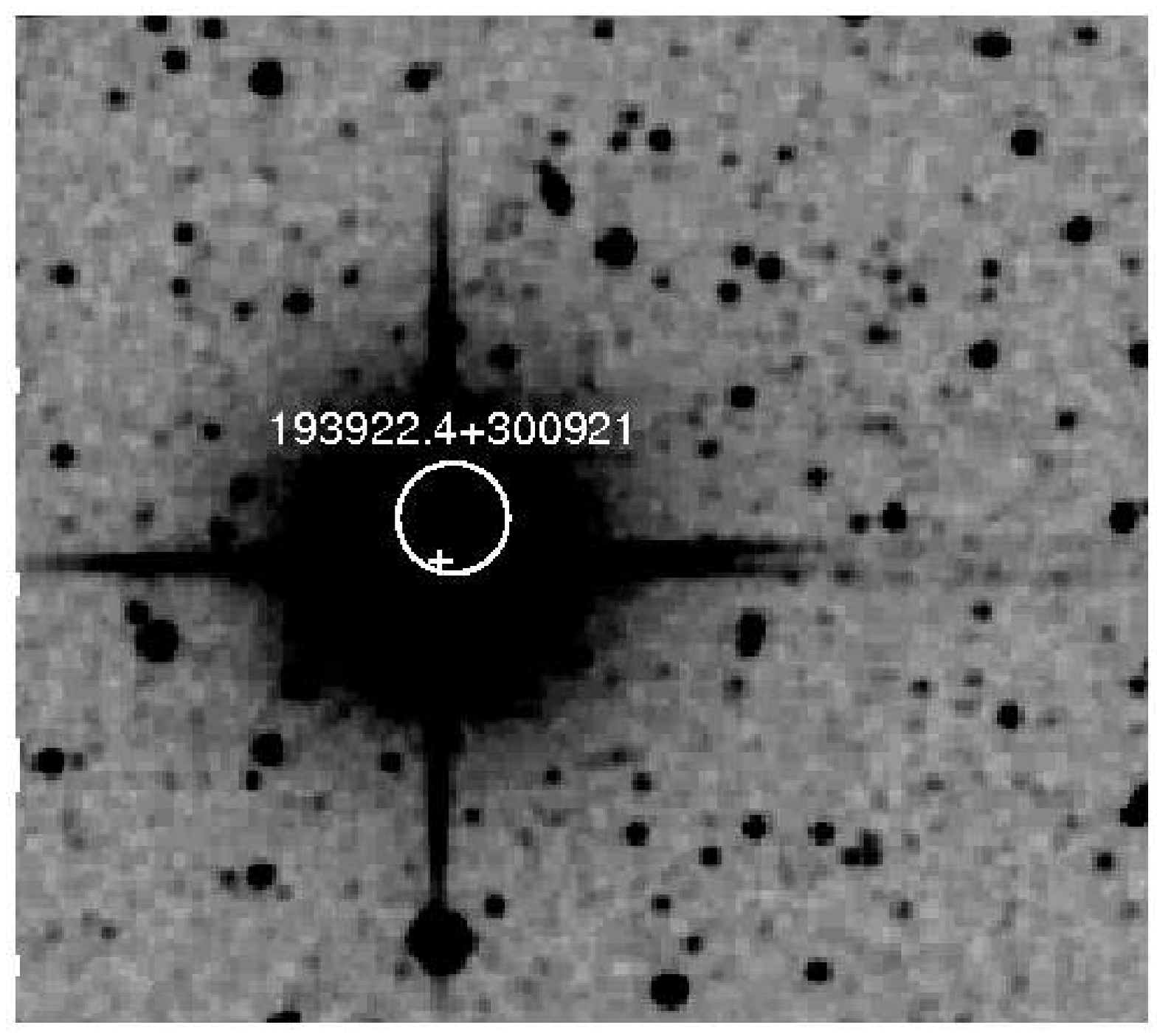}\includegraphics[width=0.5\textwidth]{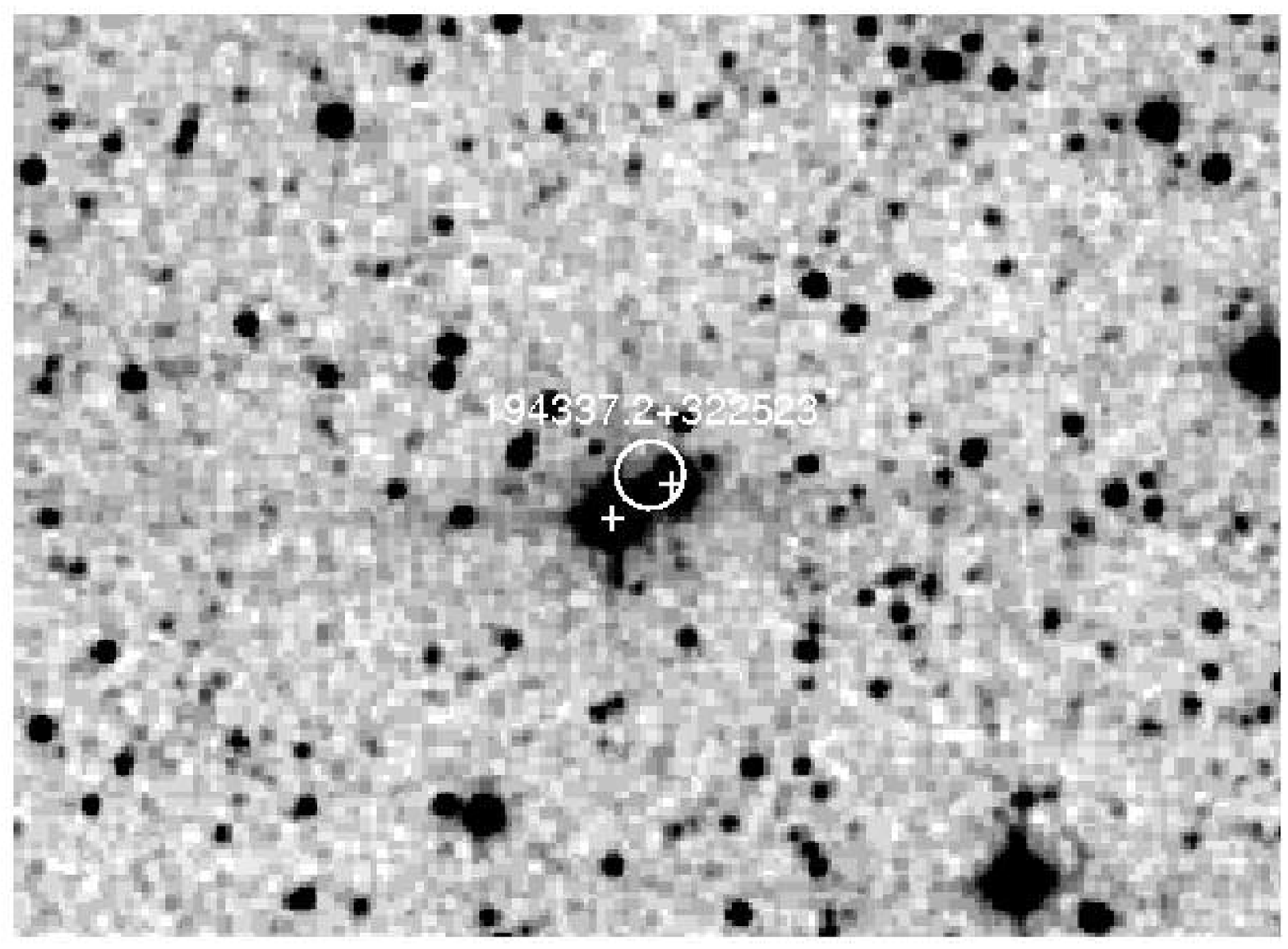}
\includegraphics[width=0.5\textwidth]{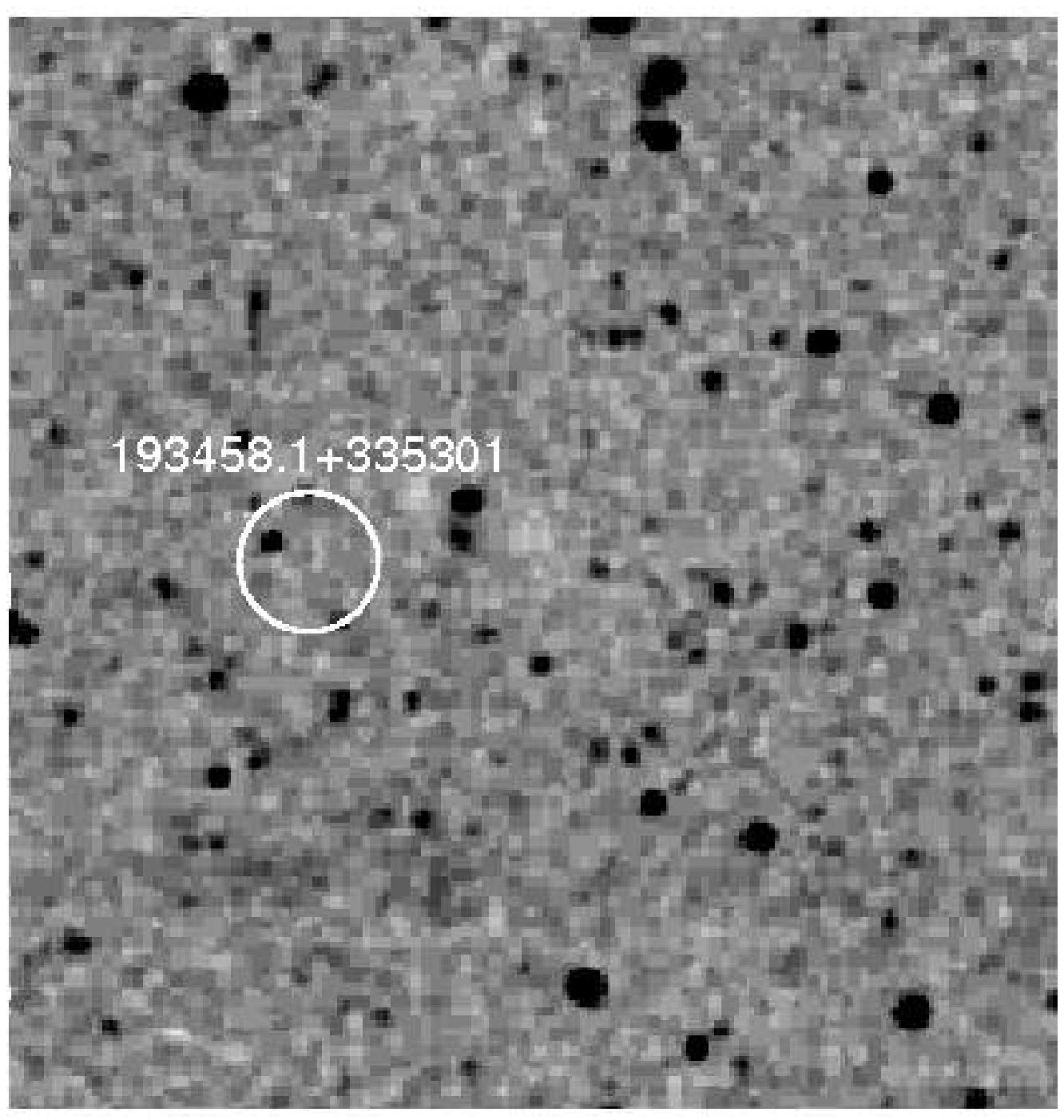}\includegraphics[width=0.5\textwidth]{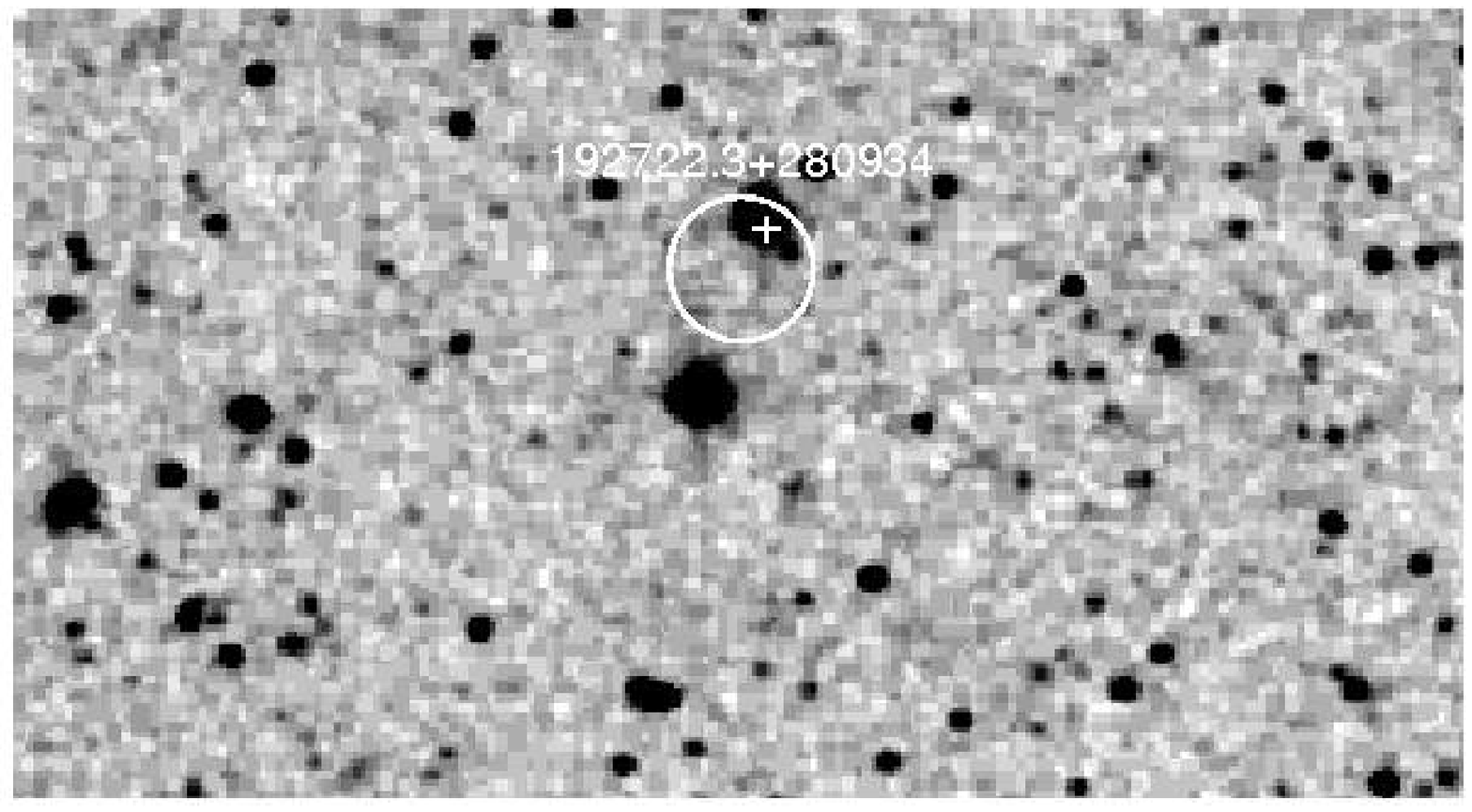}
\caption{2MASS $K_s$-band images of the sources in \snra\ from
  Table~\ref{tab:srcs}.  The images are $5\arcmin\times3.5\arcmin$, with
  North up and East to the left.  The X-ray position uncertainties are
  indicated by the circles, and the proposed optical counterparts are
  shown by the crosses.\label{fig:opt1}}
\end{figure*}

\begin{figure*}
\centering
\includegraphics[width=0.5\textwidth]{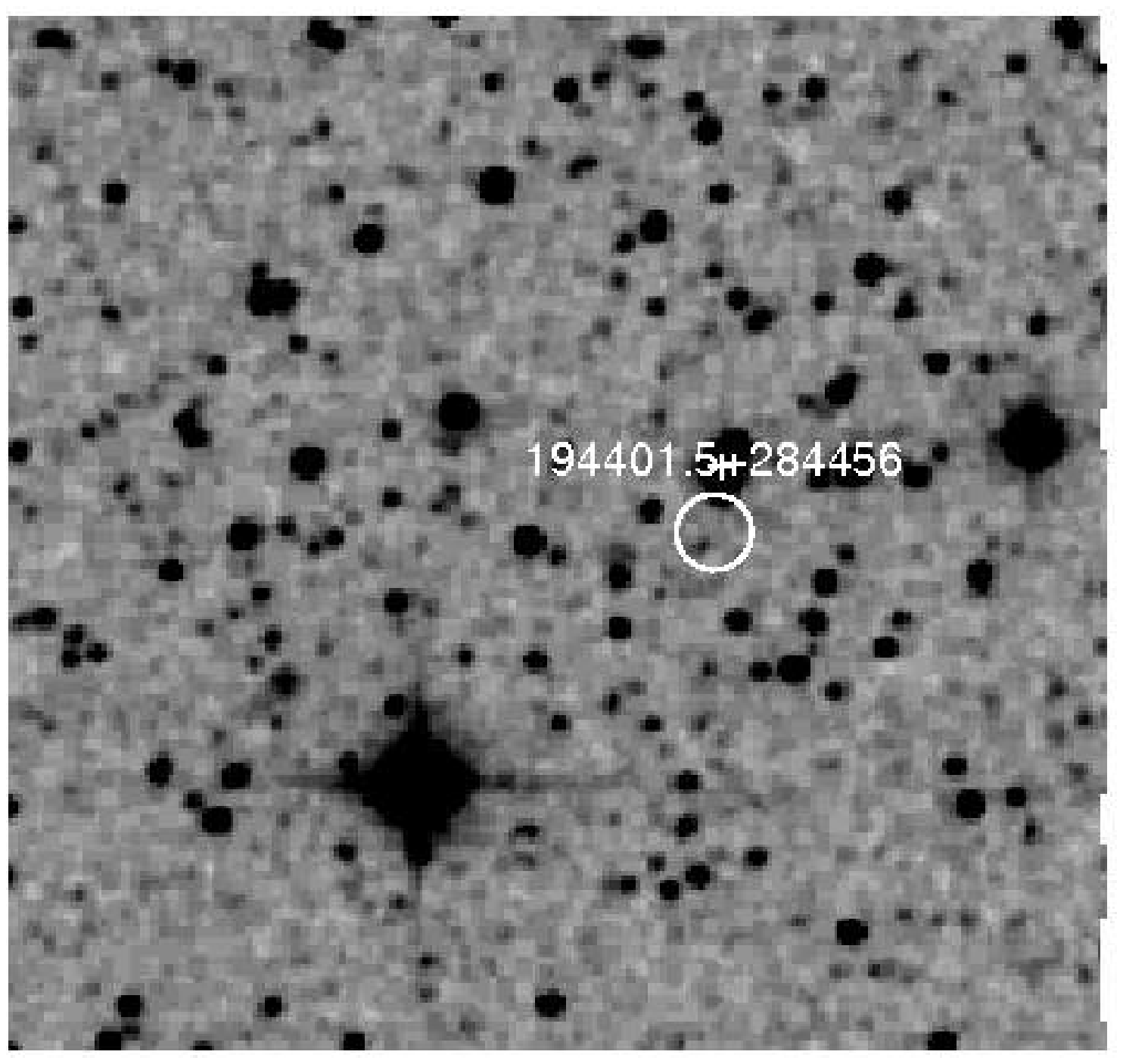}\includegraphics[width=0.5\textwidth]{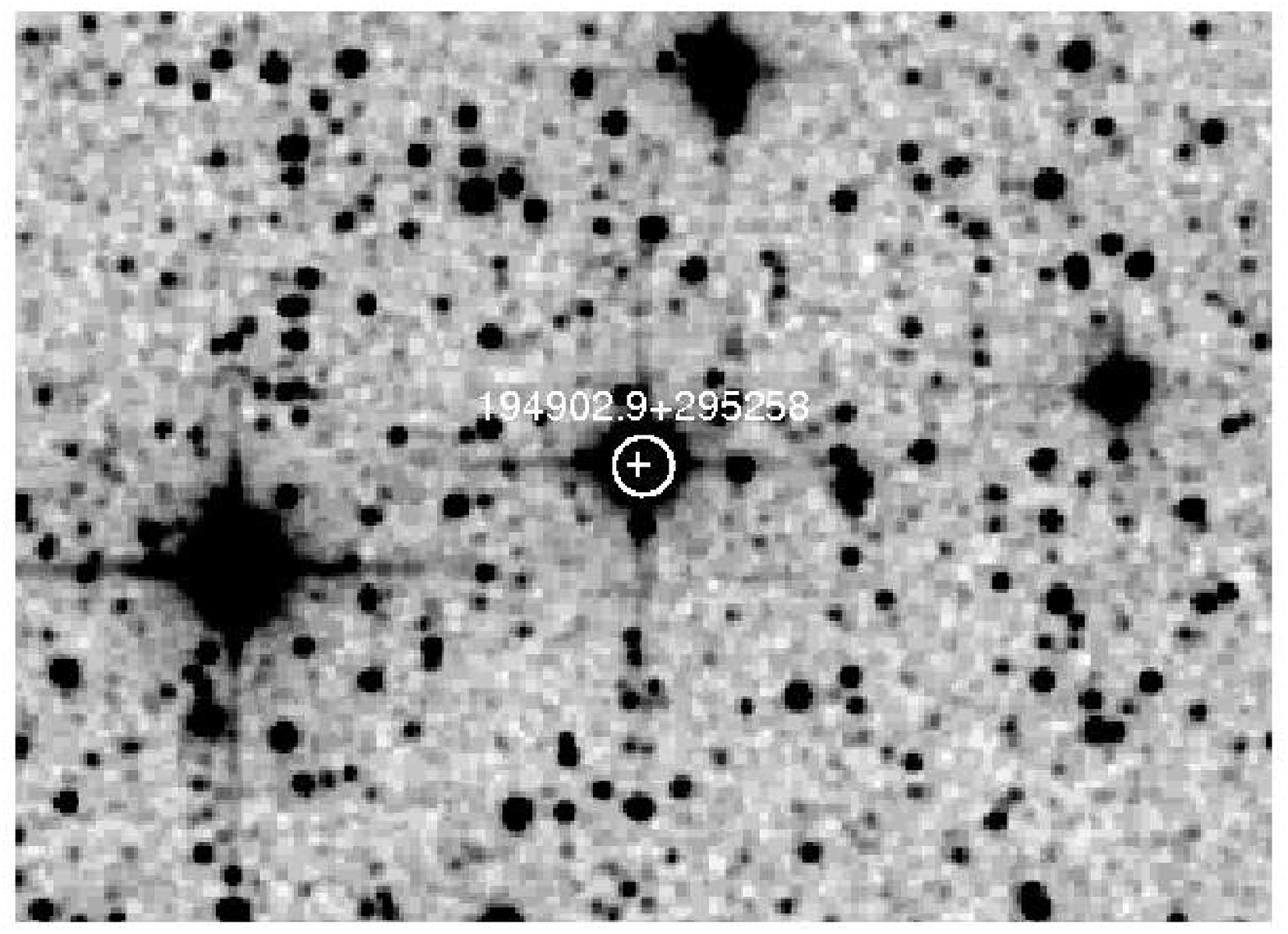}
\includegraphics[width=0.5\textwidth]{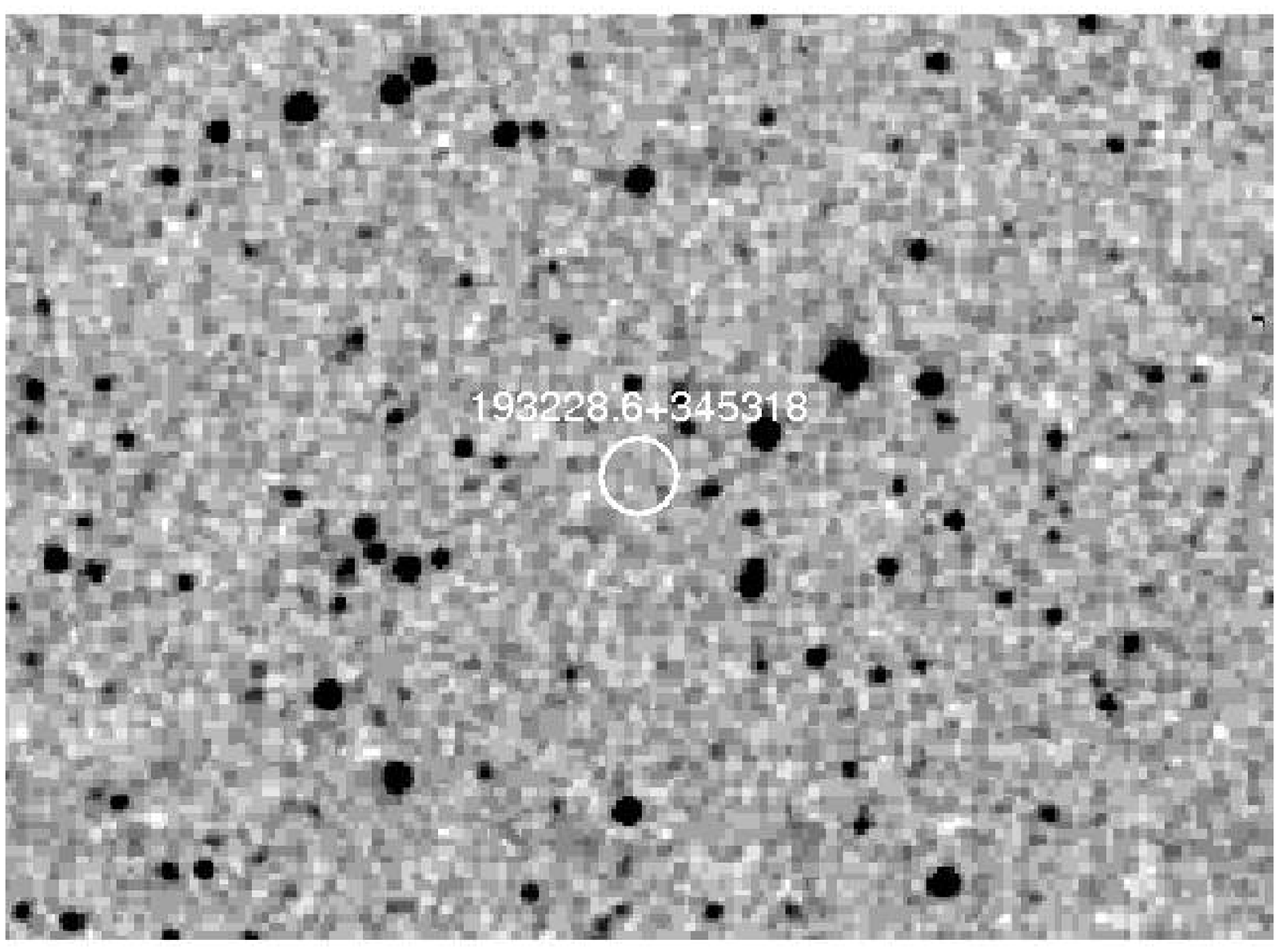}\includegraphics[width=0.5\textwidth]{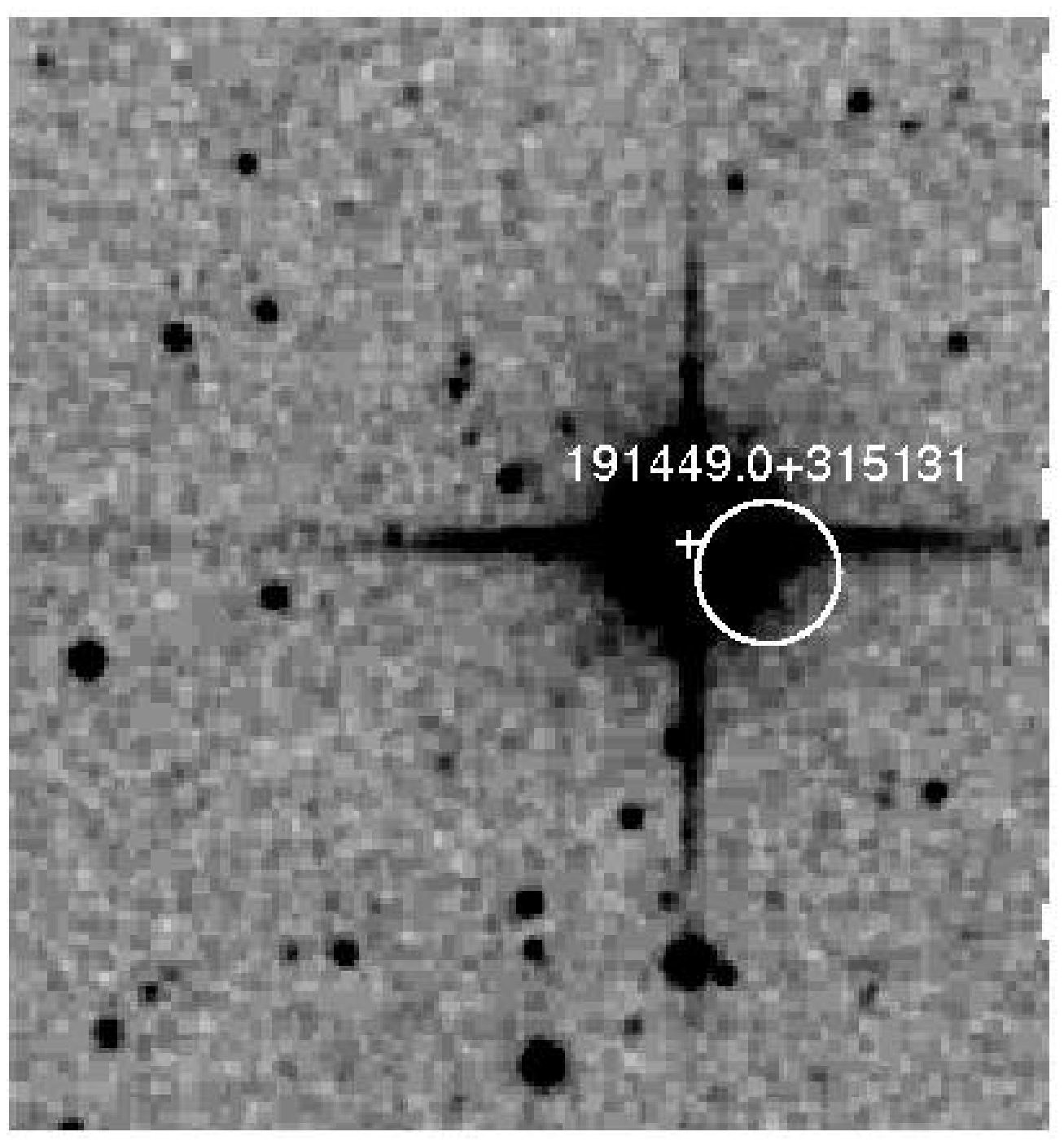}
\includegraphics[width=0.5\textwidth]{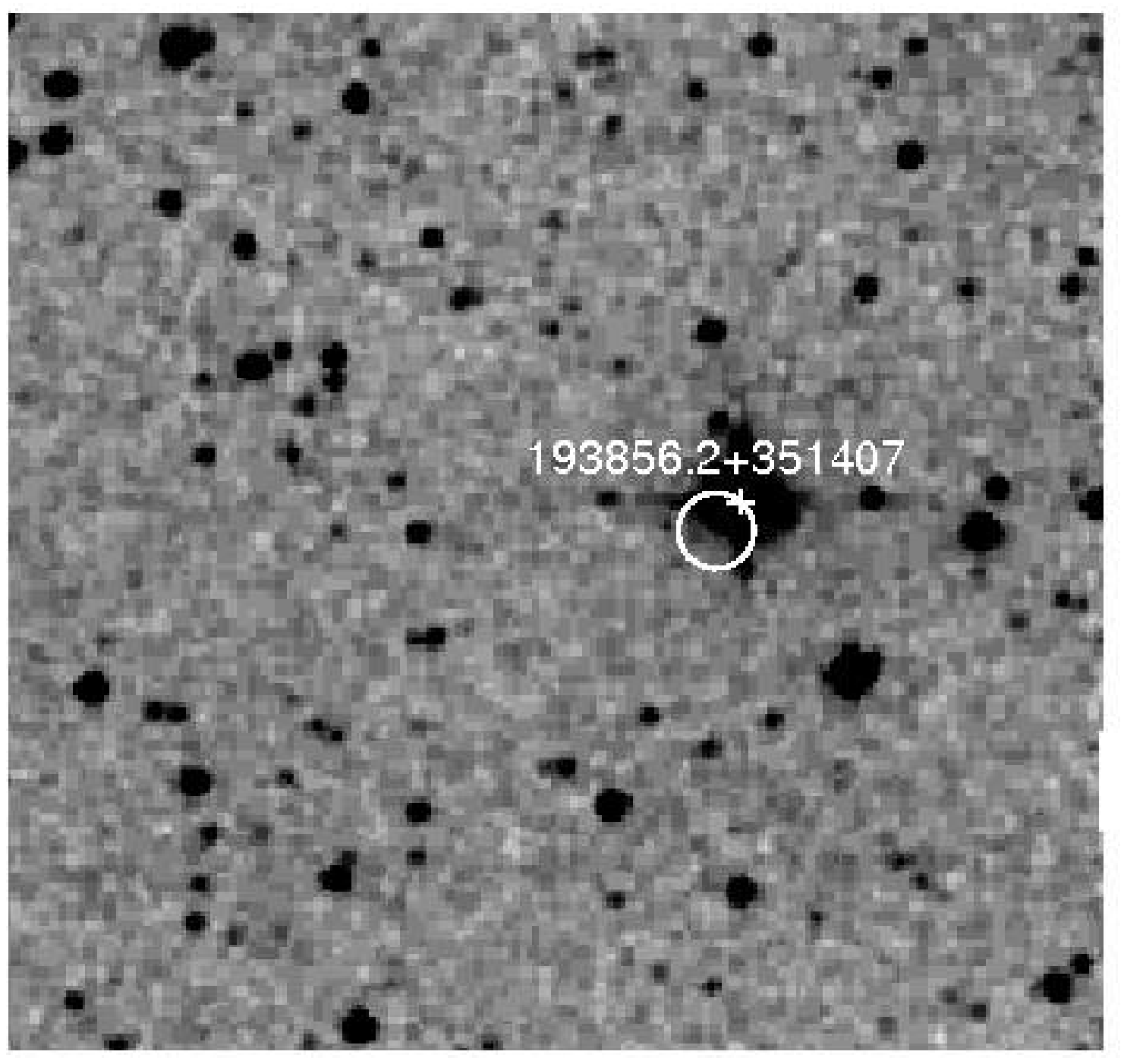}\includegraphics[width=0.5\textwidth]{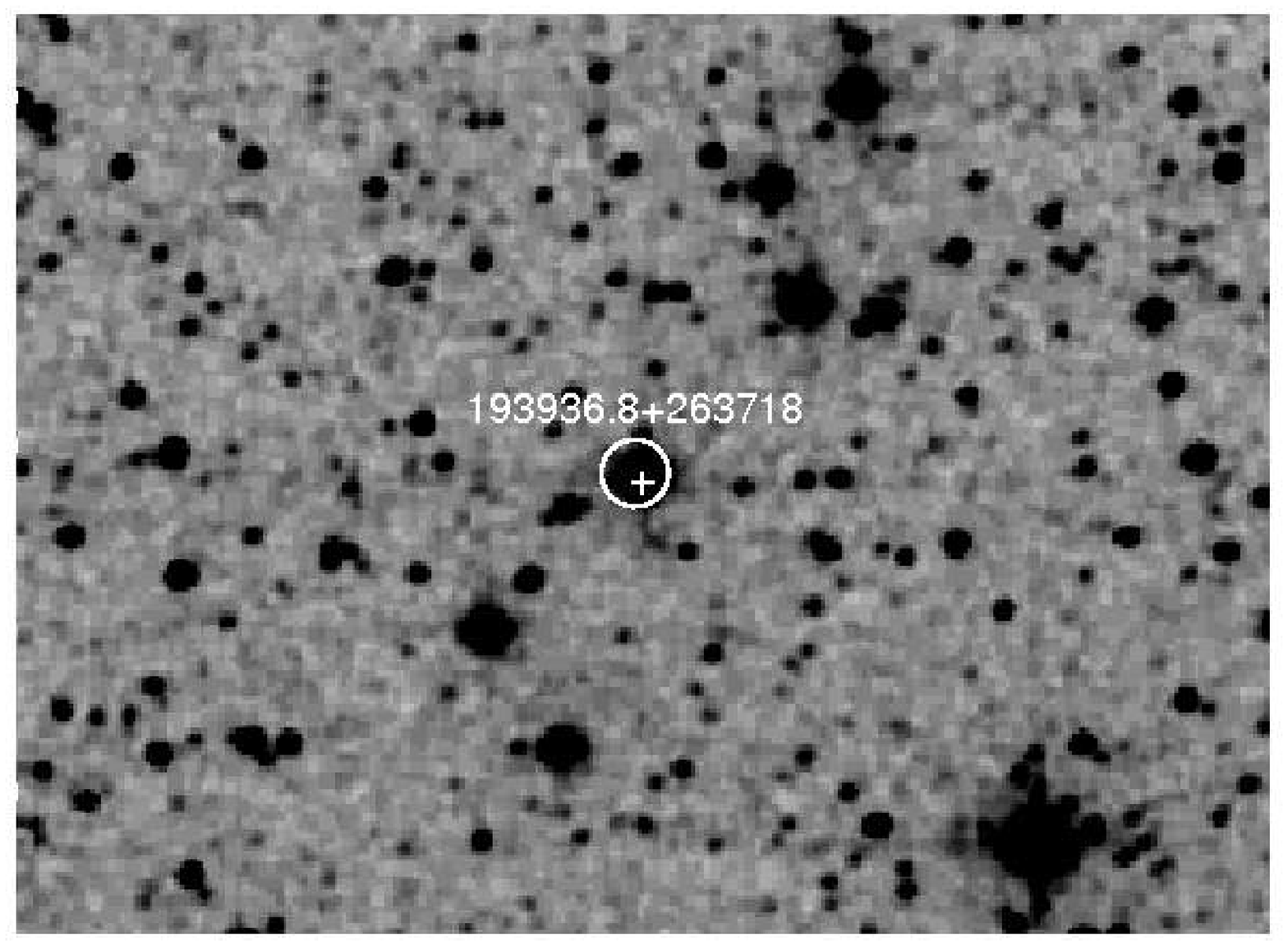}
\caption{2MASS $K_s$-band images of the sources in \snra\ from
  Table~\ref{tab:srcs} (cont.).  The images are $5\arcmin\times3.5\arcmin$, with
  North up and East to the left.  The X-ray position uncertainties are
  indicated by the circles, and the proposed optical counterparts are
  shown by the crosses.\label{fig:opt2}}
\end{figure*}

\begin{figure*}
\centering
\includegraphics[width=0.5\textwidth]{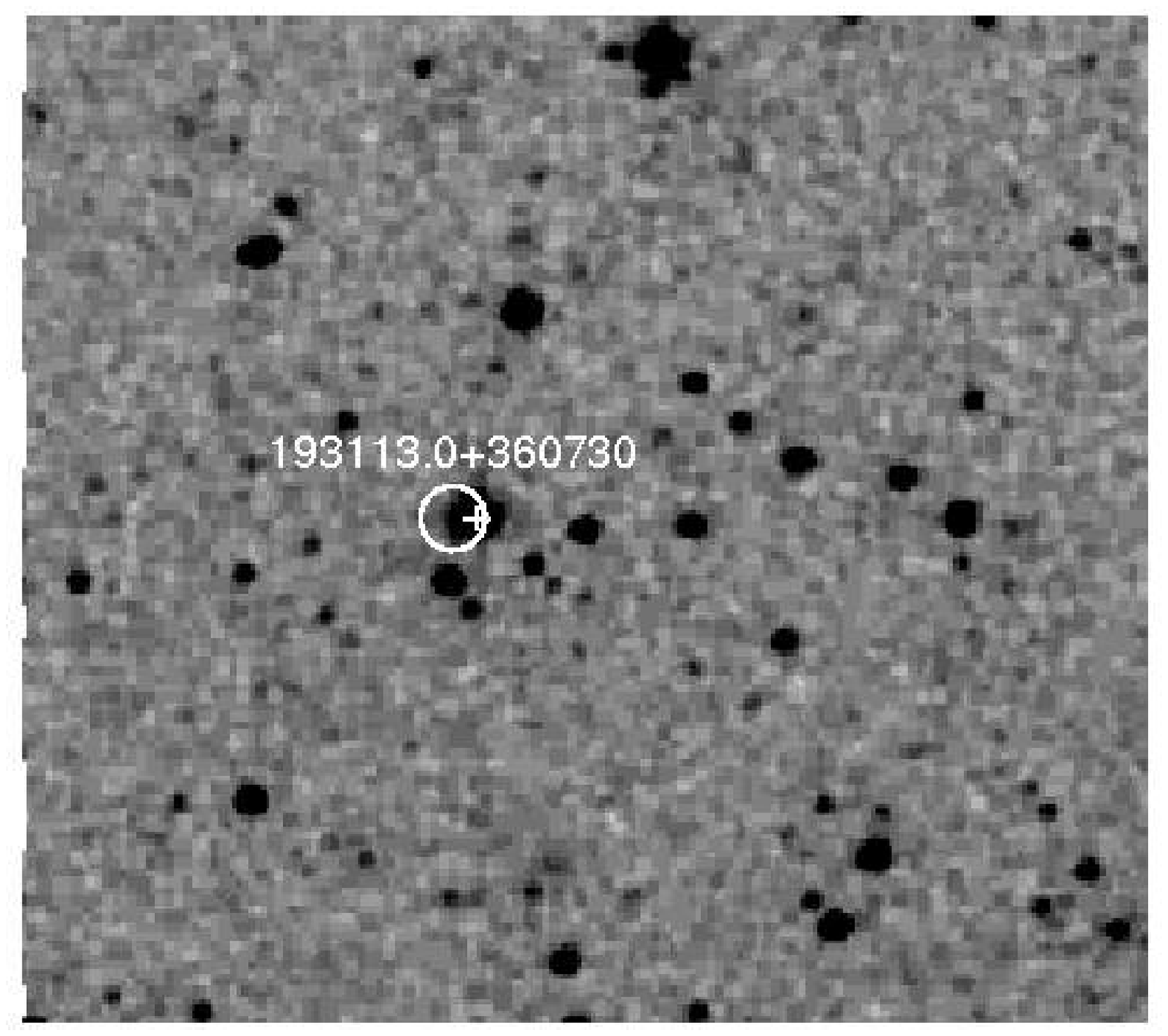}\includegraphics[width=0.5\textwidth]{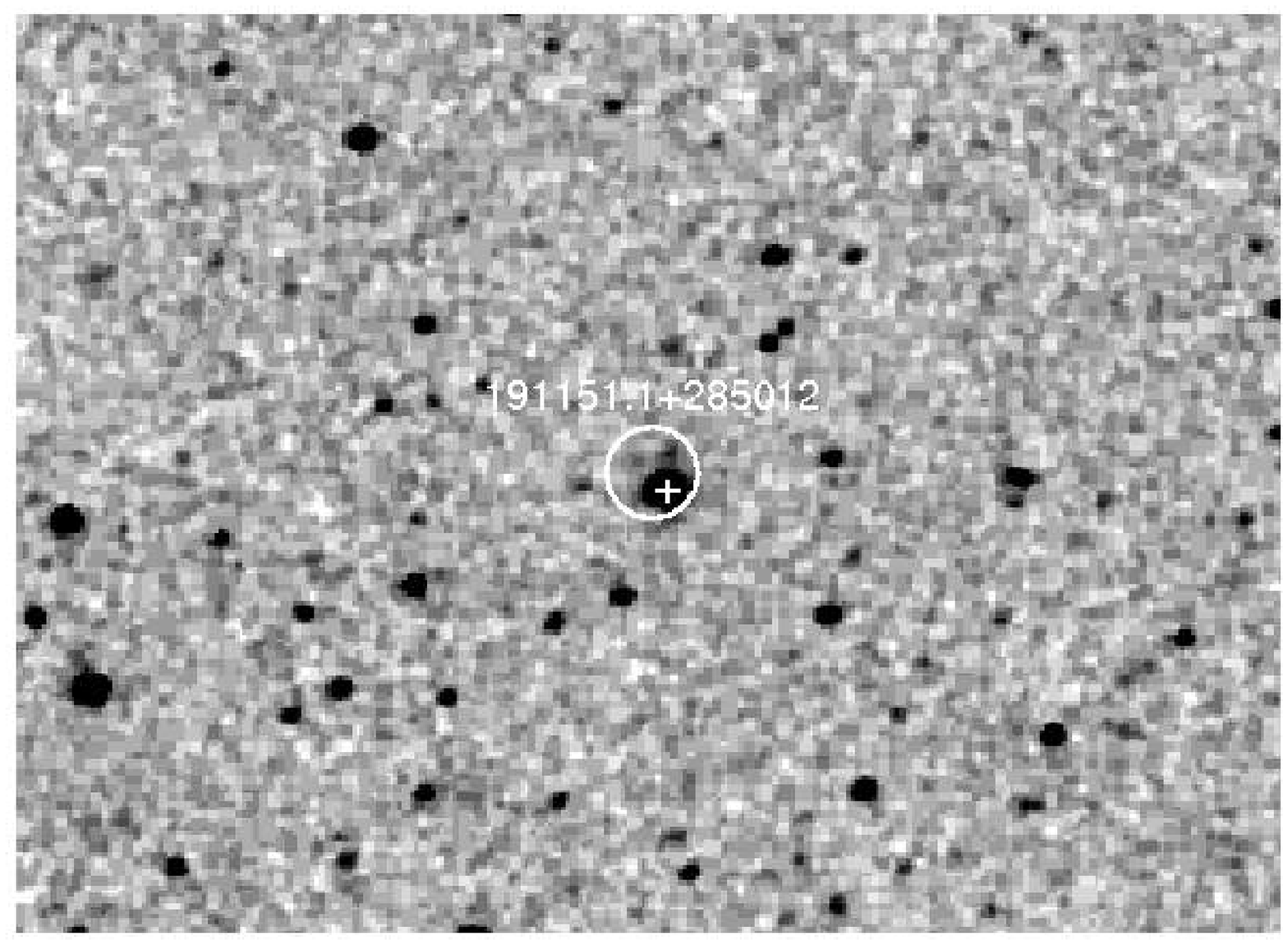}
\caption{2MASS $K_s$-band images of the sources in \snra\ from
  Table~\ref{tab:srcs} (cont.).  The images are $5\arcmin\times3.5\arcmin$, with
  North up and East to the left.  The X-ray position uncertainties are
  indicated by the circles, and the proposed optical counterparts are
  shown by the crosses.\label{fig:opt3}}
\end{figure*}

\begin{figure}
\centering
\includegraphics[width=0.5\textwidth]{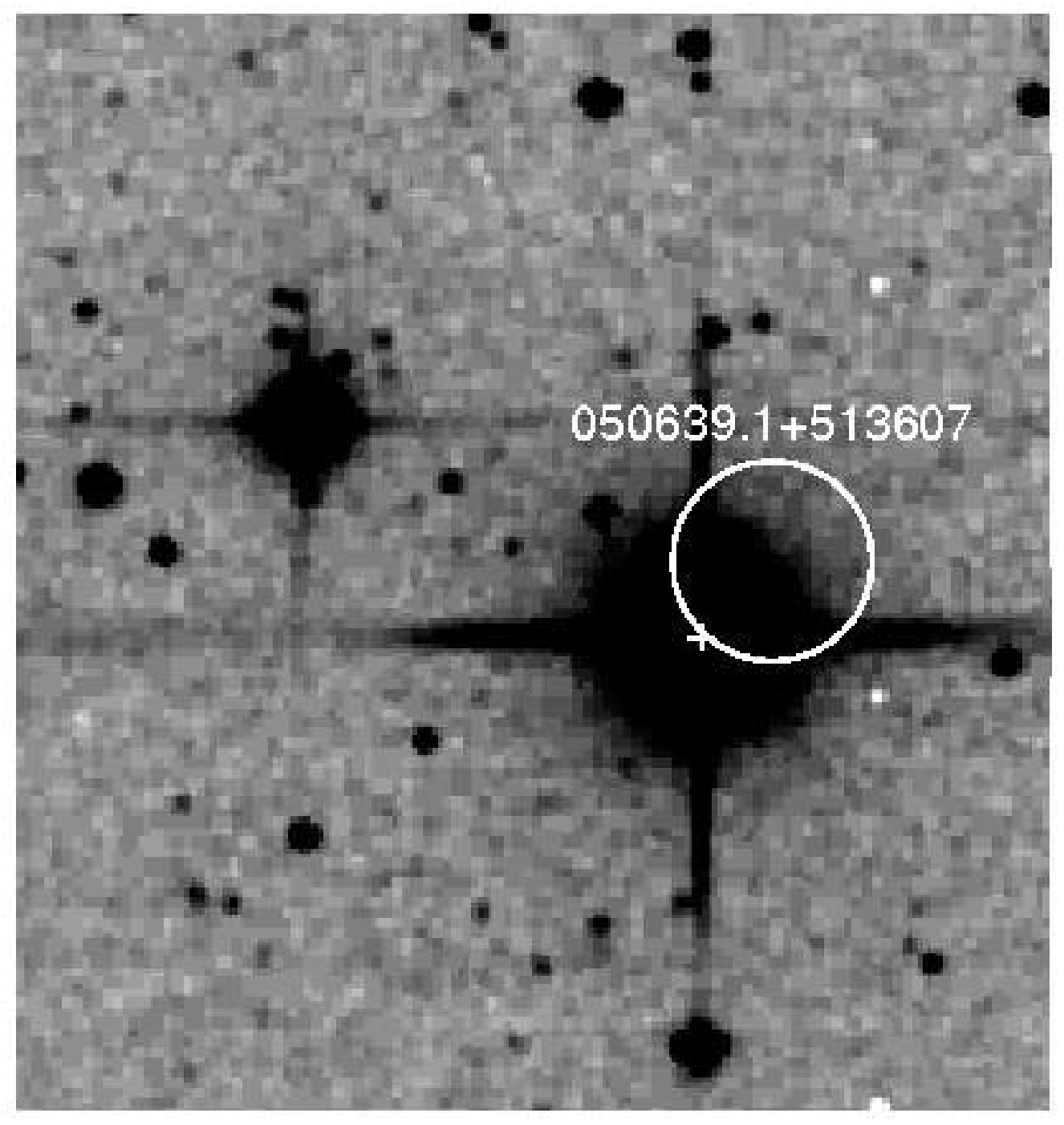}
\caption{2MASS $K_s$-band images of the source in \snrc\ from
  Table~\ref{tab:srcs}.  The images are $5\arcmin\times3.5\arcmin$, with
  North up and East to the left.  The X-ray position uncertainties are
  indicated by the circles, and the proposed optical counterparts are
  shown by the crosses.\label{fig:opt5}}
\end{figure}

\begin{figure*}
\centering
\includegraphics[width=0.5\textwidth]{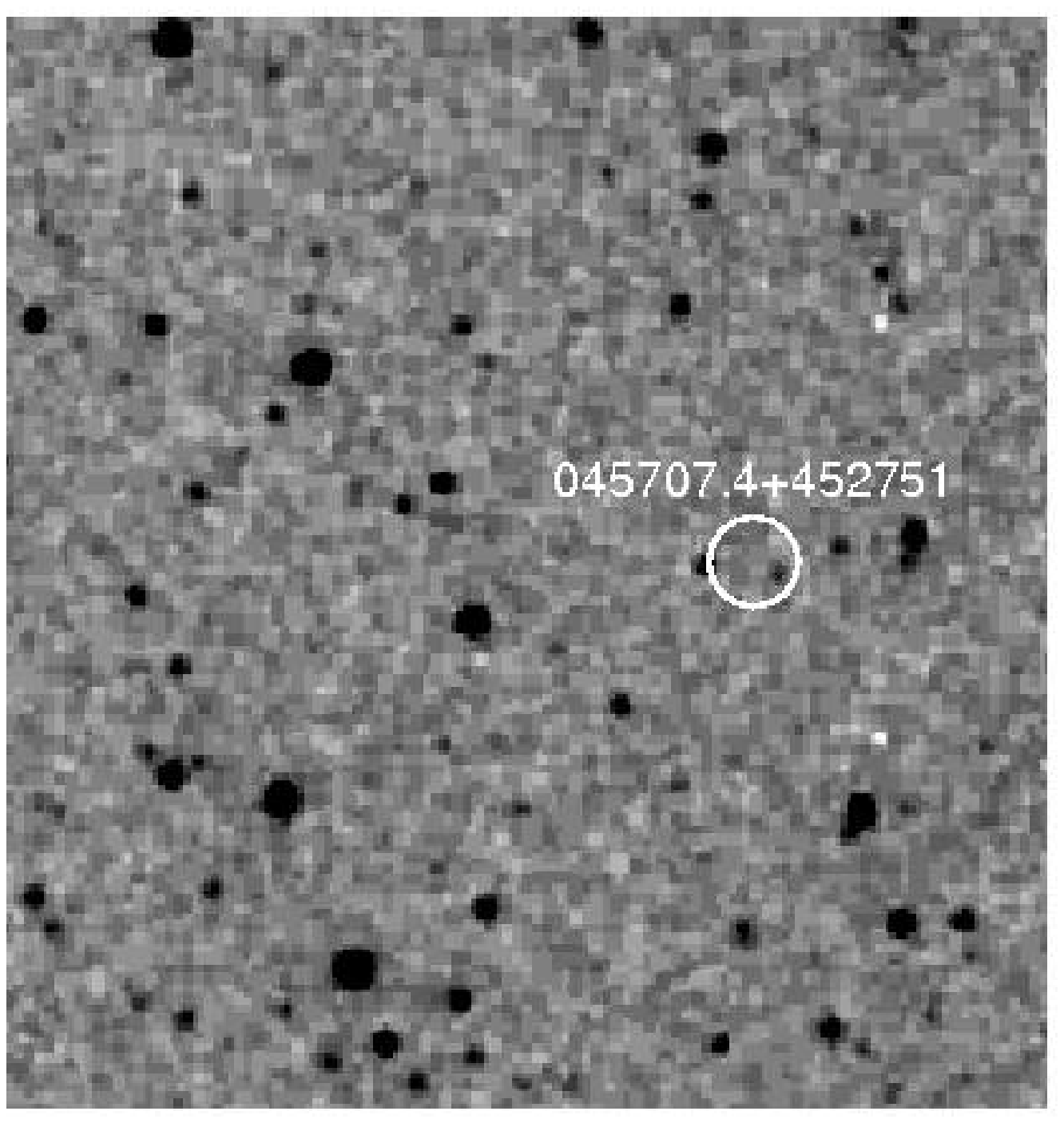}\includegraphics[width=0.5\textwidth]{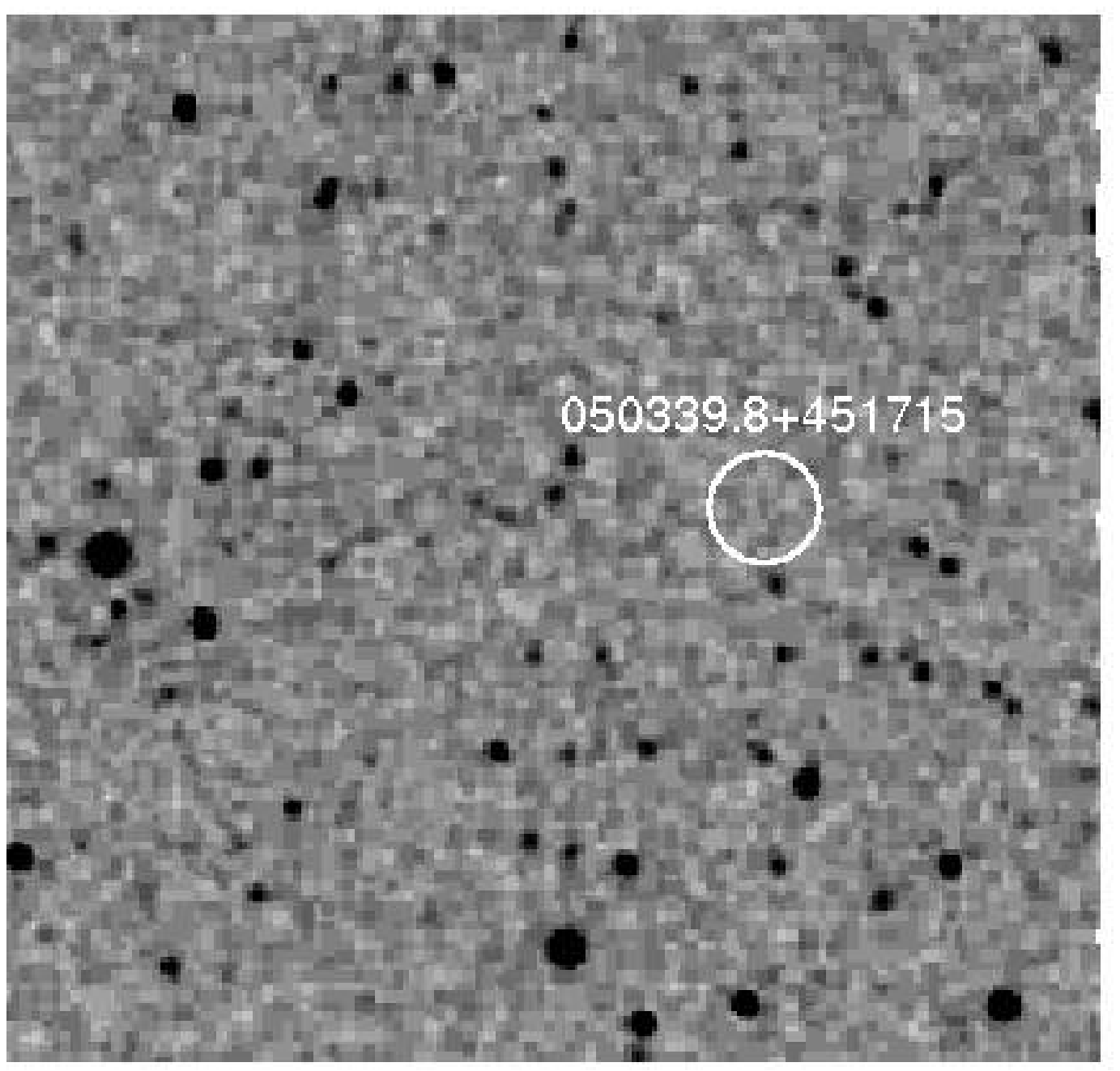}
\includegraphics[width=0.5\textwidth]{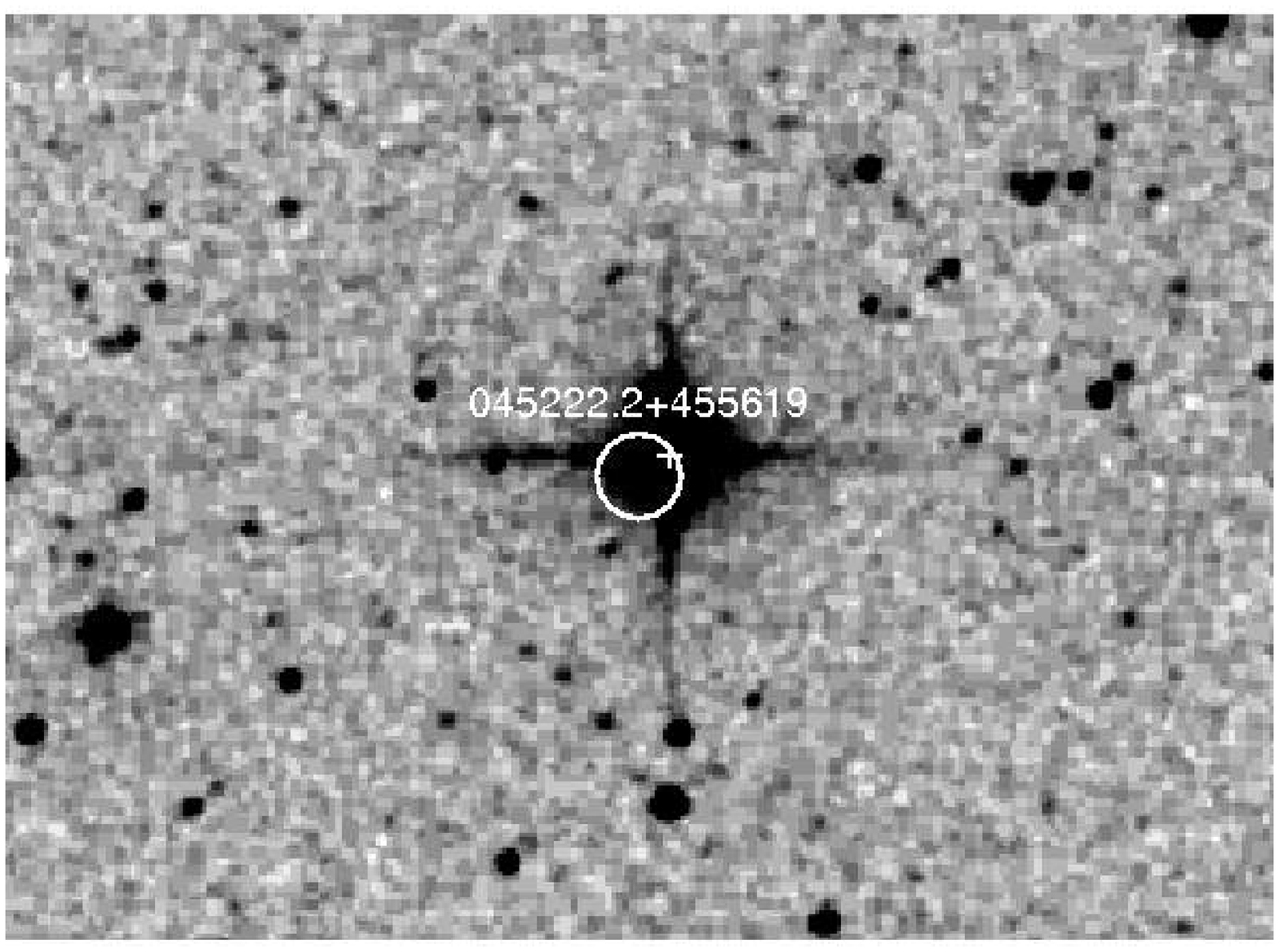}
\caption{2MASS $K_s$-band images of the sources in \snrd\ from
  Table~\ref{tab:srcs}.  The images are $5\arcmin\times3.5\arcmin$, with
  North up and East to the left.  The X-ray position uncertainties are
  indicated by the circles, and the proposed optical counterparts are
  shown by the crosses.\label{fig:opt6}}
\end{figure*}

\begin{figure*}
\centering
\includegraphics[width=0.5\textwidth]{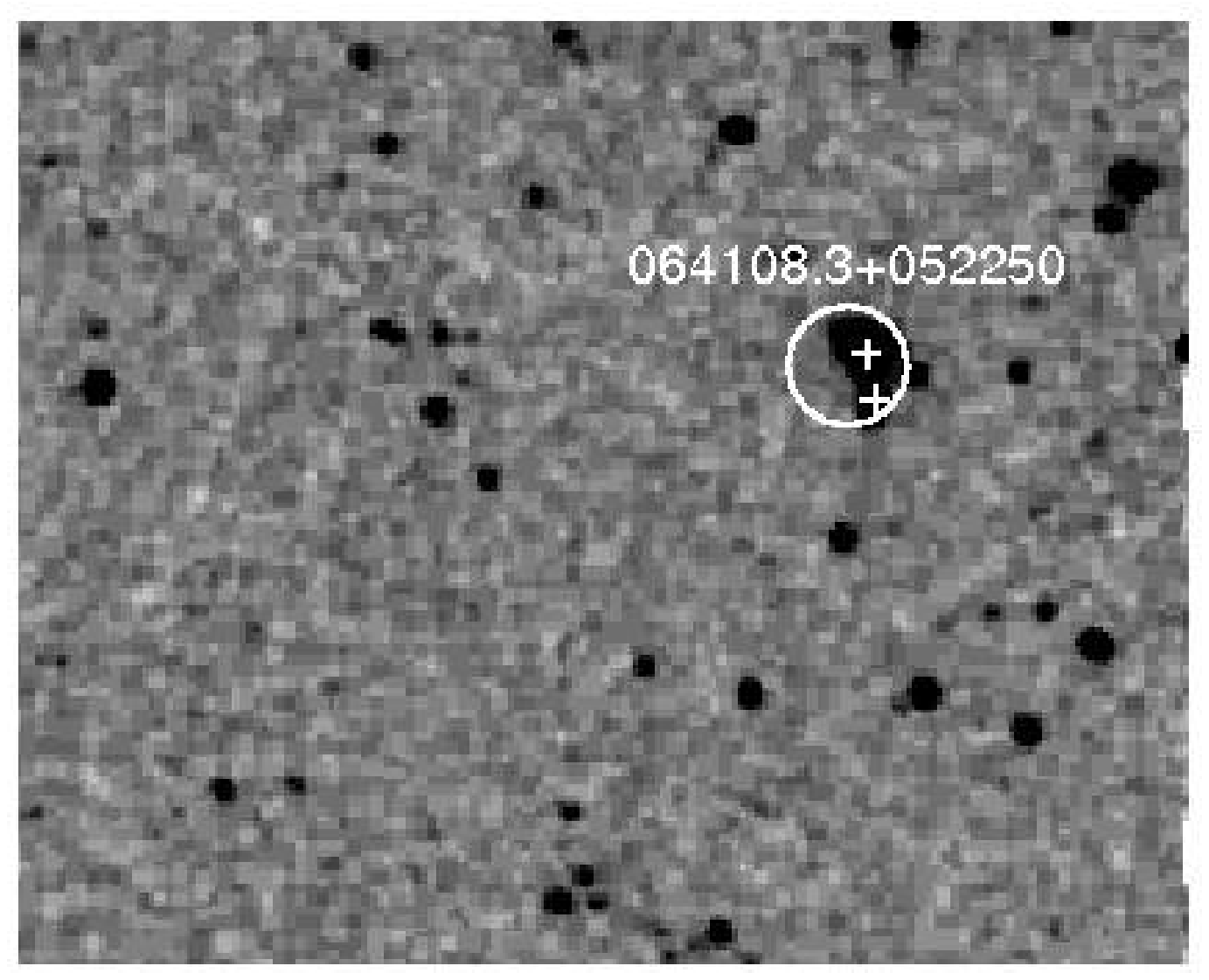}\includegraphics[width=0.5\textwidth]{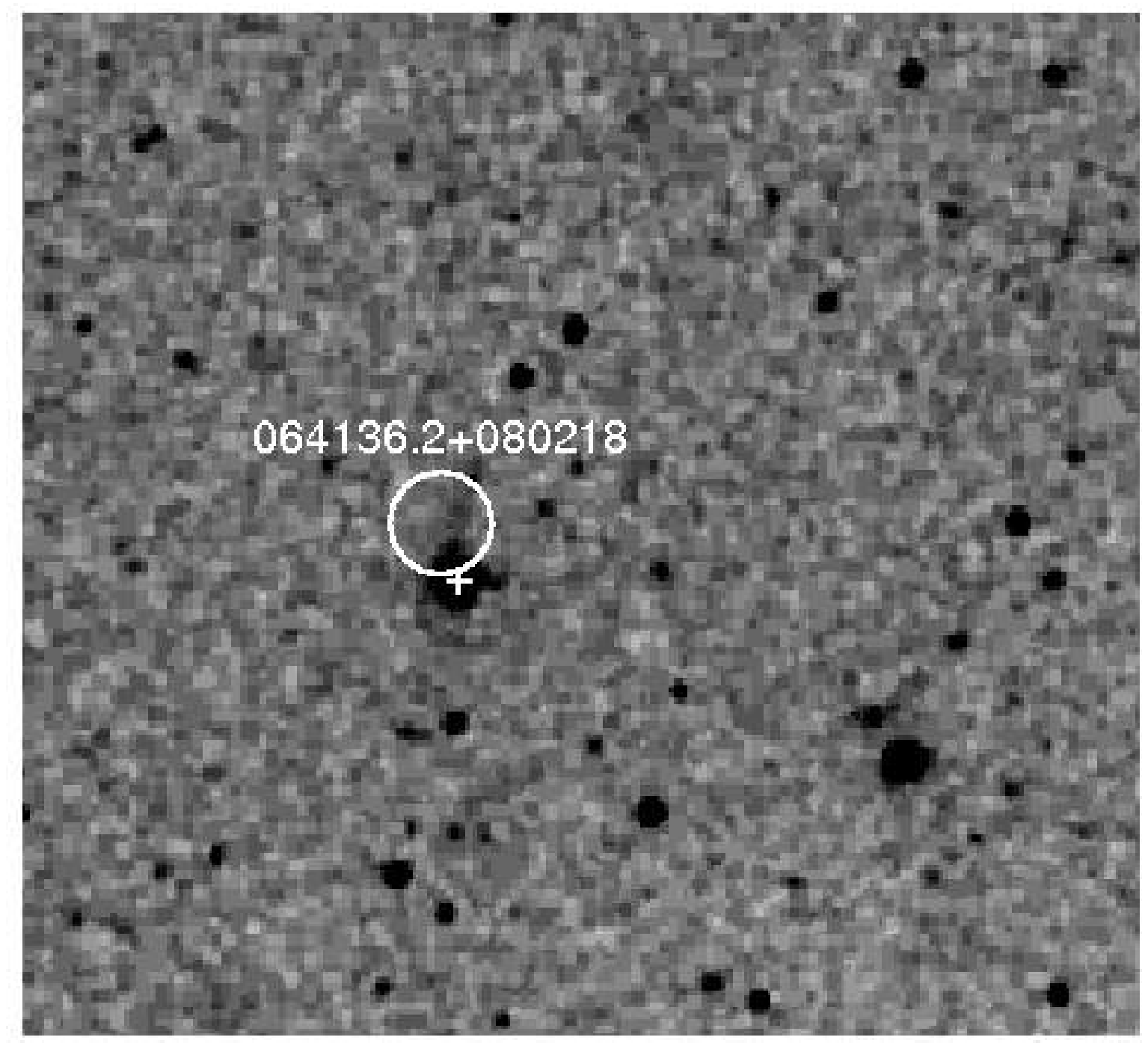}
\includegraphics[width=0.5\textwidth]{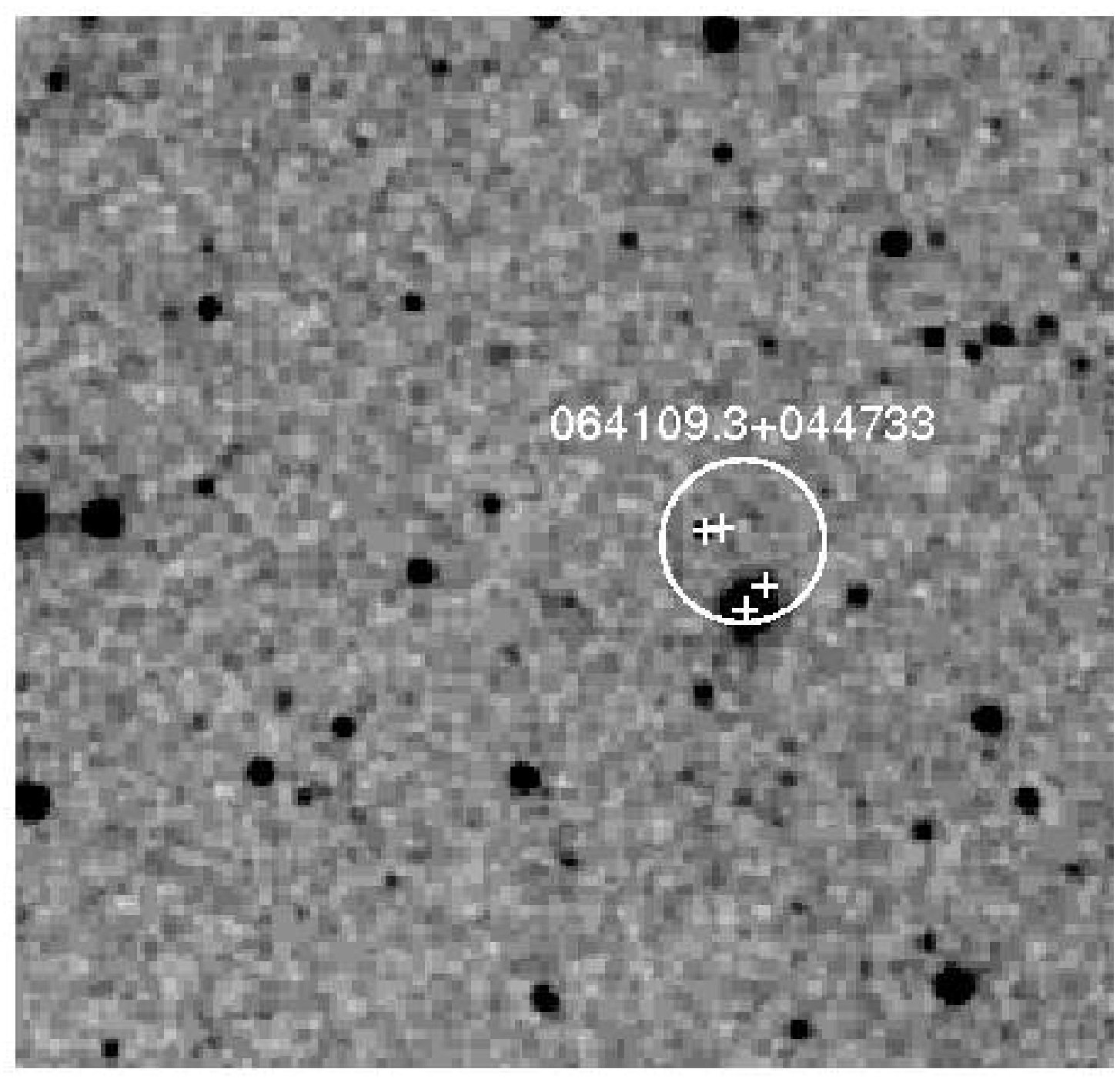}\includegraphics[width=0.5\textwidth]{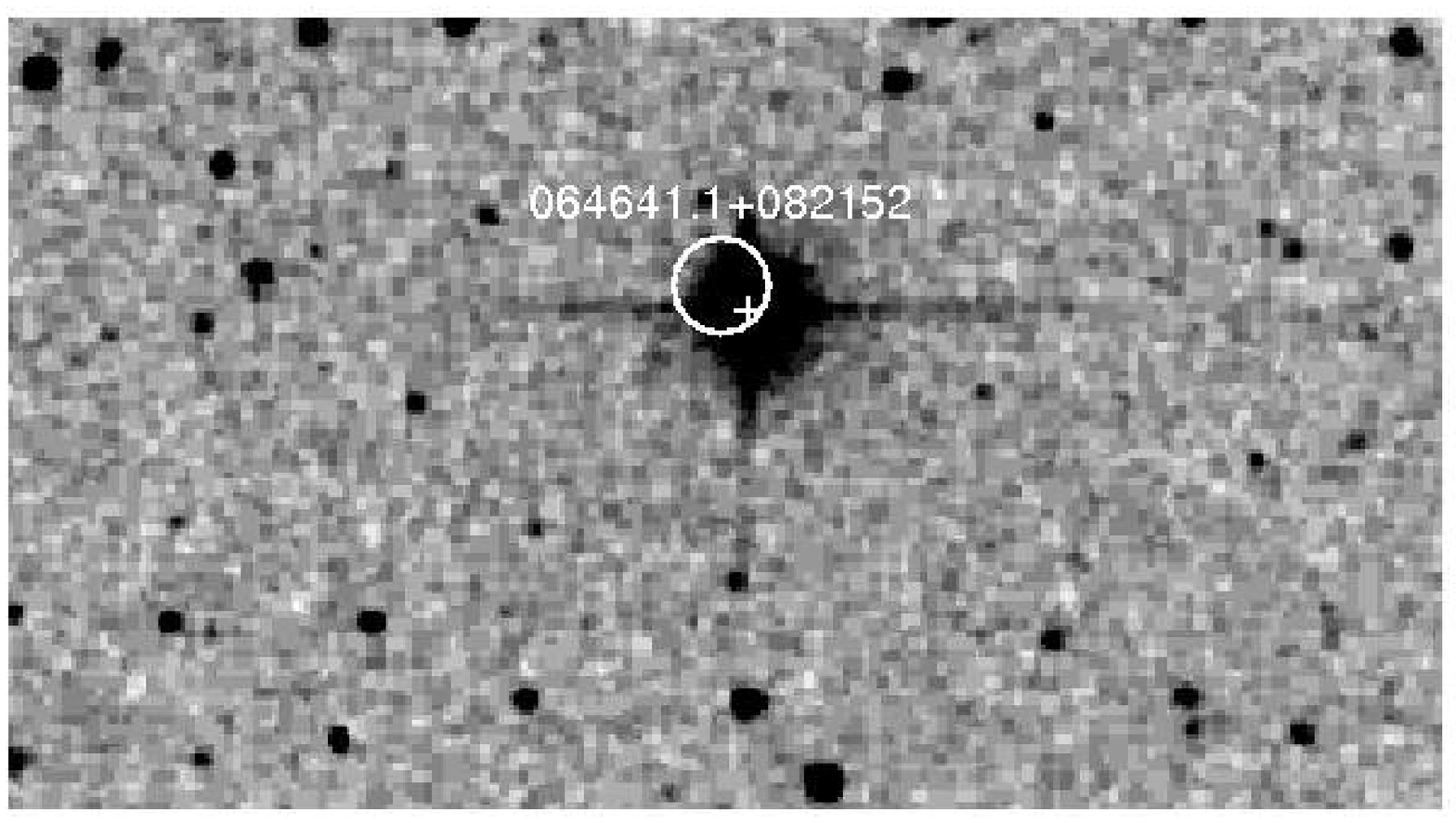}
\includegraphics[width=0.5\textwidth]{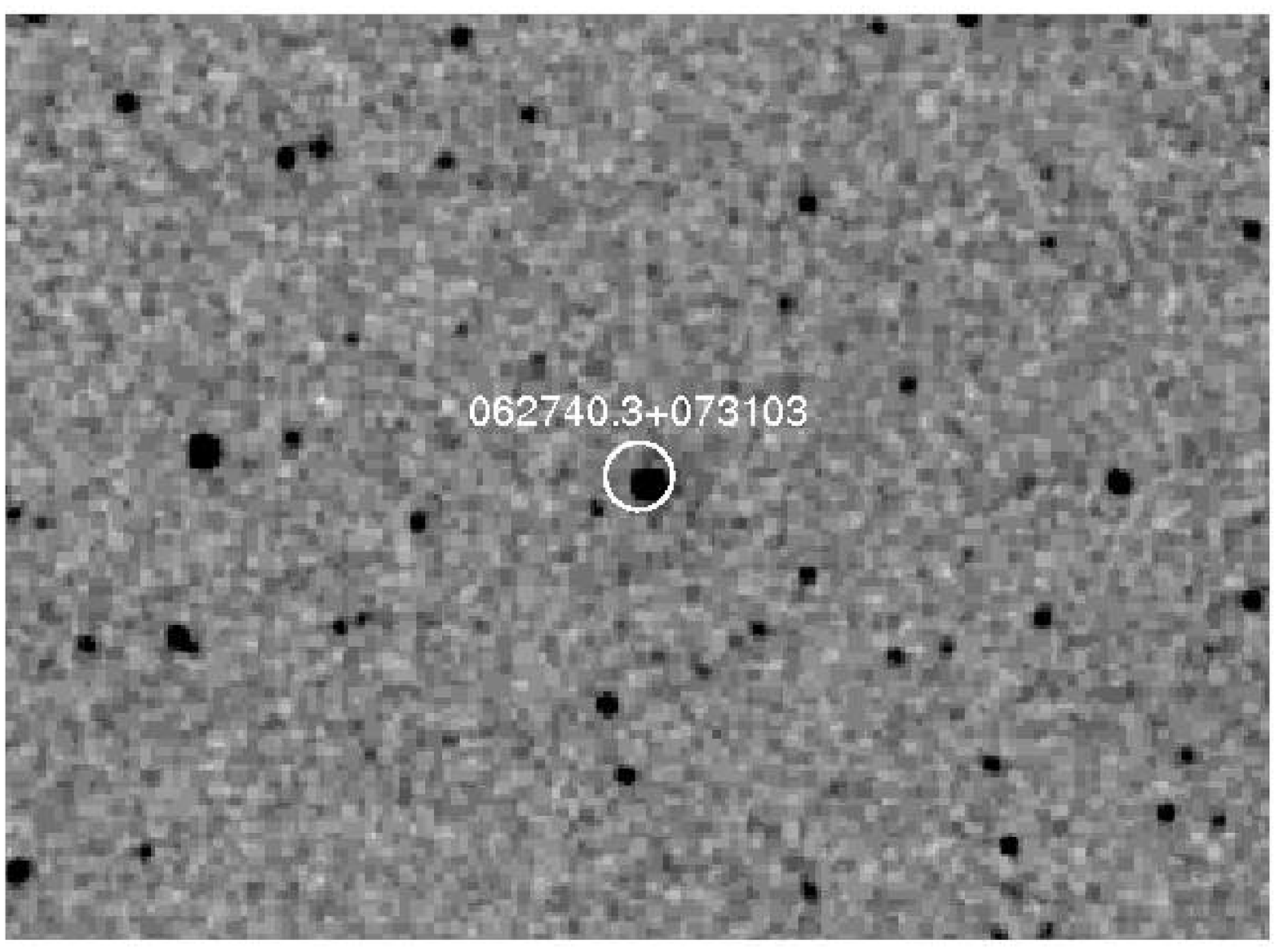}\includegraphics[width=0.5\textwidth]{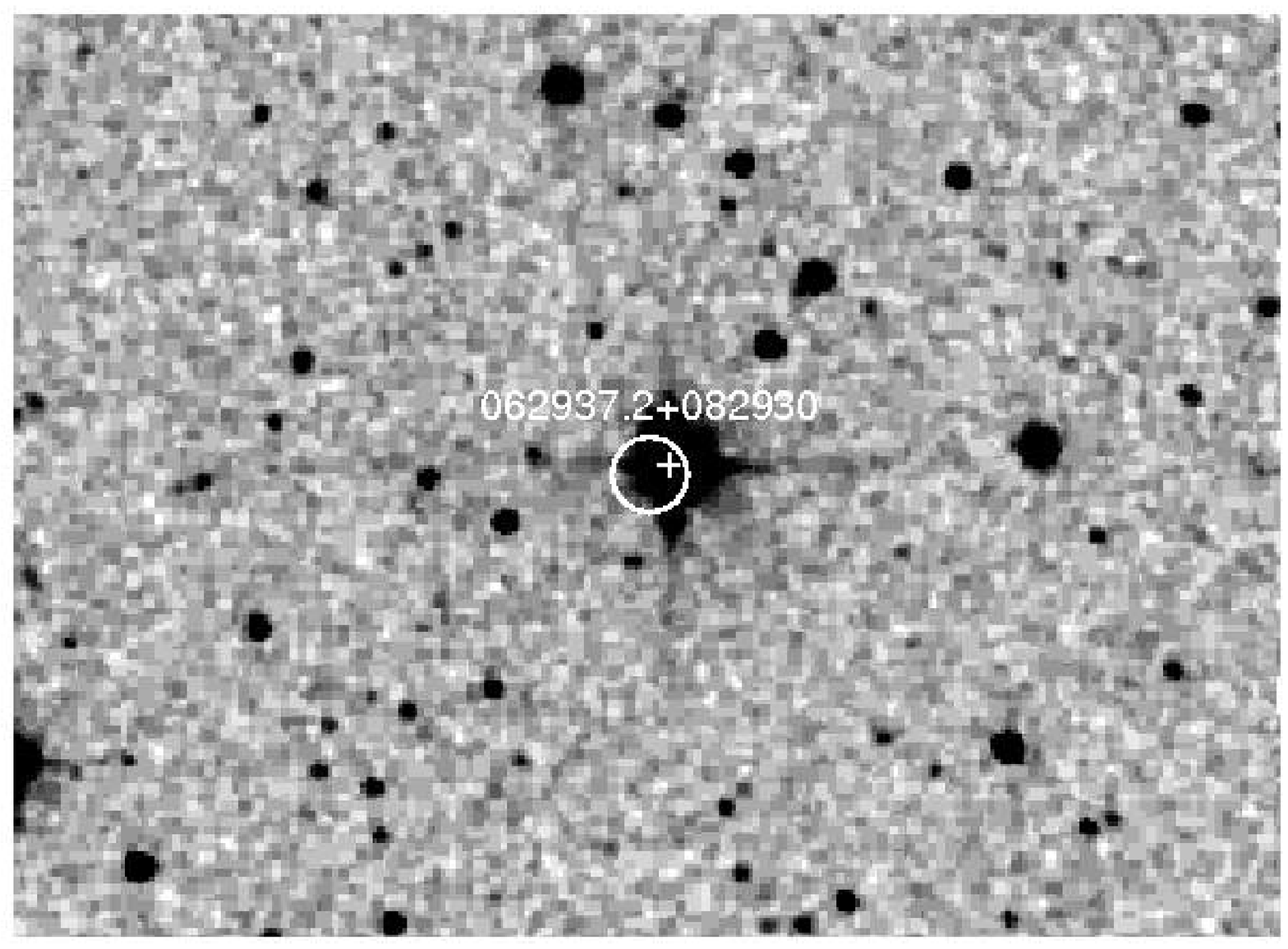}
\caption{2MASS $K_s$-band images of the sources in \snre\ from
  Table~\ref{tab:srcs}.  The images are $5\arcmin\times3.5\arcmin$, with
  North up and East to the left.  The X-ray position uncertainties are
  indicated by the circles, and the proposed optical counterparts are
  shown by the crosses.\label{fig:opt7}}
\end{figure*}

\begin{figure*}
\centering
\includegraphics[width=0.5\textwidth]{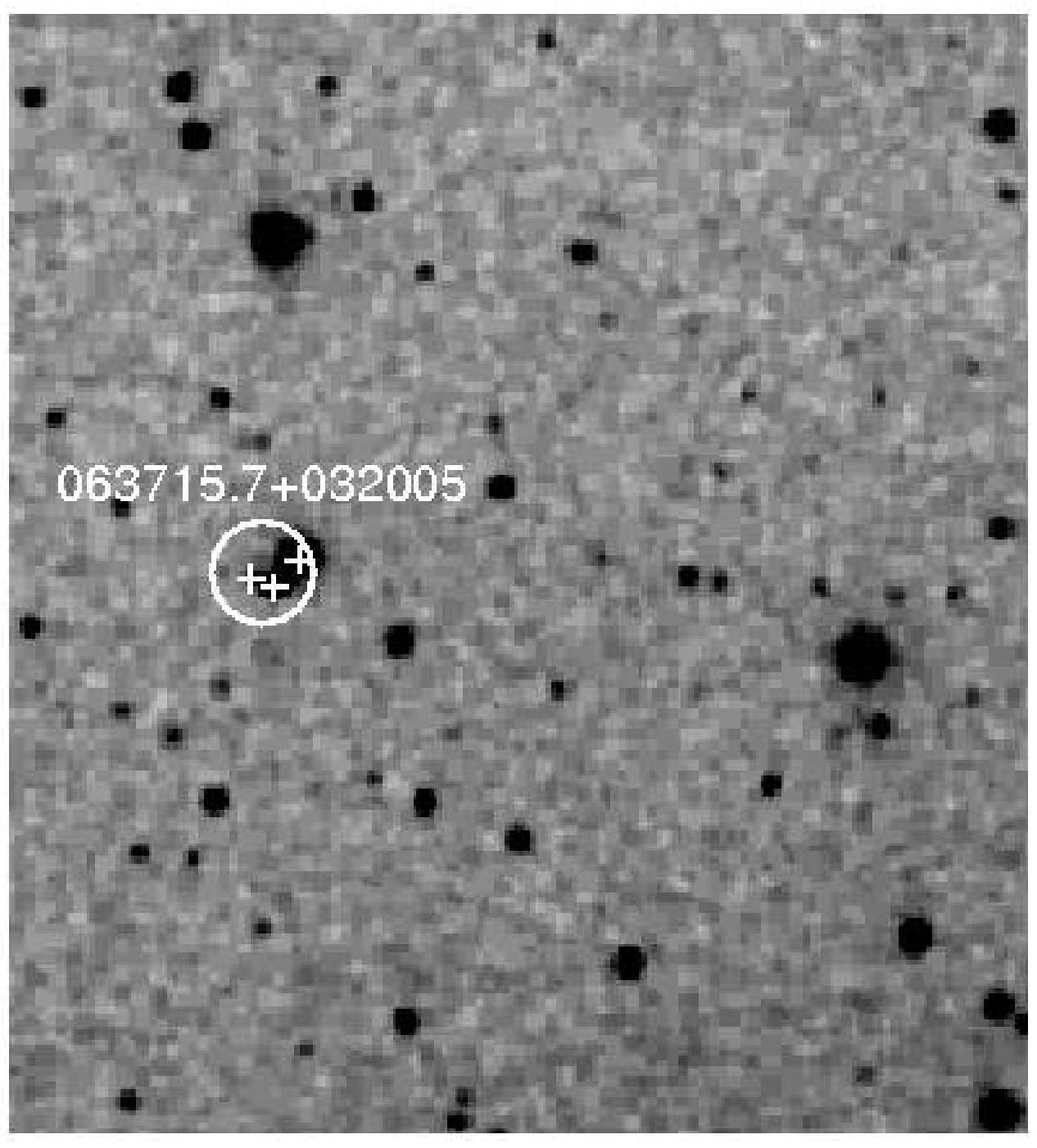}\includegraphics[width=0.5\textwidth]{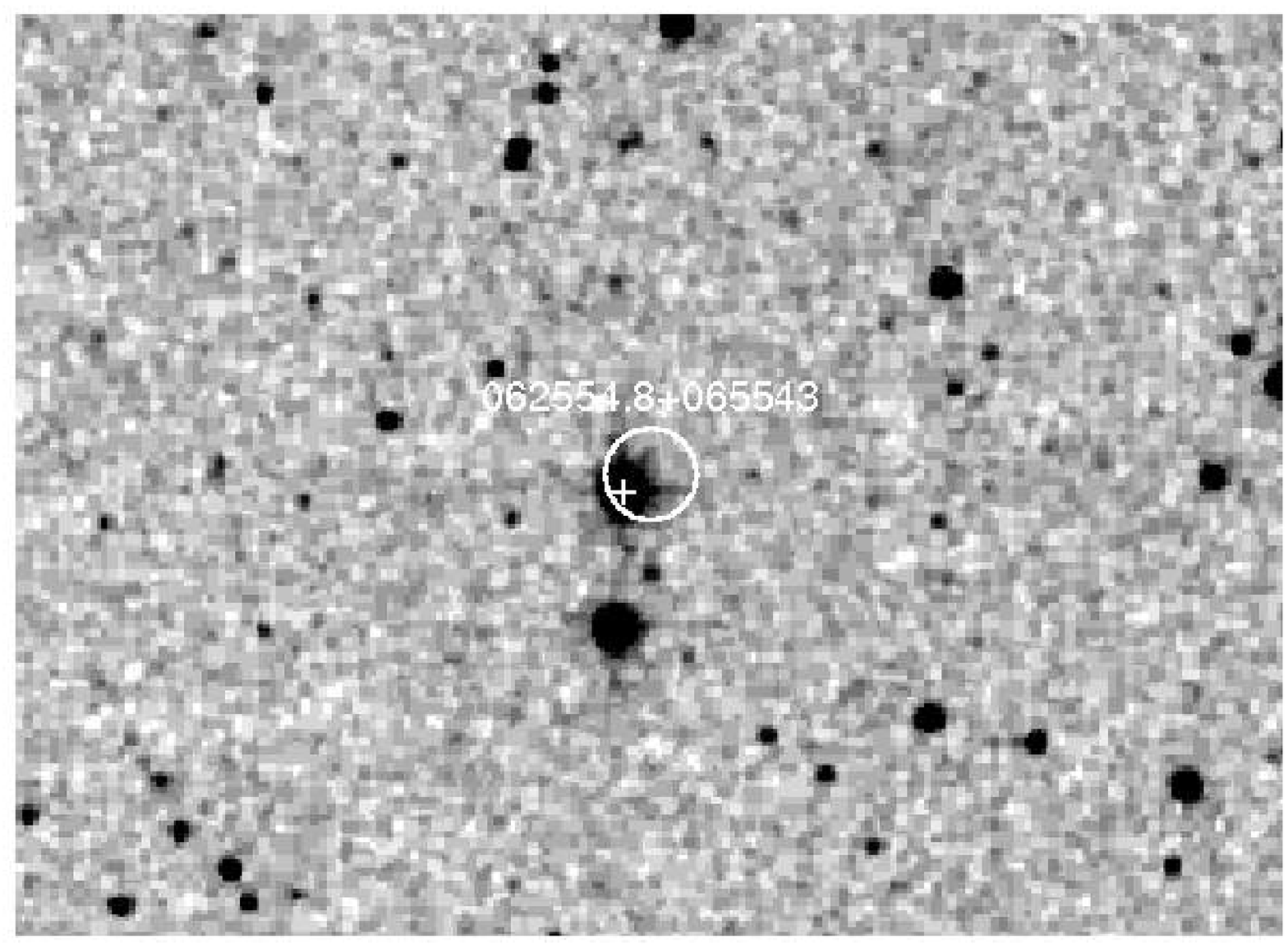}
\caption{2MASS $K_s$-band images of the sources in \snre\ from
  Table~\ref{tab:srcs} (cont.).  The images are $5\arcmin\times3.5\arcmin$, with
  North up and East to the left.  The X-ray position uncertainties are
  indicated by the circles, and the proposed optical counterparts are
  shown by the crosses.\label{fig:opt8}}
\end{figure*}

\begin{figure*}
\centering
\includegraphics[width=0.5\textwidth]{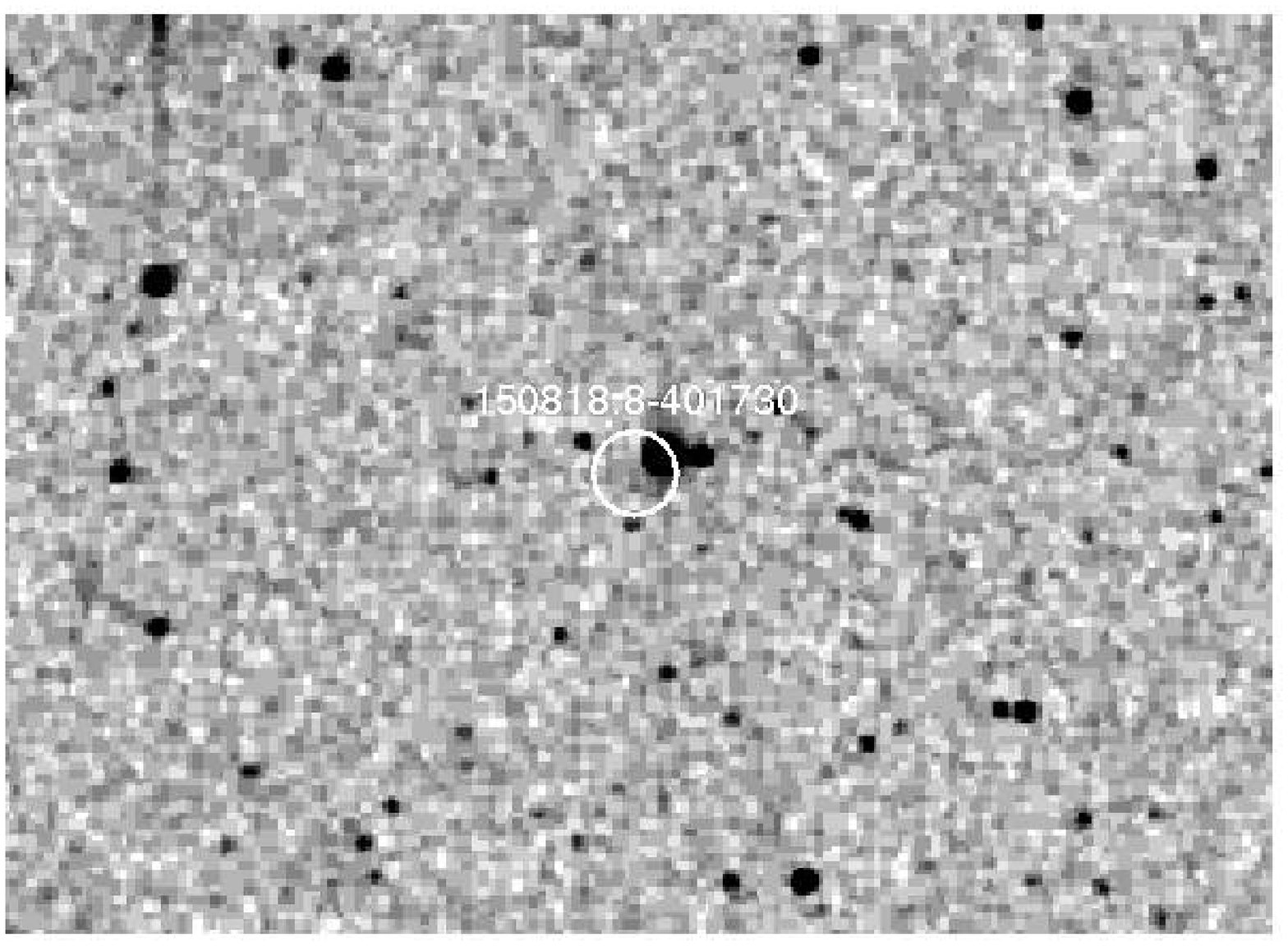}\includegraphics[width=0.5\textwidth]{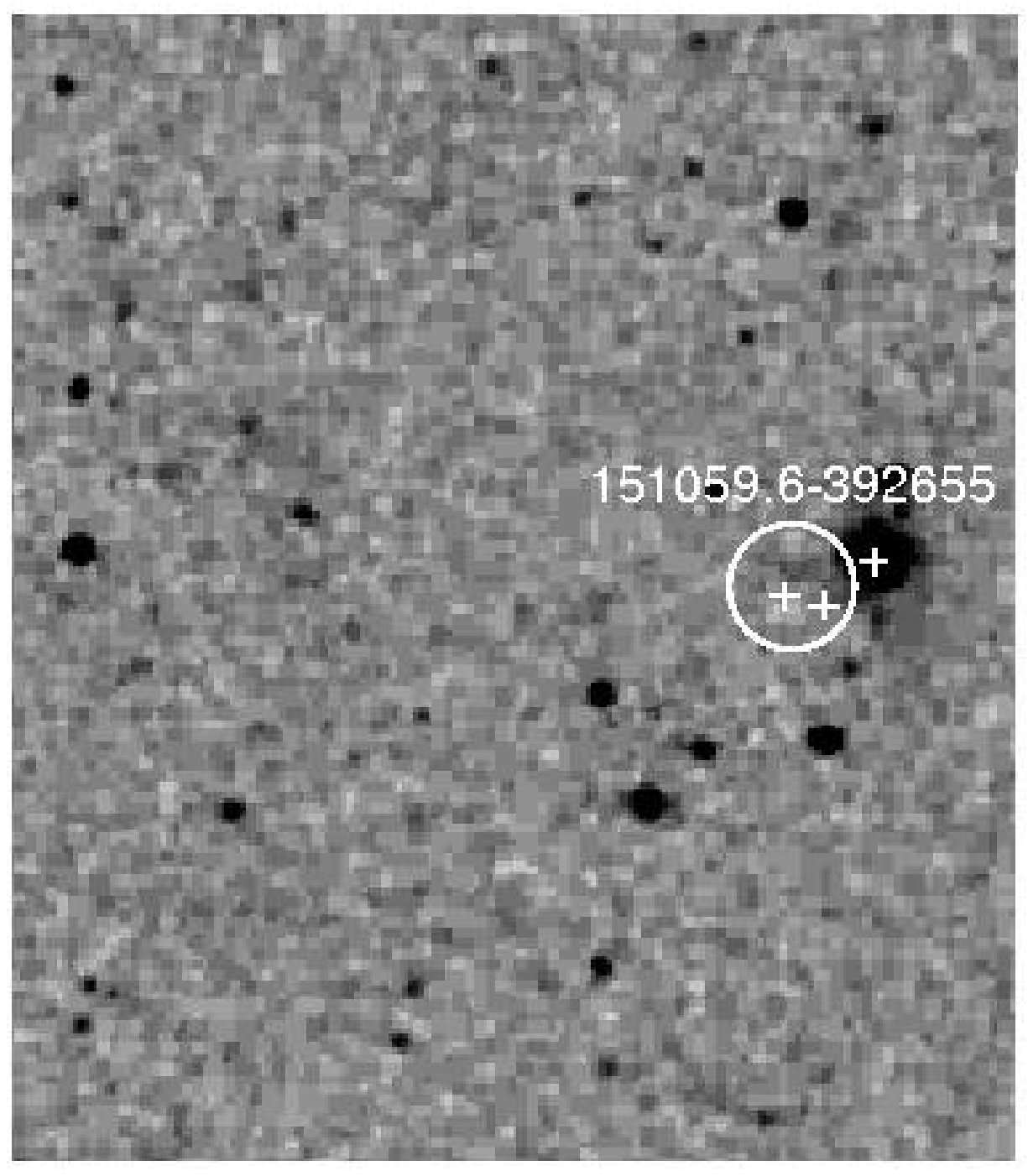}
\includegraphics[width=0.5\textwidth]{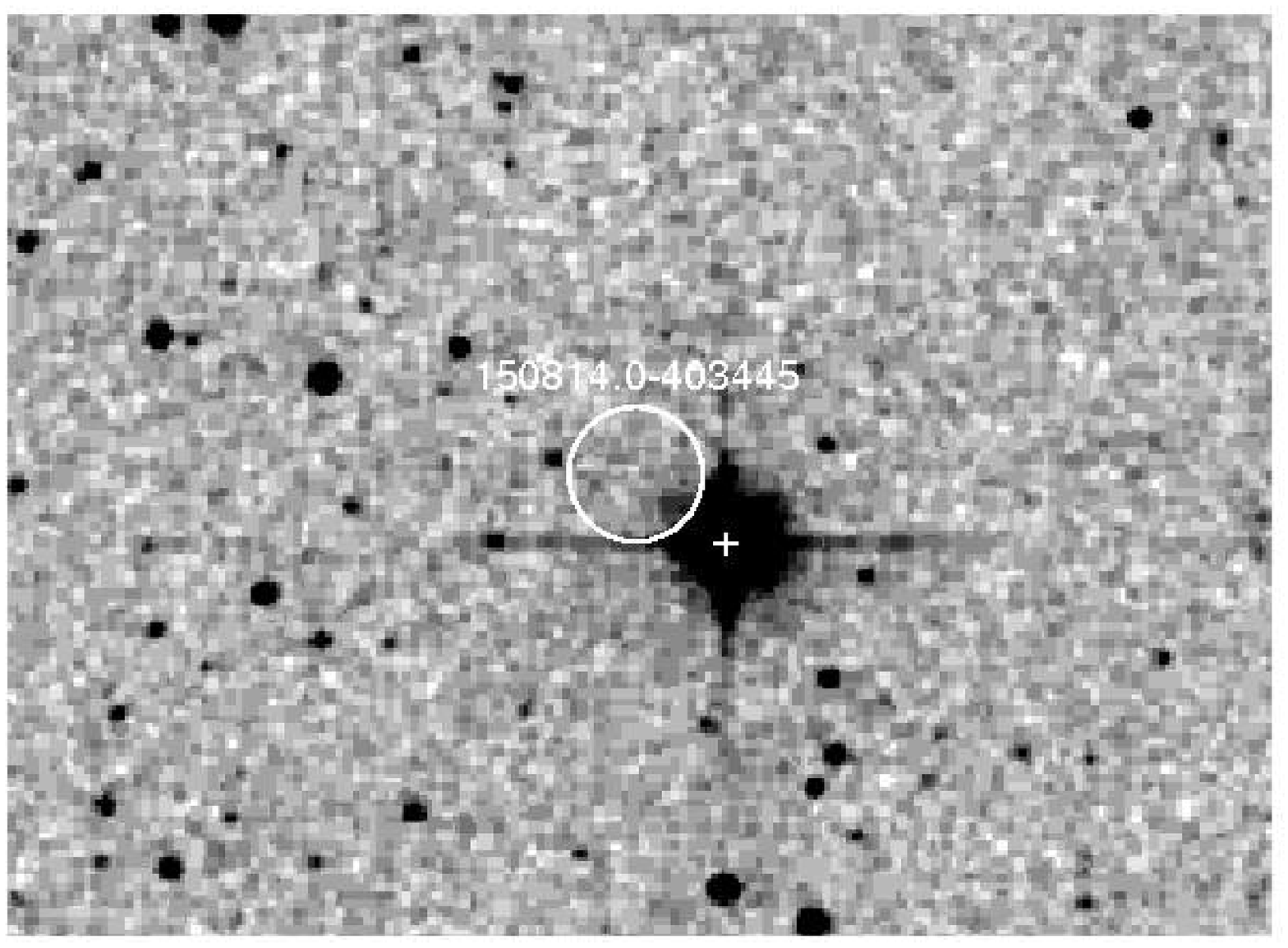}\includegraphics[width=0.5\textwidth]{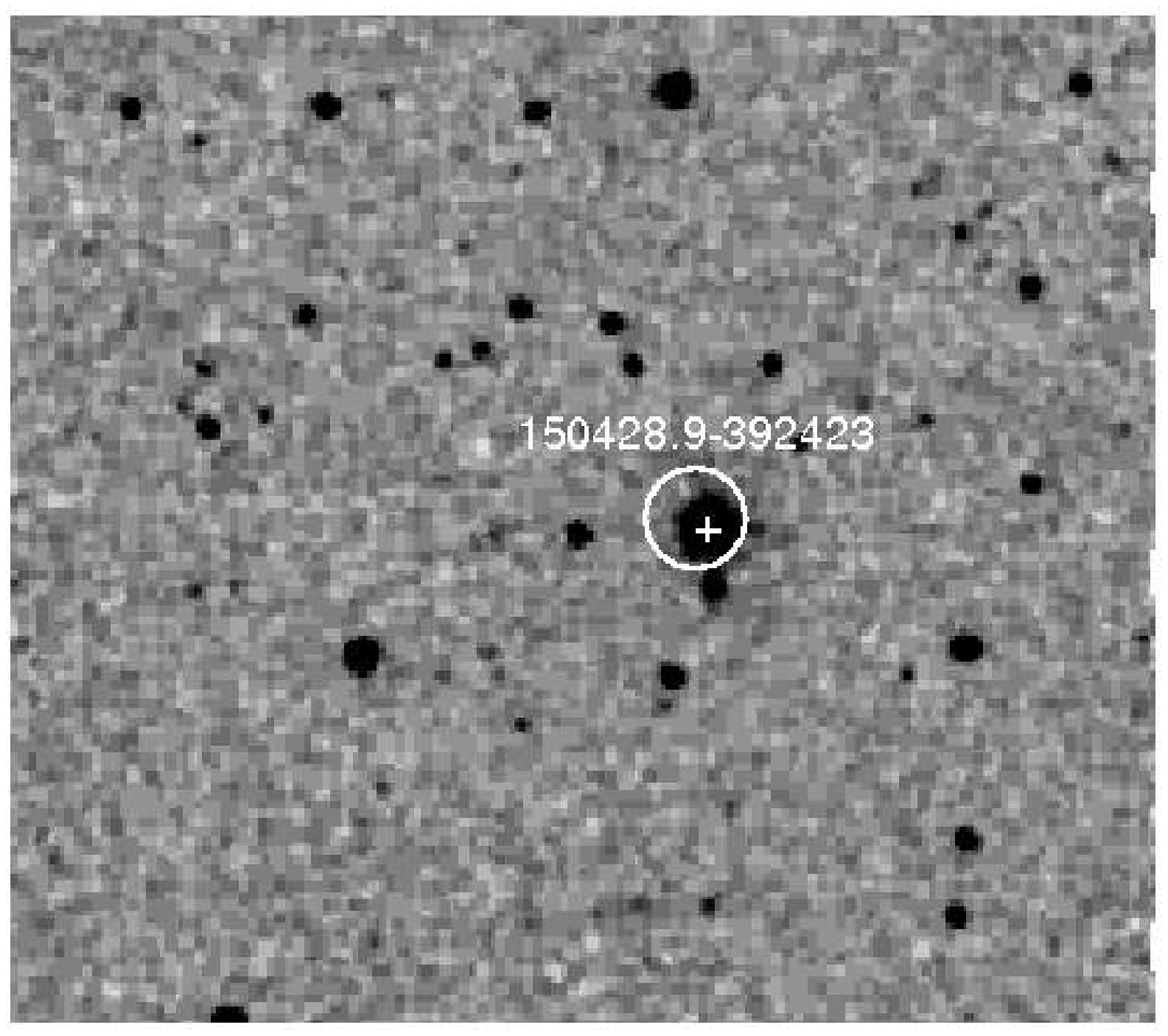}
\includegraphics[width=0.5\textwidth]{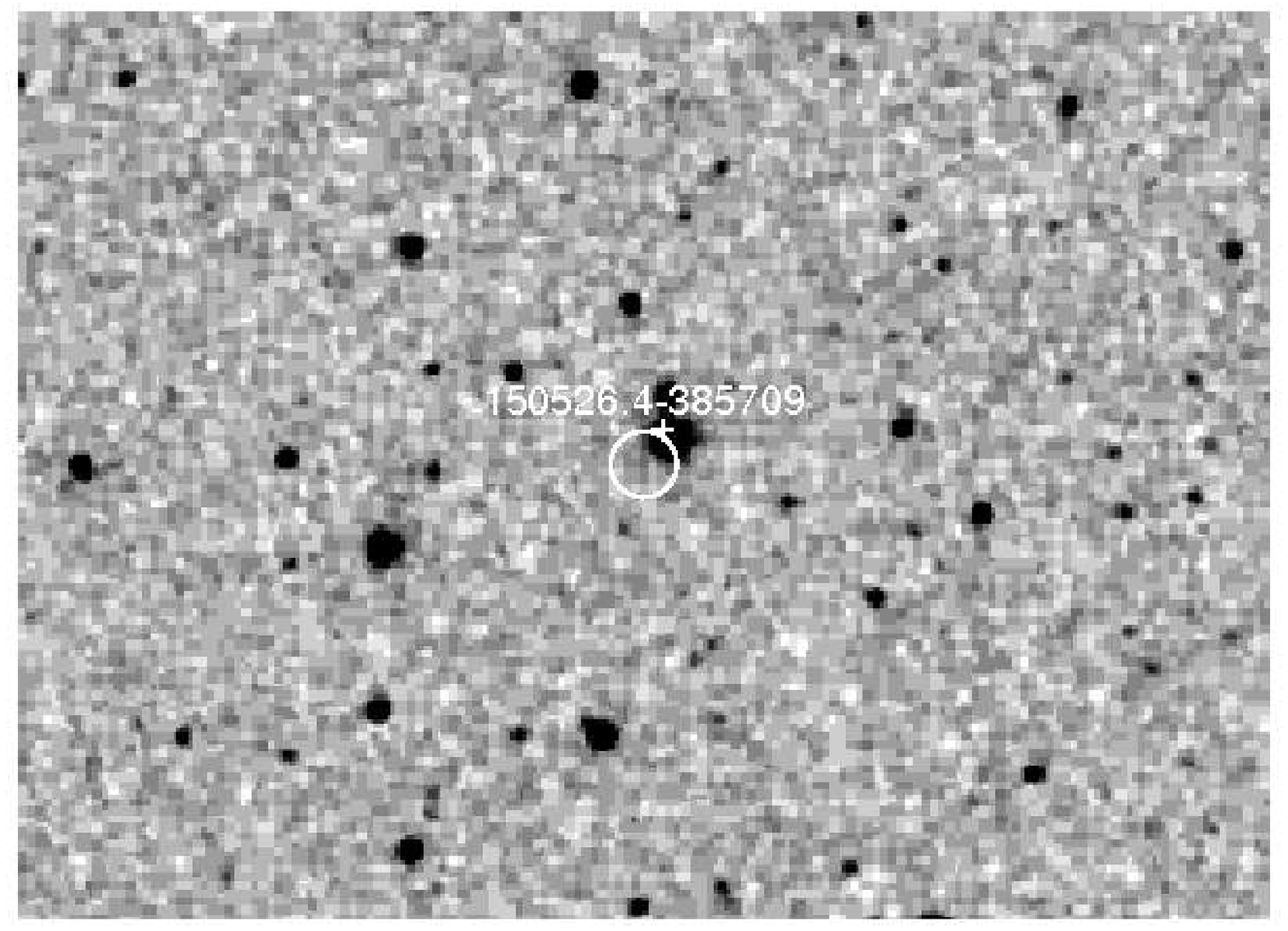}\includegraphics[width=0.5\textwidth]{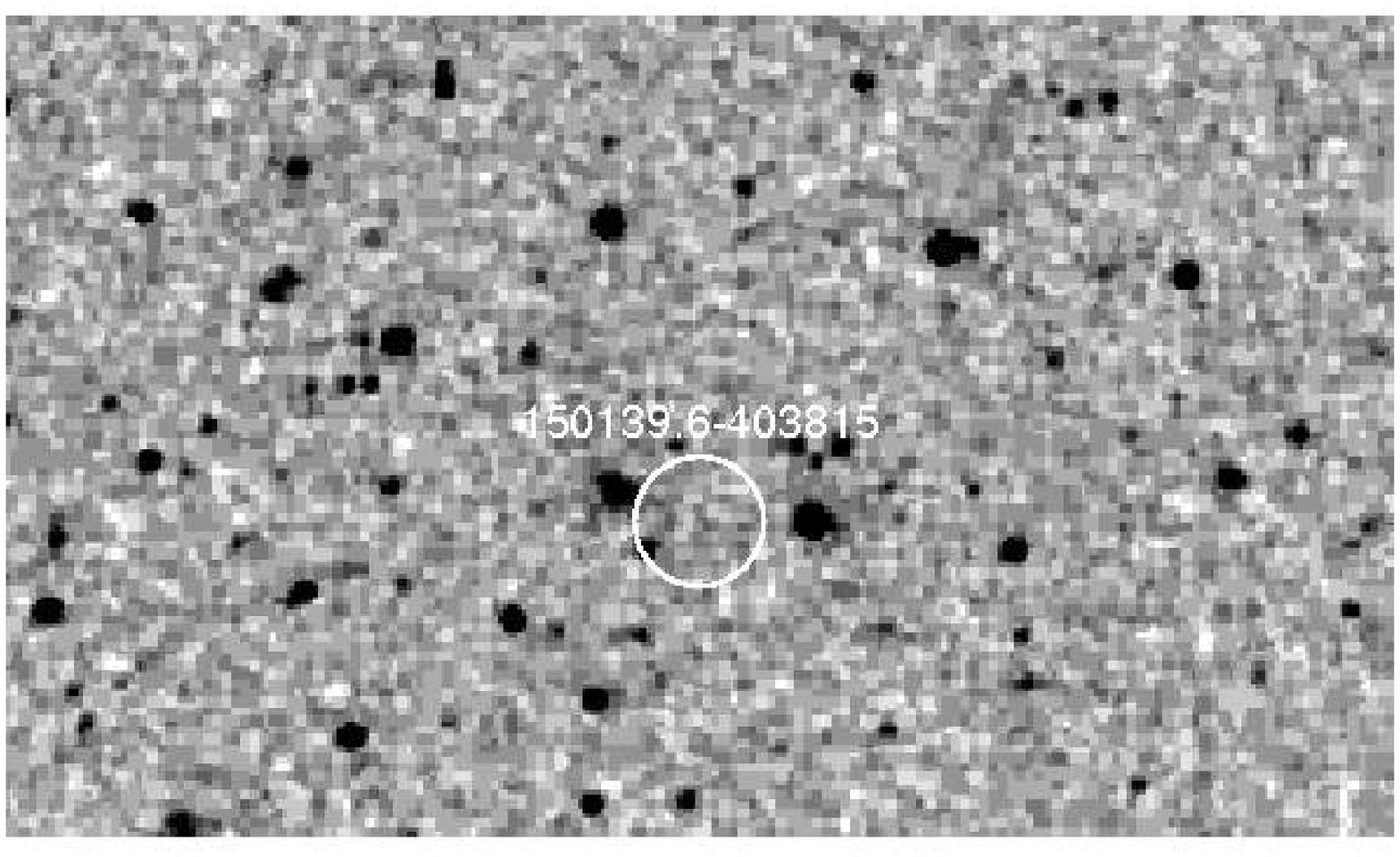}
\caption{2MASS $K_s$-band images of the sources in \snrf\ from
  Table~\ref{tab:srcs}.  The images are $5\arcmin\times3.5\arcmin$, with
  North up and East to the left.  The X-ray position uncertainties are
  indicated by the circles, and the proposed optical counterparts are
  shown by the crosses.\label{fig:opt9}}
\end{figure*}

\begin{figure*}
\centering
\includegraphics[width=0.5\textwidth]{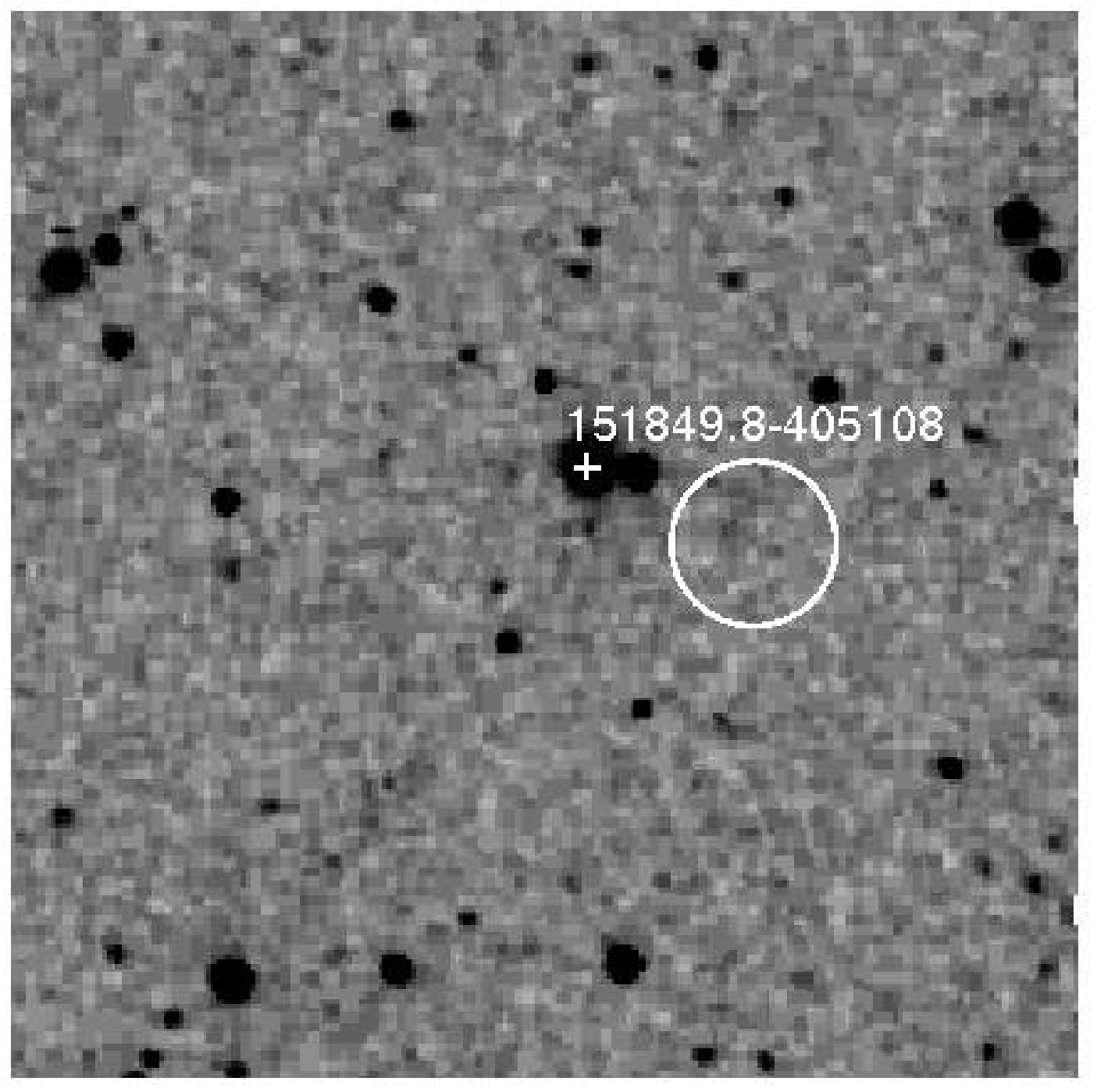}\includegraphics[width=0.5\textwidth]{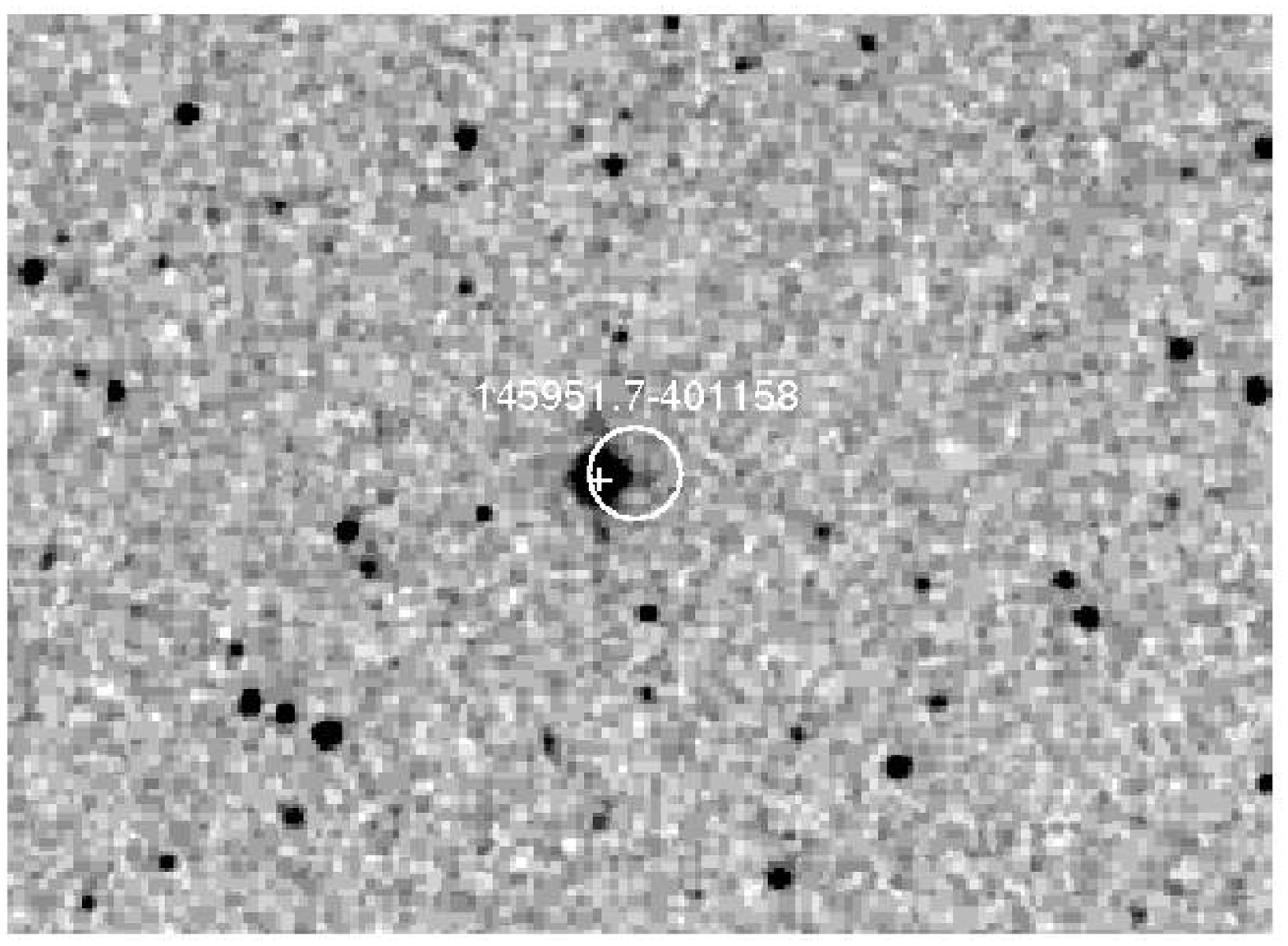}
\includegraphics[width=0.5\textwidth]{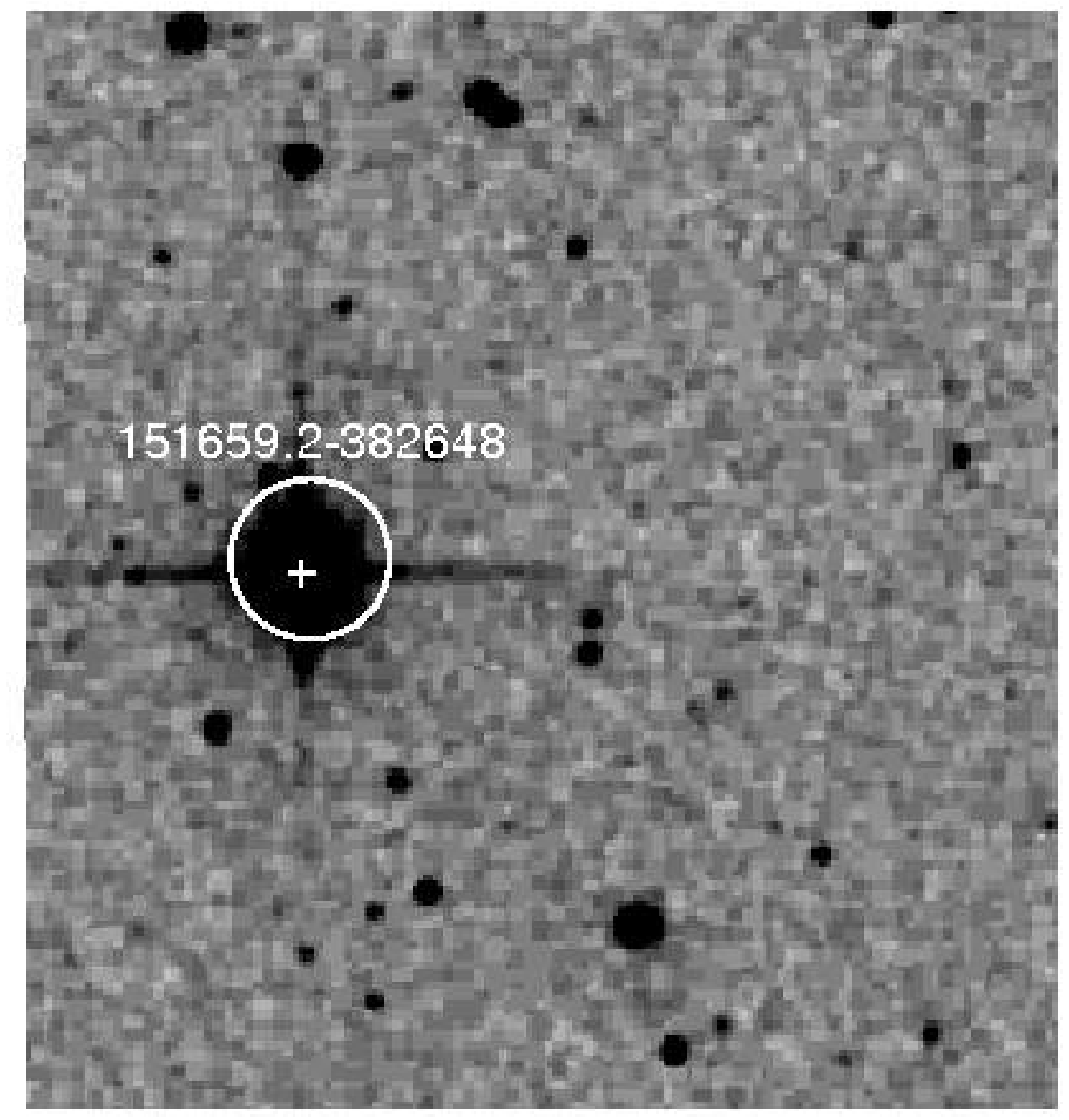}\includegraphics[width=0.5\textwidth]{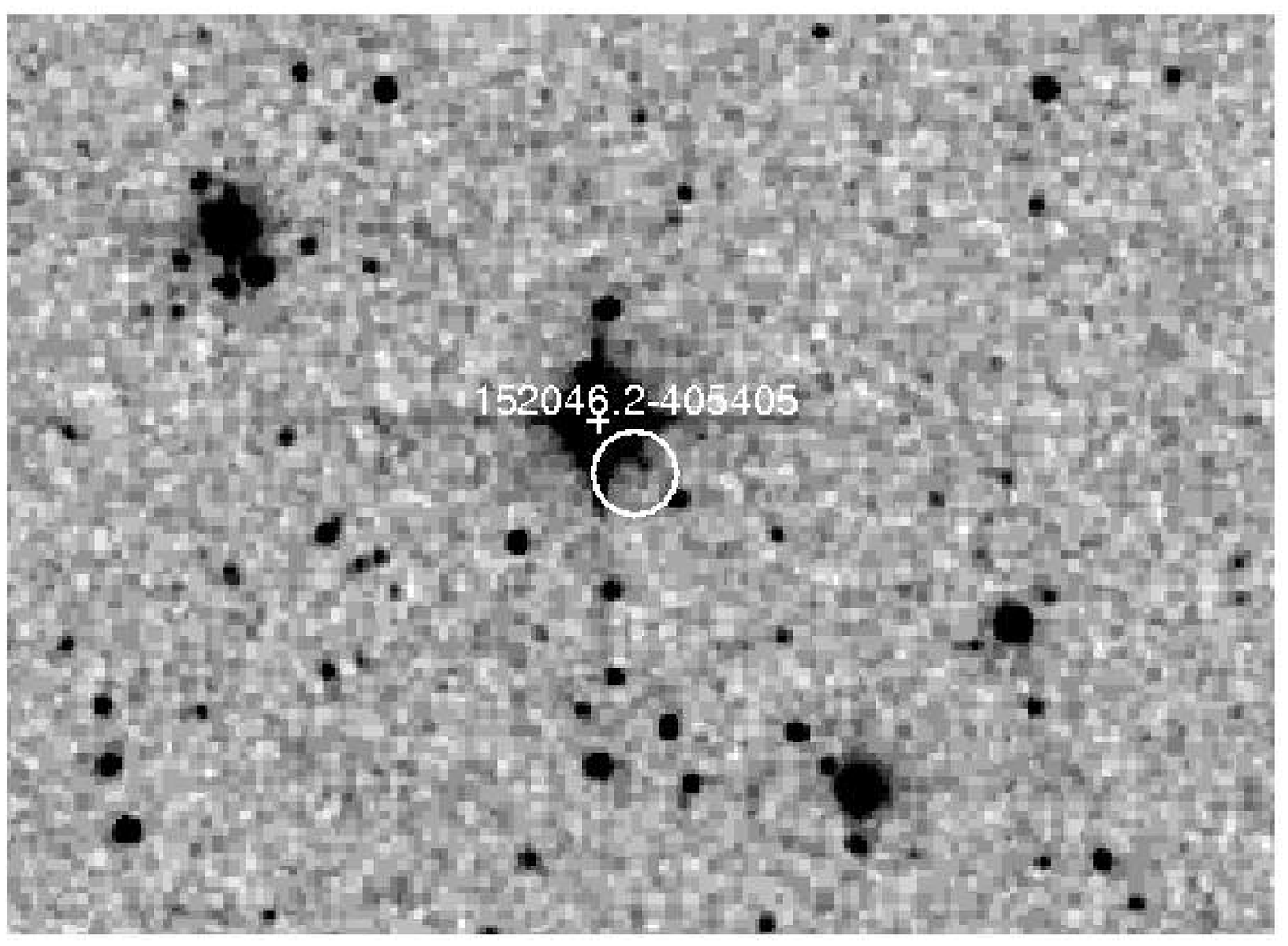}
\includegraphics[width=0.5\textwidth]{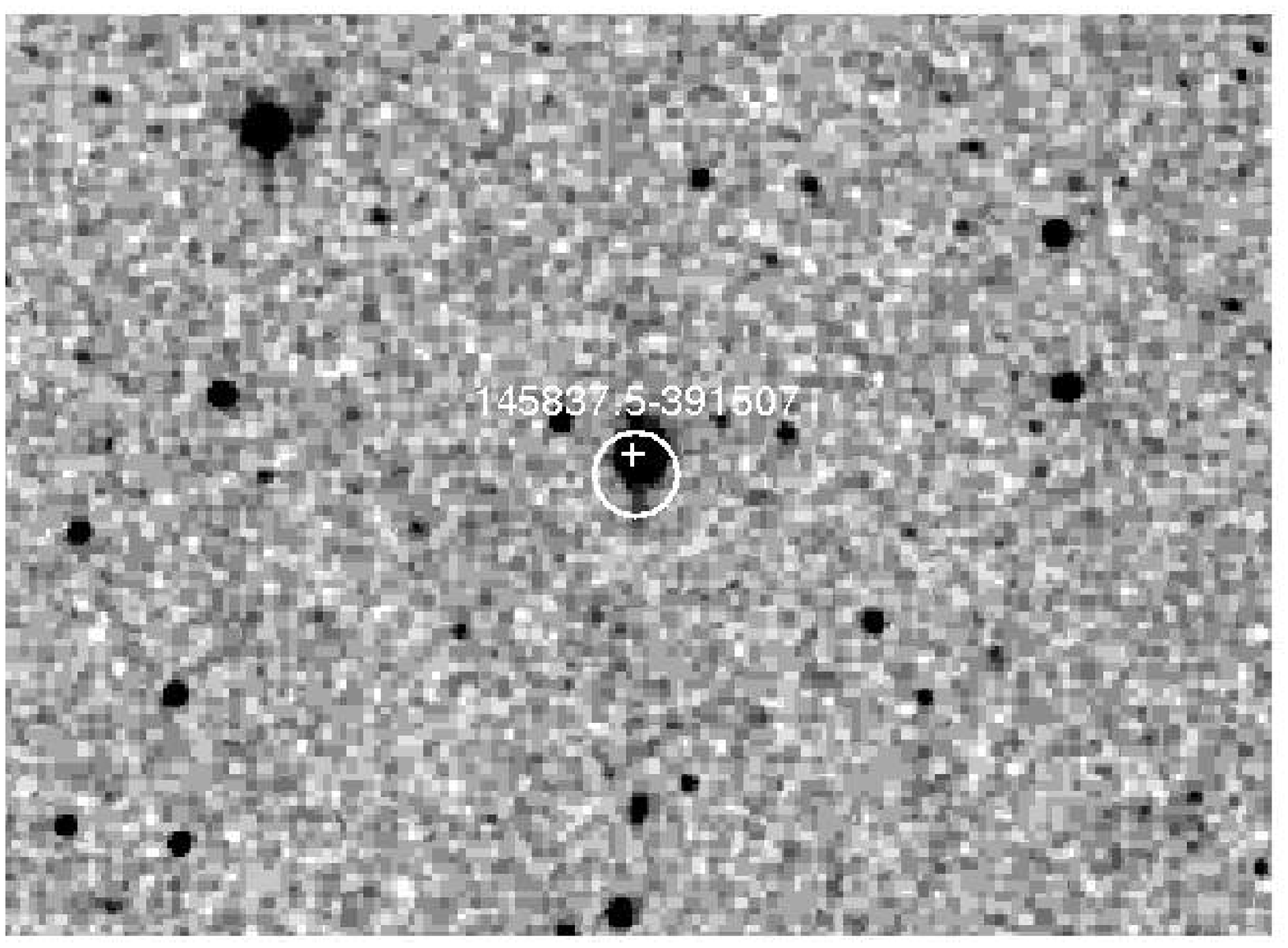}\includegraphics[width=0.5\textwidth]{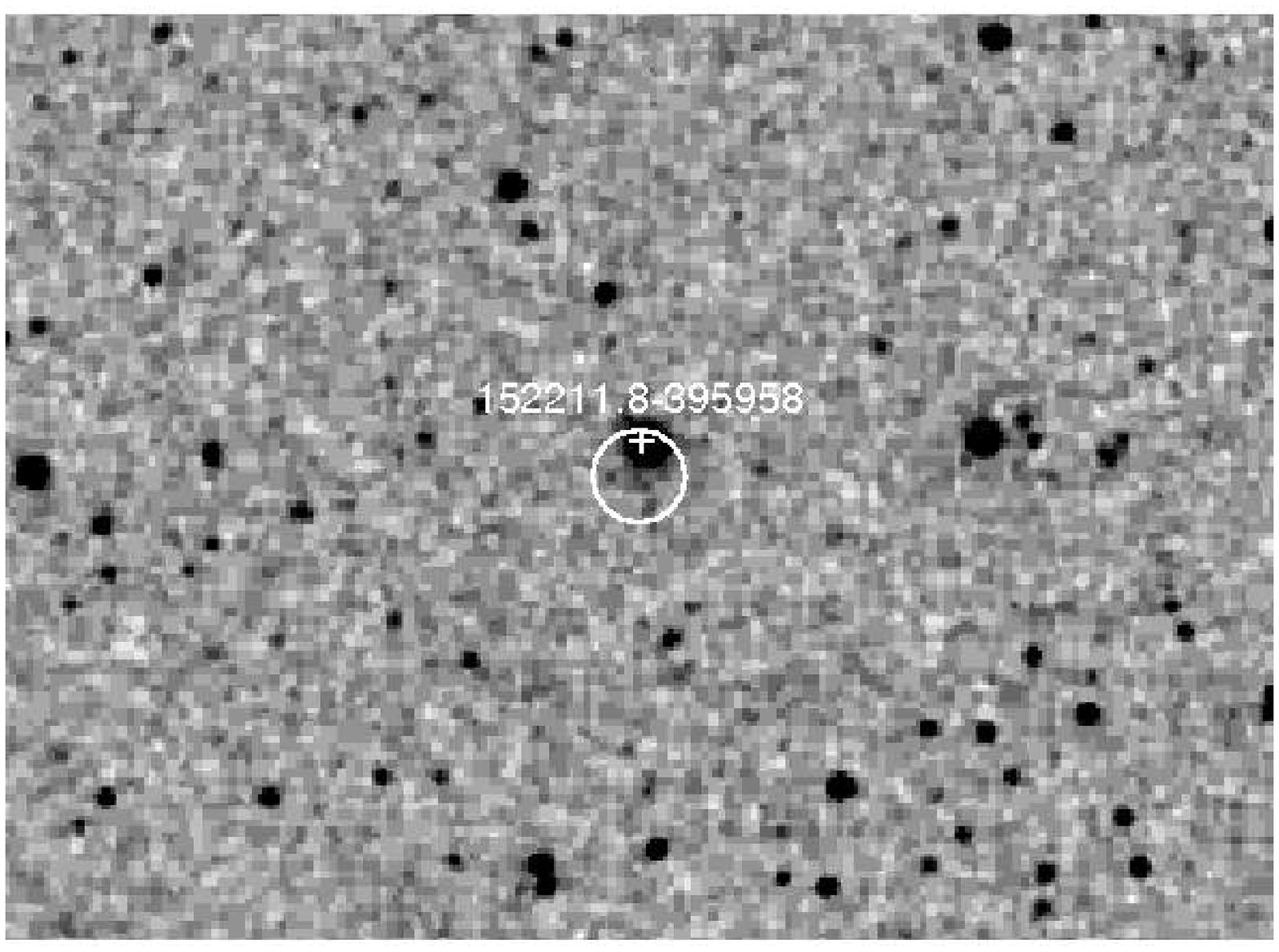}

\caption{2MASS $K_s$-band images of the sources in \snrf\ from
  Table~\ref{tab:srcs} (cont.).  The images are $5\arcmin\times3.5\arcmin$, with
  North up and East to the left.  The X-ray position uncertainties are
  indicated by the circles, and the proposed optical counterparts are
  shown by the crosses.\label{fig:opt10}}
\end{figure*}

\begin{figure*}
\centering
\includegraphics[width=0.5\textwidth]{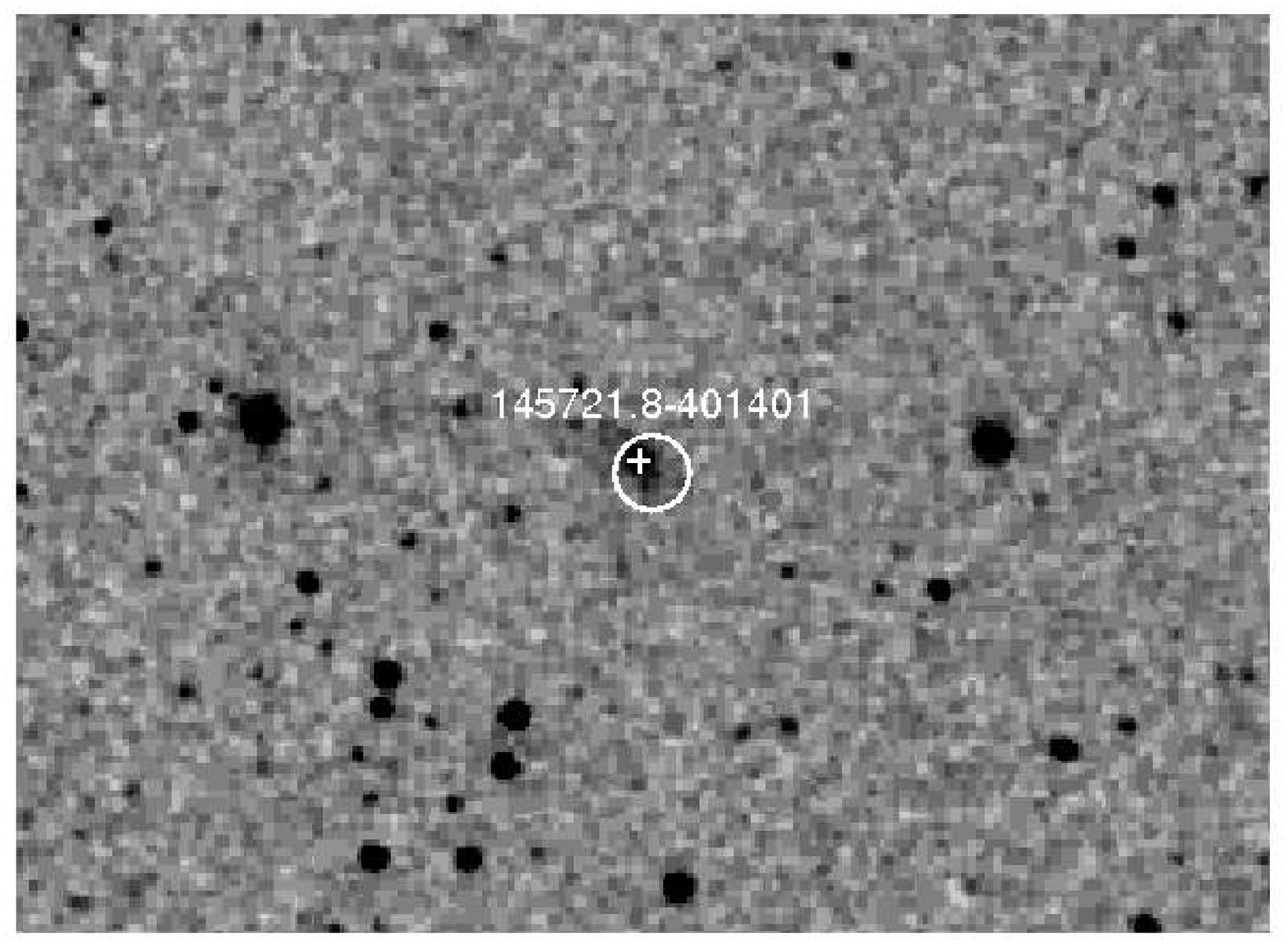}\includegraphics[width=0.5\textwidth]{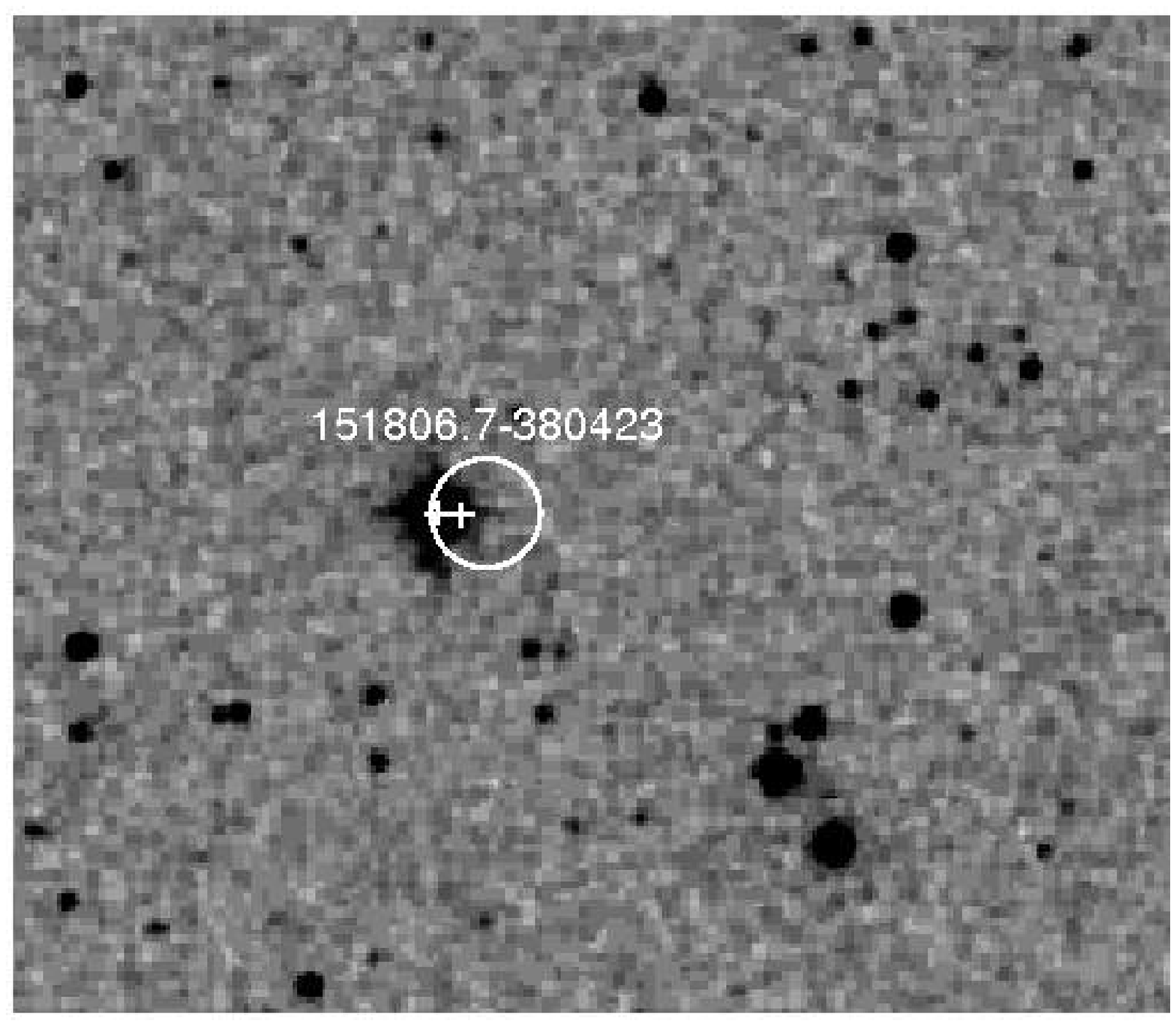}
\includegraphics[width=0.5\textwidth]{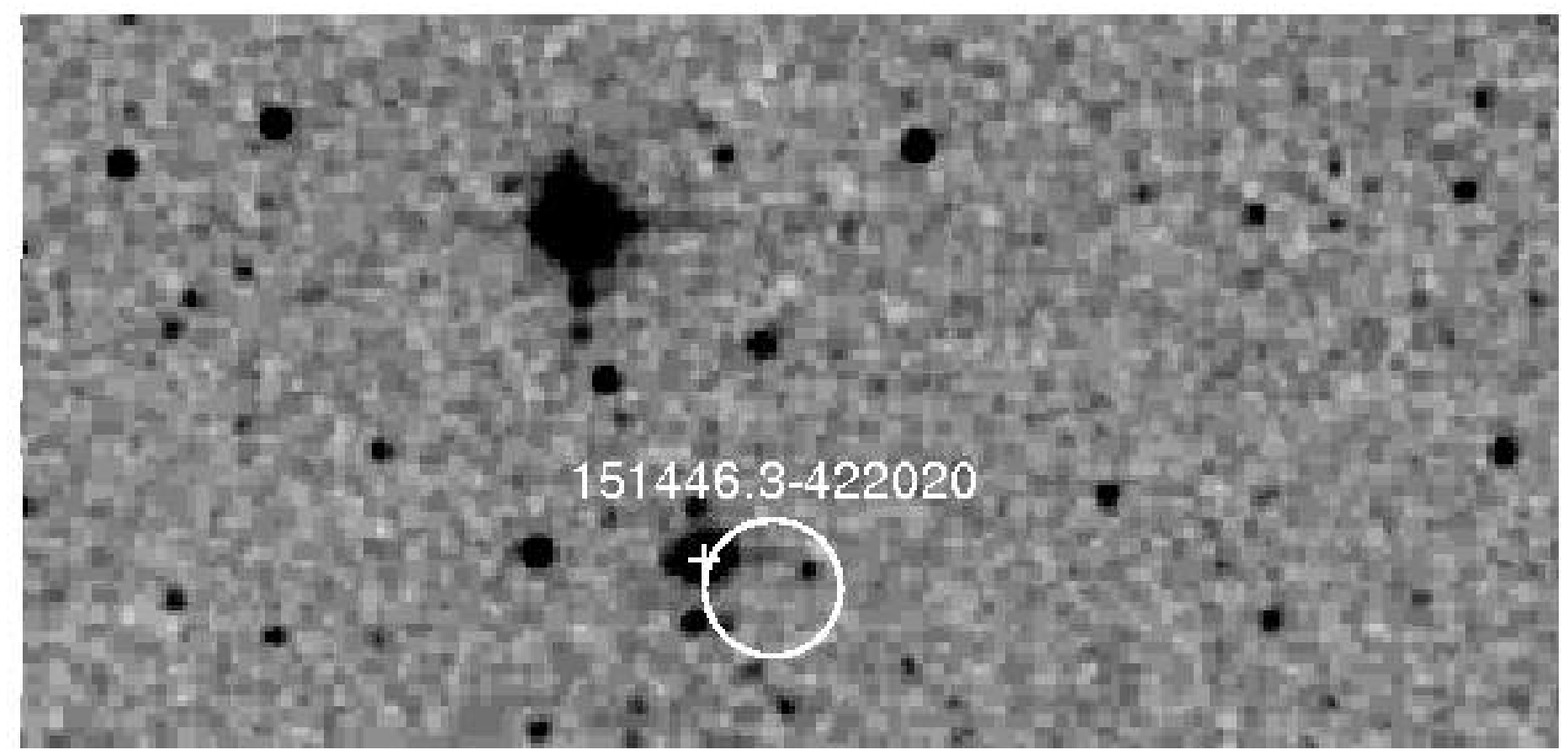}\includegraphics[width=0.5\textwidth]{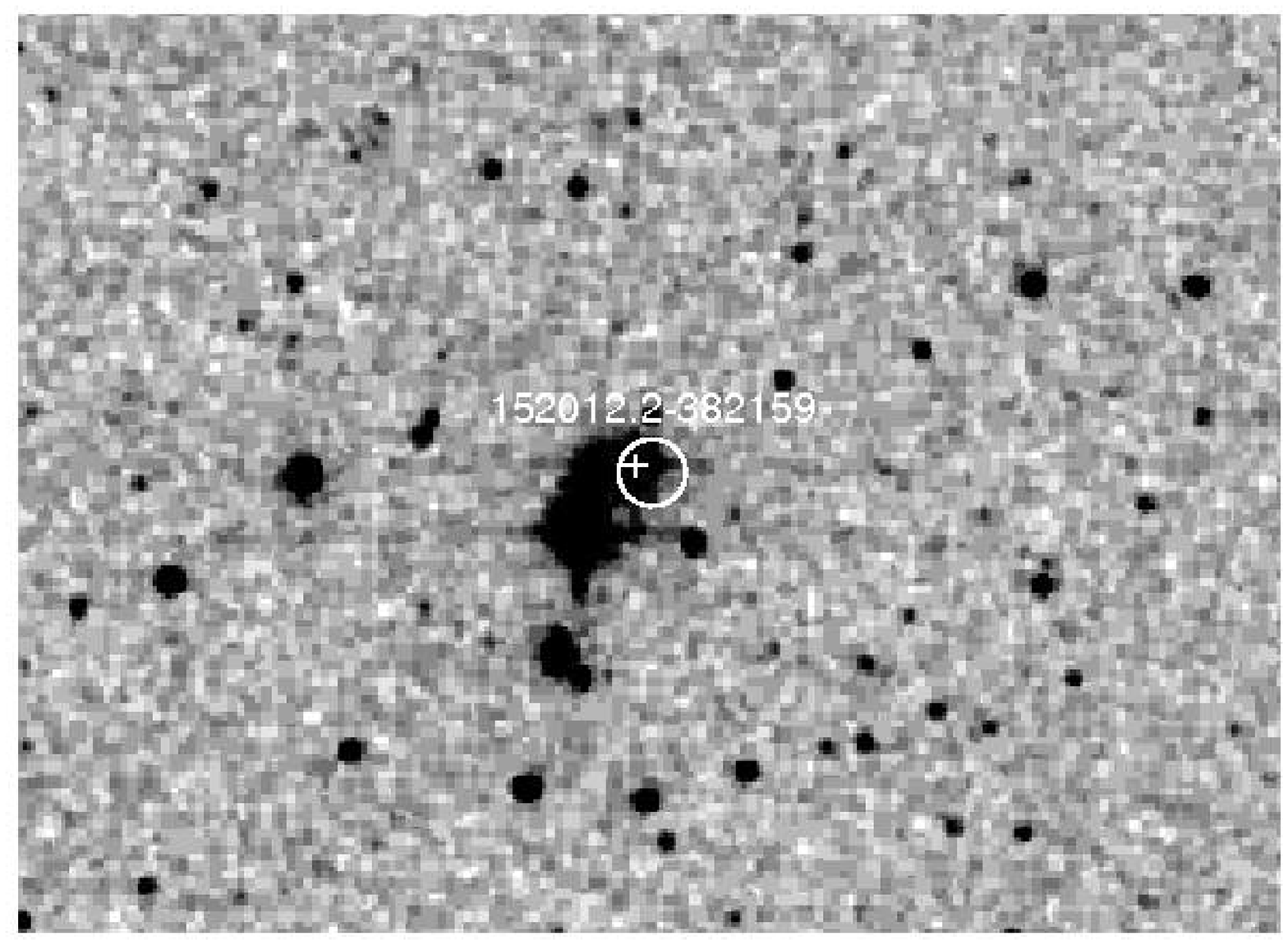}
\includegraphics[width=0.5\textwidth]{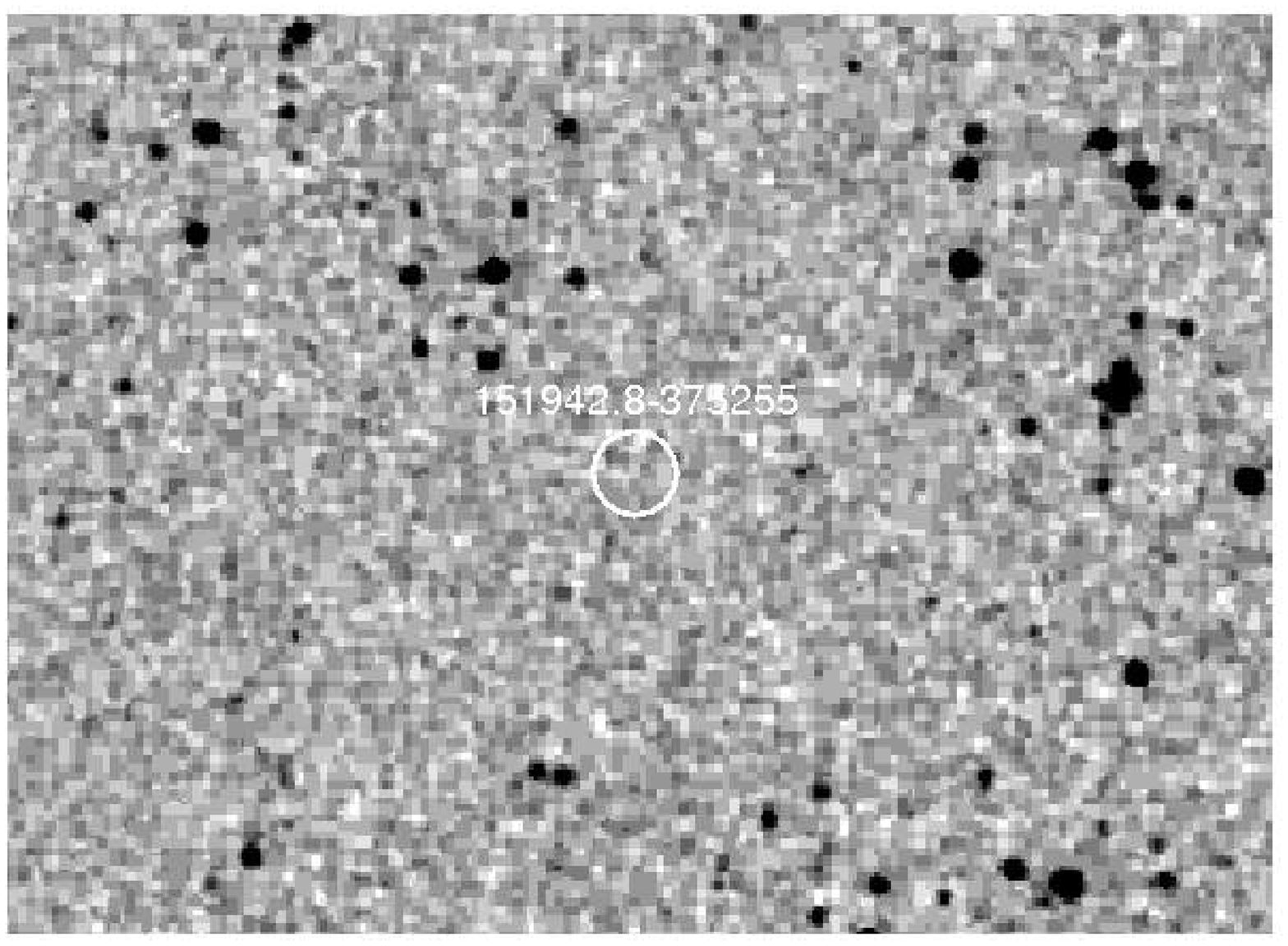}\includegraphics[width=0.5\textwidth]{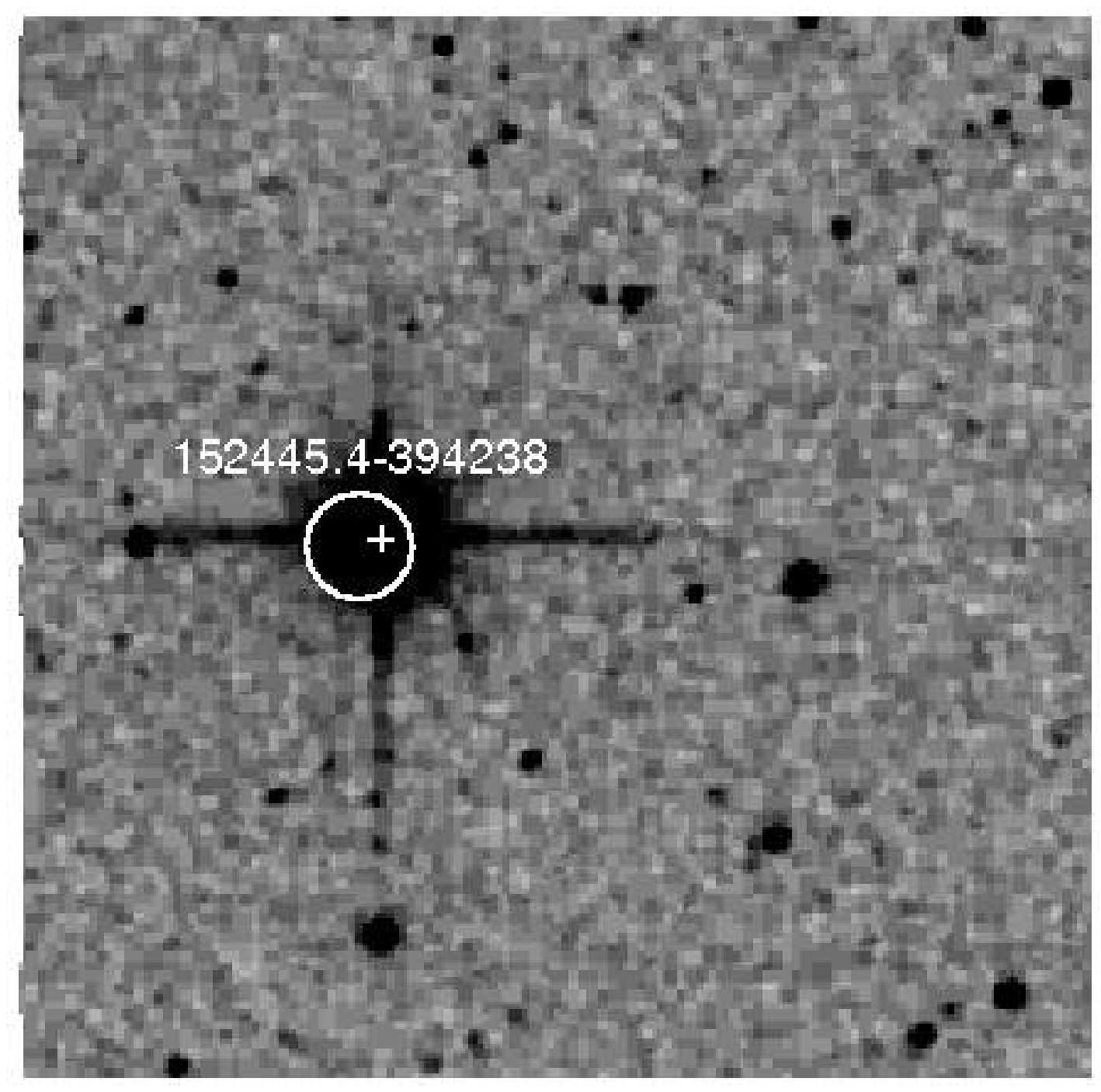}

\caption{2MASS $K_s$-band images of the sources in \snrf\ from
  Table~\ref{tab:srcs} (cont.).  The images are $5\arcmin\times3.5\arcmin$, with
  North up and East to the left.  The X-ray position uncertainties are
  indicated by the circles, and the proposed optical counterparts are
  shown by the crosses.\label{fig:opt11}}
\end{figure*}

\begin{figure}
\centering
\includegraphics[width=0.5\textwidth]{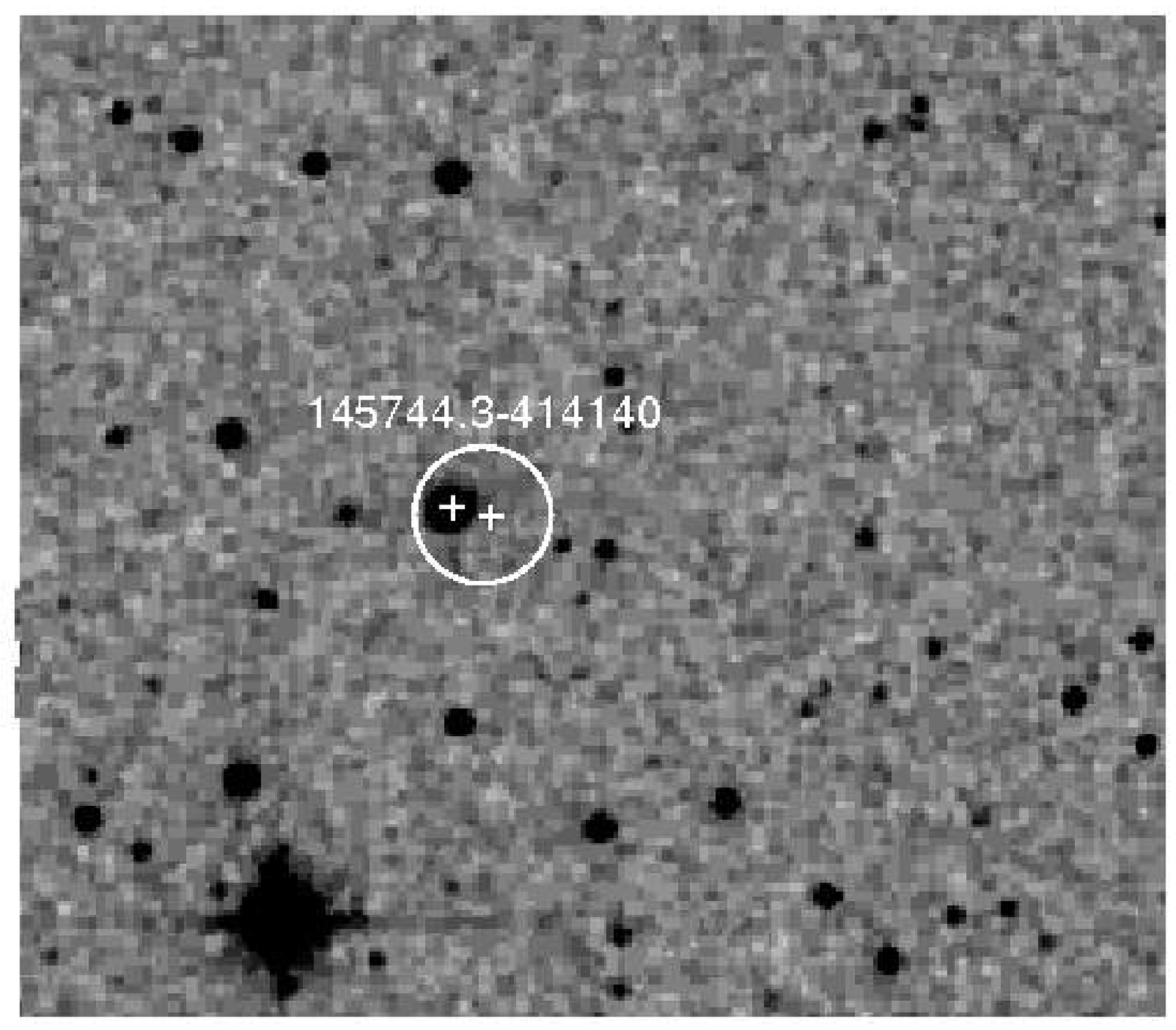}
\caption{2MASS $K_s$-band images of the sources in \snrf\ from
  Table~\ref{tab:srcs} (cont.).  The images are $5\arcmin\times3.5\arcmin$, with
  North up and East to the left.  The X-ray position uncertainties are
  indicated by the circles, and the proposed optical counterparts are
  shown by the crosses.\label{fig:opt12}}
\end{figure}

Given the uncertainties in spectrum and foreground column density,
virtually all of the X-ray sources are consistent with being stars (as
opposed to active galaxies; Fig.~\ref{fig:fxk}).  We show 2MASS images
of the X-ray sources with optical counterparts indicated in
Figures~\ref{fig:opt1}--\ref{fig:opt12}.

\section{\chandra\ Observations}
\label{sec:cxo}
The nine BSC sources that had no obvious optical counterparts
(excluding \rxsd, as mentioned above), plus the four sources from
\S~\ref{sec:addl}, were selected for \chandra\ followup observations.
Here, as in \citetalias{kfg+04}, we selected the exposure times
(3--6~ksec) based on the known column densities to the SNRs
(Tab.~\ref{tab:snrs}) and a blackbody spectrum with $kT=0.25$~keV.
The positions are known to sufficient accuracy to allow use of the
ACIS-S3 CCD \citep{gbf+03}.  Depending on the source brightnesses,
however, we were concerned about photon
pileup for some of the sources, so we used the $1/2$- or
$1/4$-subarray modes (which also provides improved timing
information), depending on the positional uncertainties.  A log of the
observations is in Table~\ref{tab:cxo}.

In most of the cases, the \chandra\ observations revealed nothing
extraordinary.  In the case of the BSC sources, the \chandra\ data
typically showed that the BSC position was significantly off and/or
the counterpart was faint (Figs.~\ref{fig:cxoopt1} and
\ref{fig:cxoopt2}).  The additional four sources from the literature were
all coincident with stellar sources, once we had \chandra\ positions.
Of the 13 sources with \chandra\ followup, nine had point-like
\chandra\ sources with obvious IR counterparts
(\S~\ref{sec:notescxo}).  Of the other four sources: two show extended
X-ray emission with \chandra, one has no obvious 2MASS counterpart,
and one source was not detected in the \chandra\ observation. We discuss
these sources in more detail in Section~\ref{sec:remain}.

\subsection{Notes on \chandra\ Sources}
\label{sec:notescxo}
\begin{figure*}
\centering
\includegraphics[width=0.5\textwidth]{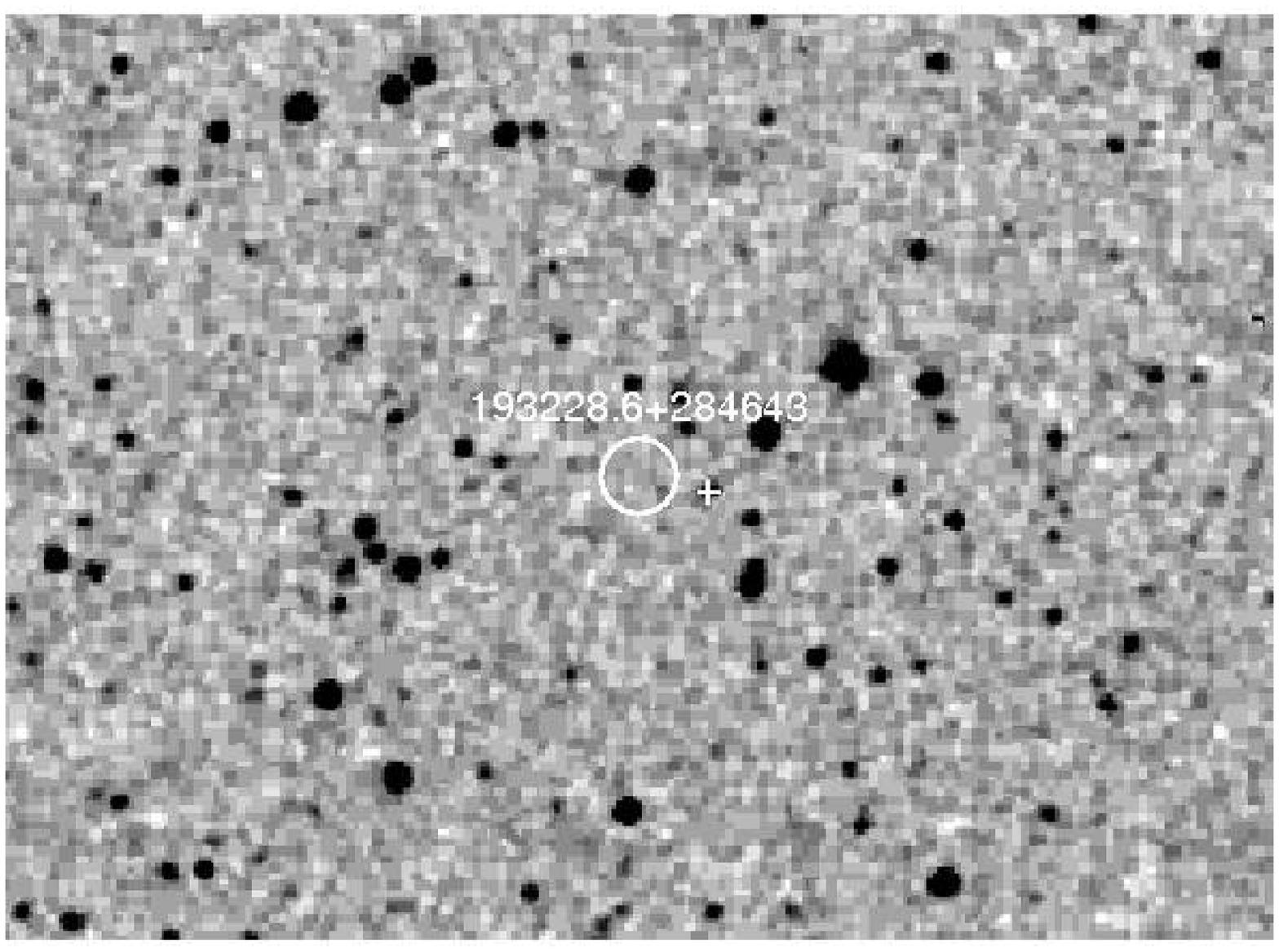}\includegraphics[width=0.5\textwidth]{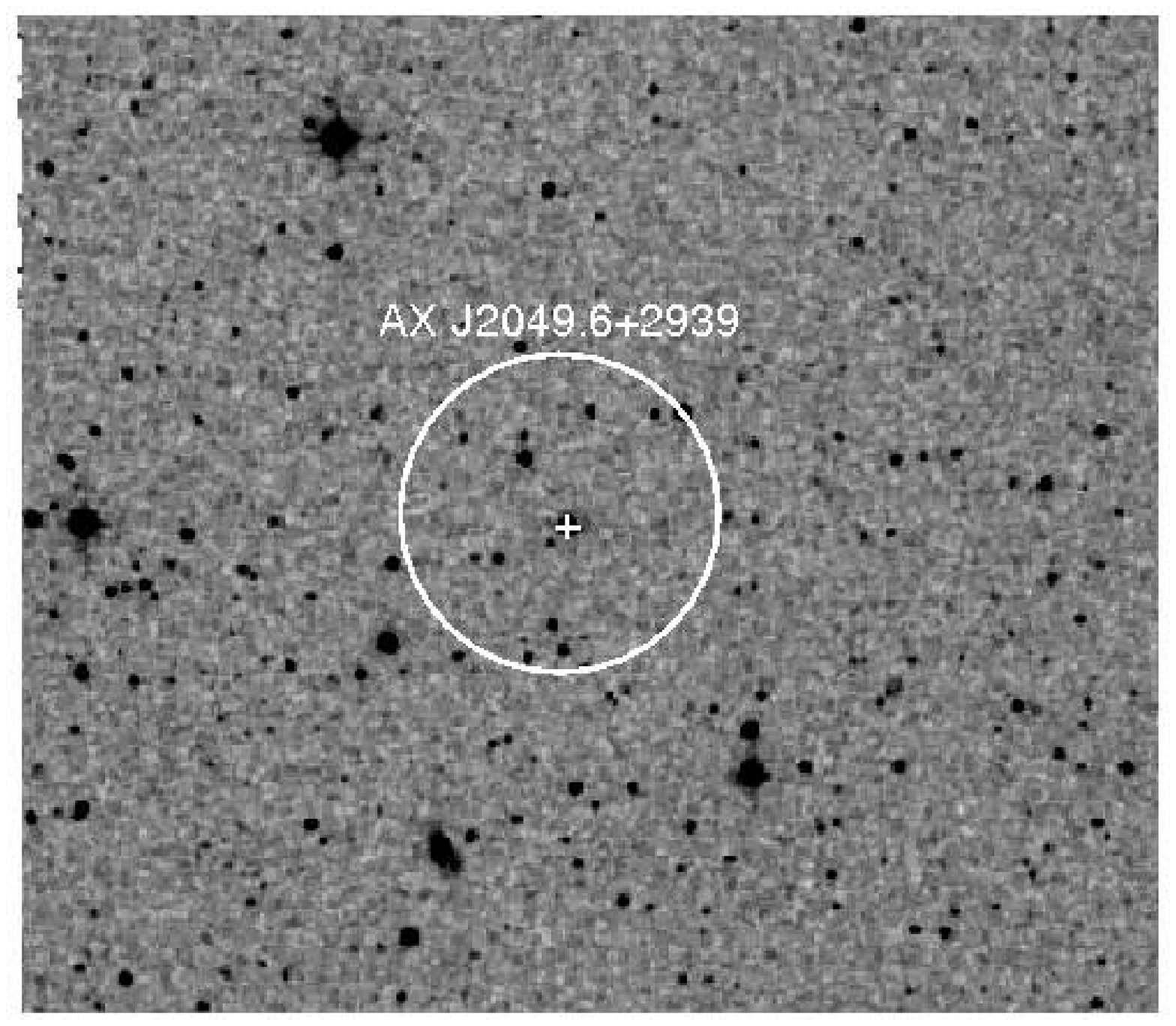}
\includegraphics[width=0.5\textwidth]{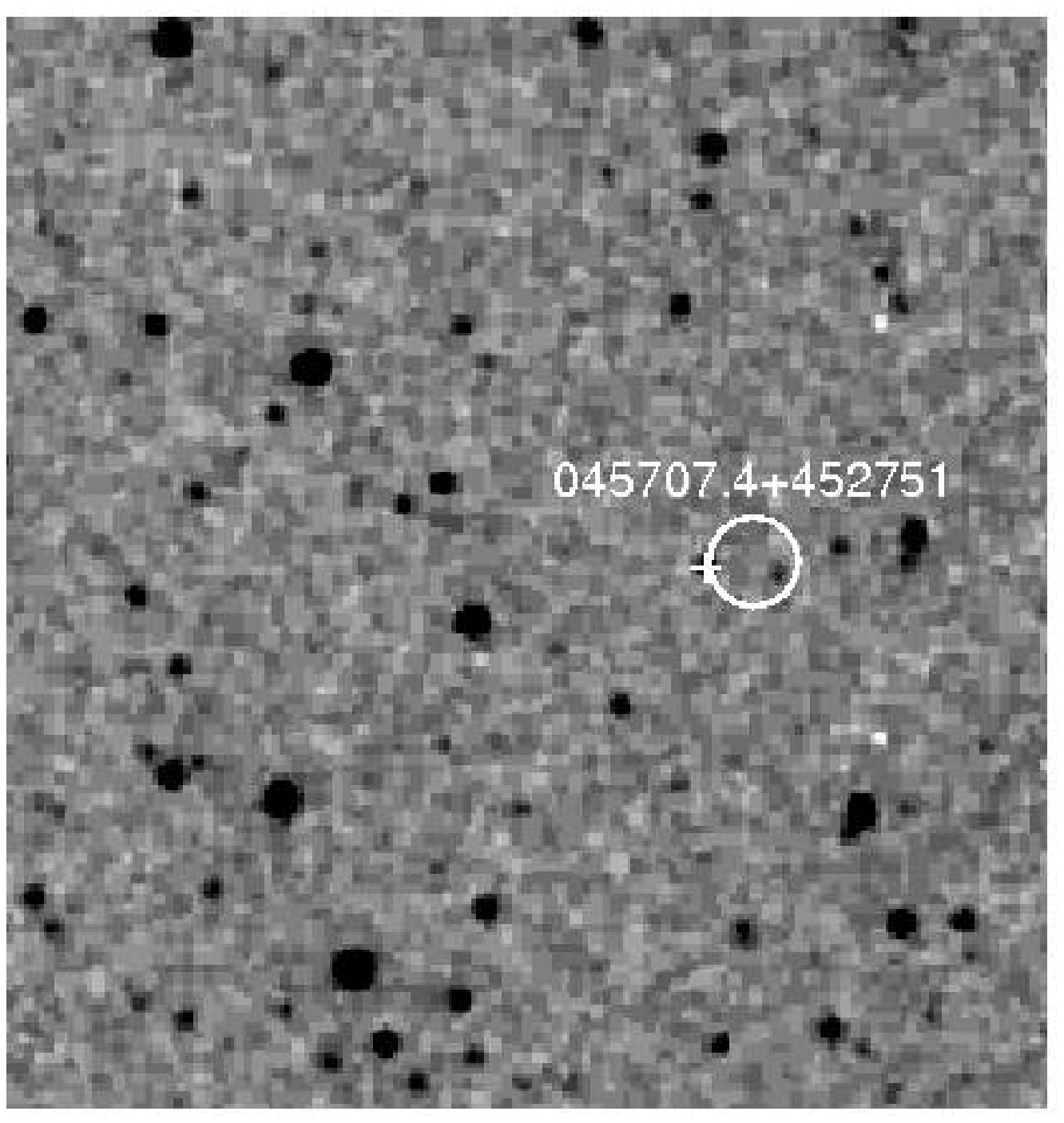}\includegraphics[width=0.5\textwidth]{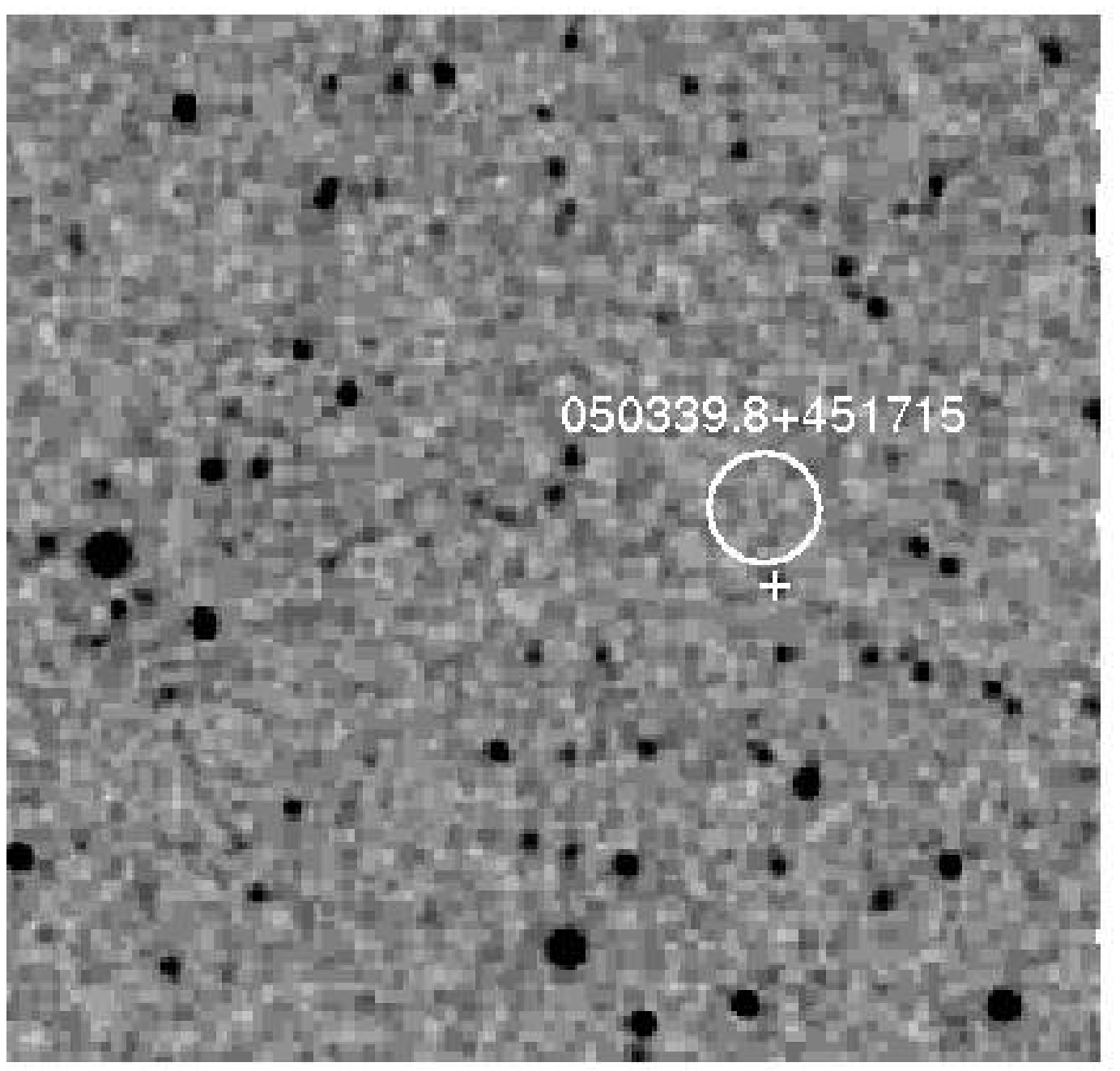}
\includegraphics[width=0.5\textwidth]{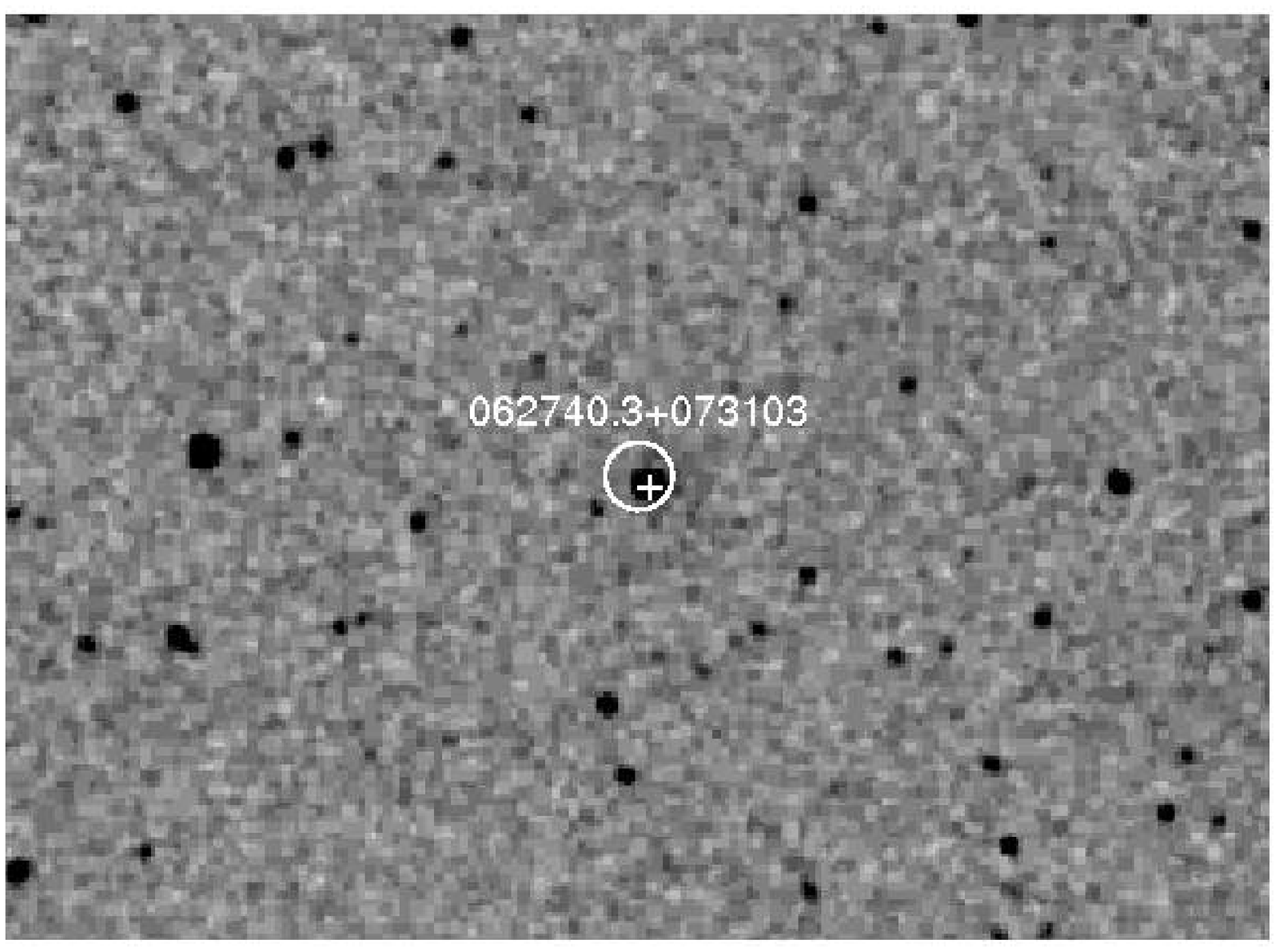}\includegraphics[width=0.5\textwidth]{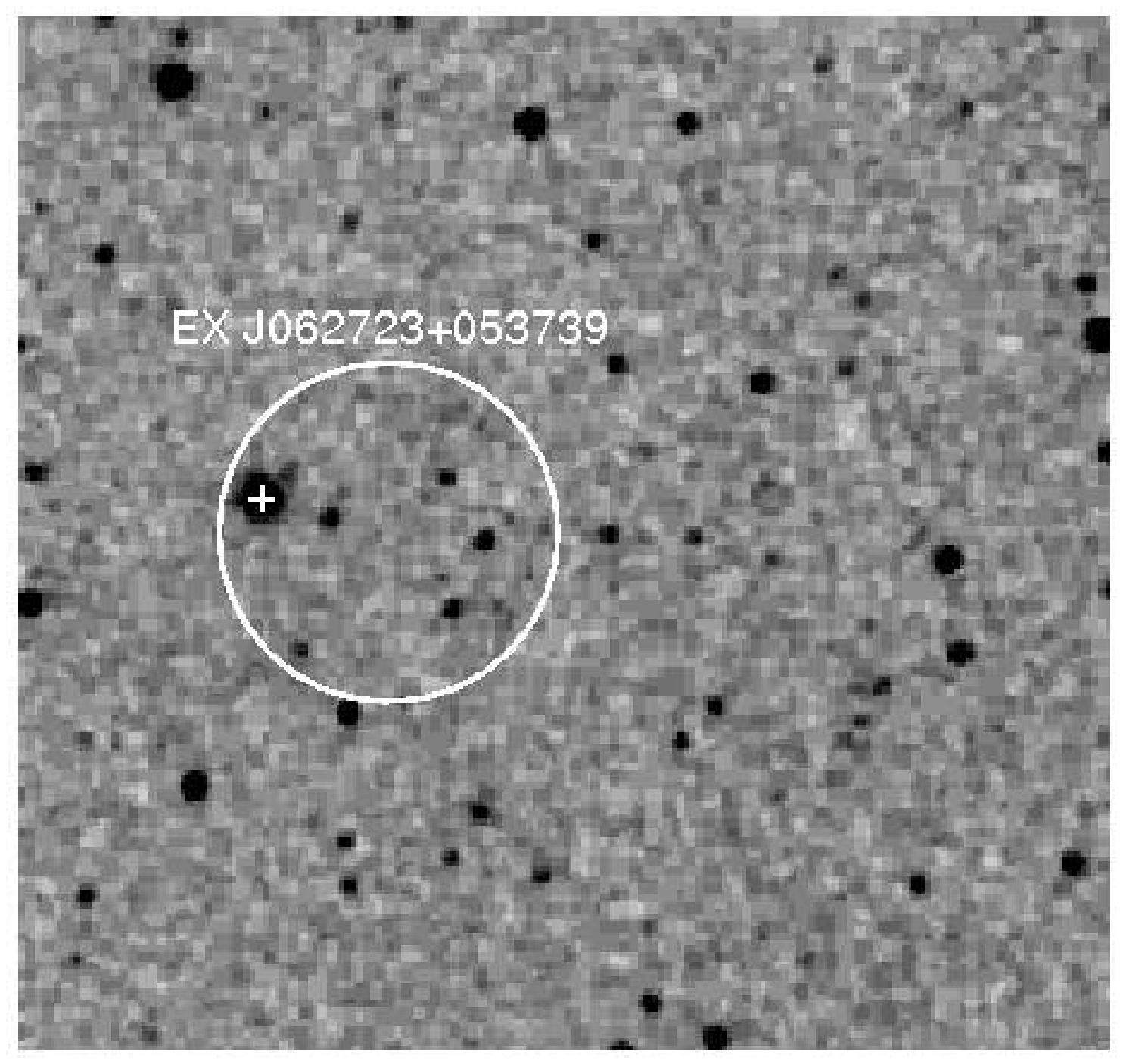}

\caption{2MASS $K_s$-band images of the sources from
  Table~\ref{tab:cxo} with point-like X-ray sources and secure
  counterpart identifications.  The images are
  $5\arcmin\times3.5\arcmin$ (except for that of AX~J2049.6+2939,
  which is $10\arcmin\times7\arcmin$), with North up and East to the
  left.  The BSC/\textit{ASCA}/\textit{Einstein} X-ray position
  uncertainties are indicated by the circles, and the \chandra\
  positions and optical counterparts are shown by the crosses.
\label{fig:cxoopt1}}
\end{figure*}

\begin{figure*}
\centering
\includegraphics[width=0.5\textwidth]{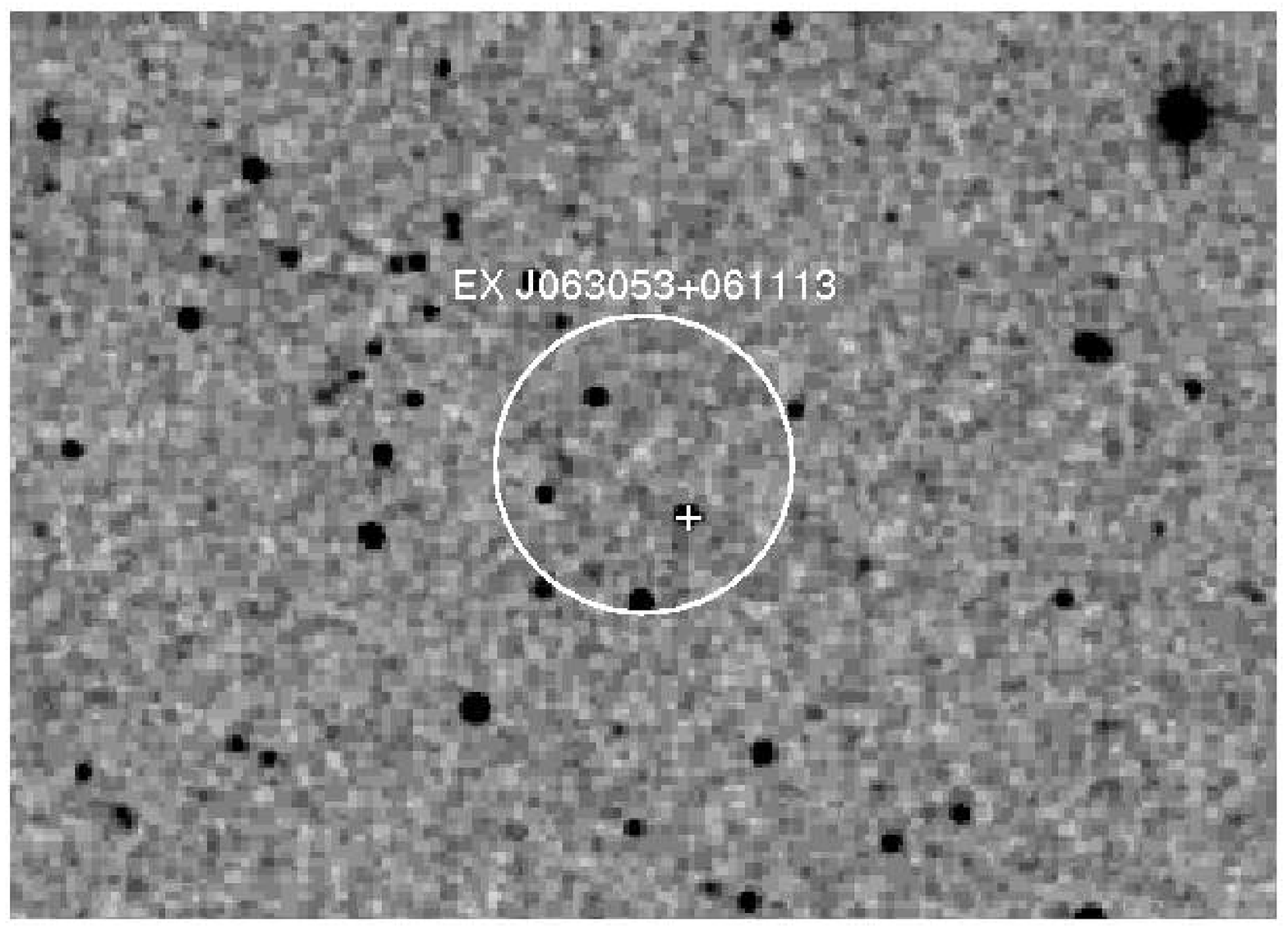}\includegraphics[width=0.5\textwidth]{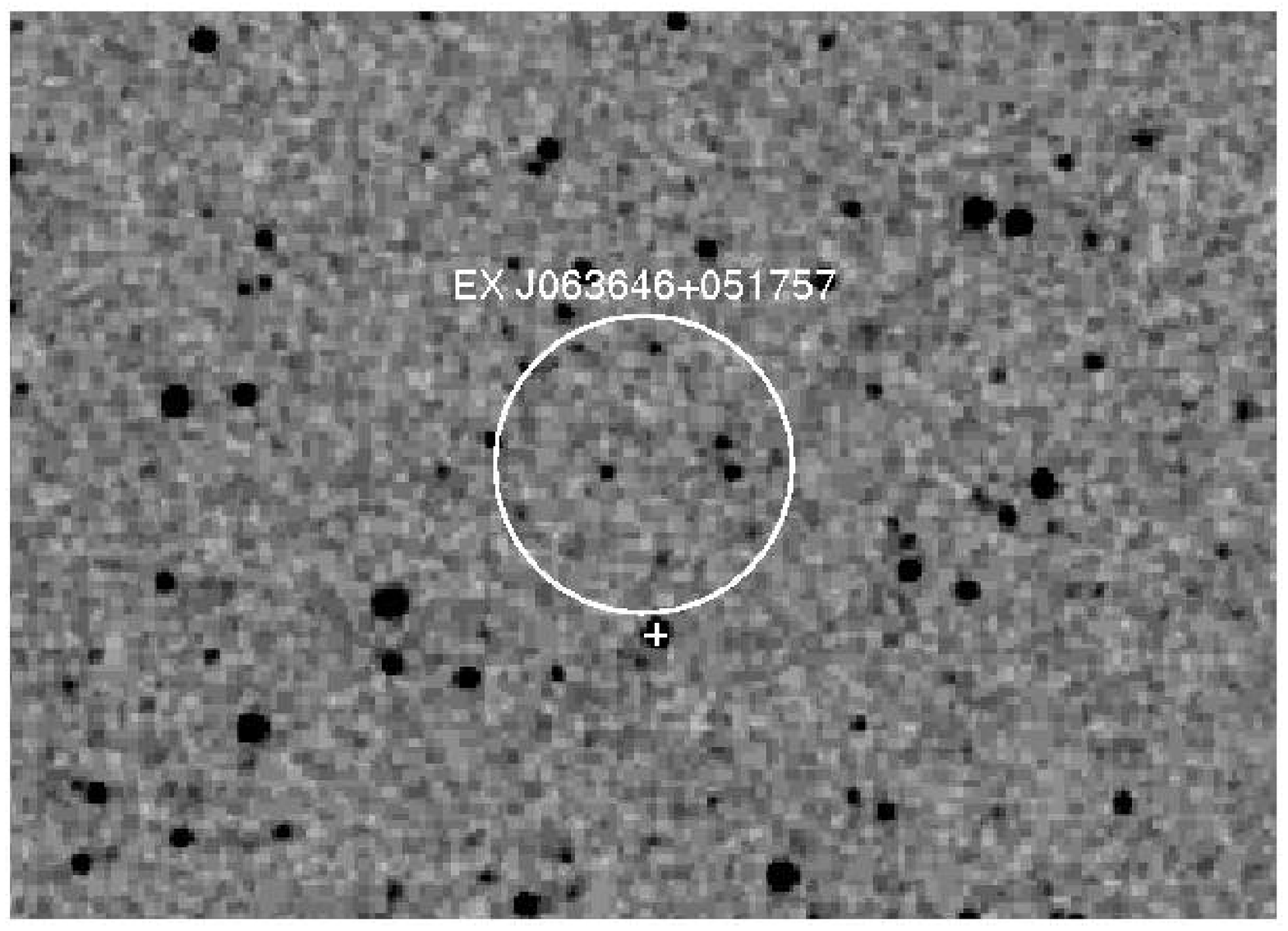}
\includegraphics[width=0.5\textwidth]{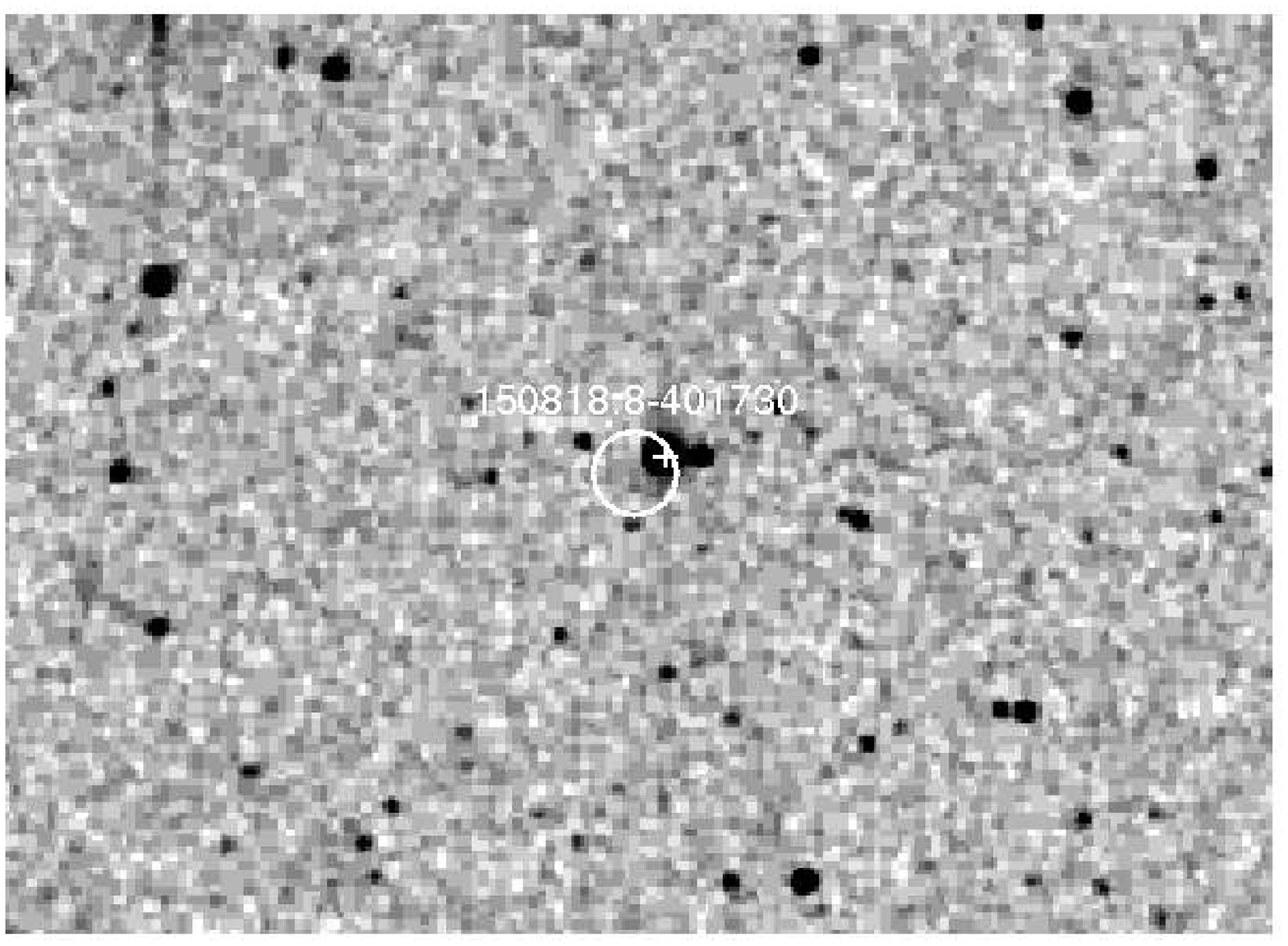}
\caption{2MASS $K_s$-band images of the sources from
  Table~\ref{tab:cxo} with point-like X-ray sources and secure
  counterpart identifications (cont.).  The images are
  $5\arcmin\times3.5\arcmin$, with North up and East to the left.  The
  BSC/\textit{ASCA}/\textit{Einstein} X-ray position uncertainties are
  indicated by the circles, and the \chandra\ positions and optical
  counterparts are shown by the crosses.
\label{fig:cxoopt2}}
\end{figure*}

In the cases where \chandra\ point-sources were detected, the
counterpart identifications are essentially secure.  This is due to
the very small positional uncertainty of the \chandra\ positions
($<1\arcsec$) coupled with the brightnesses of the counterparts (see
Fig.~\ref{fig:stars} and \citetalias{kfg+04}).
Figures~\ref{fig:cxoopt1} and \ref{fig:cxoopt2} contain images of
those sources with counterpart identifications.  With these
identifications we can eliminate these sources as candidate compact
objects using the X-ray-to-optical flux ratio (see Fig.~\ref{fig:fxk}
and \citetalias{kfg+04}).   Here we comment on all of the sources
observed with \chandra.

\begin{figure*}
\plottwo{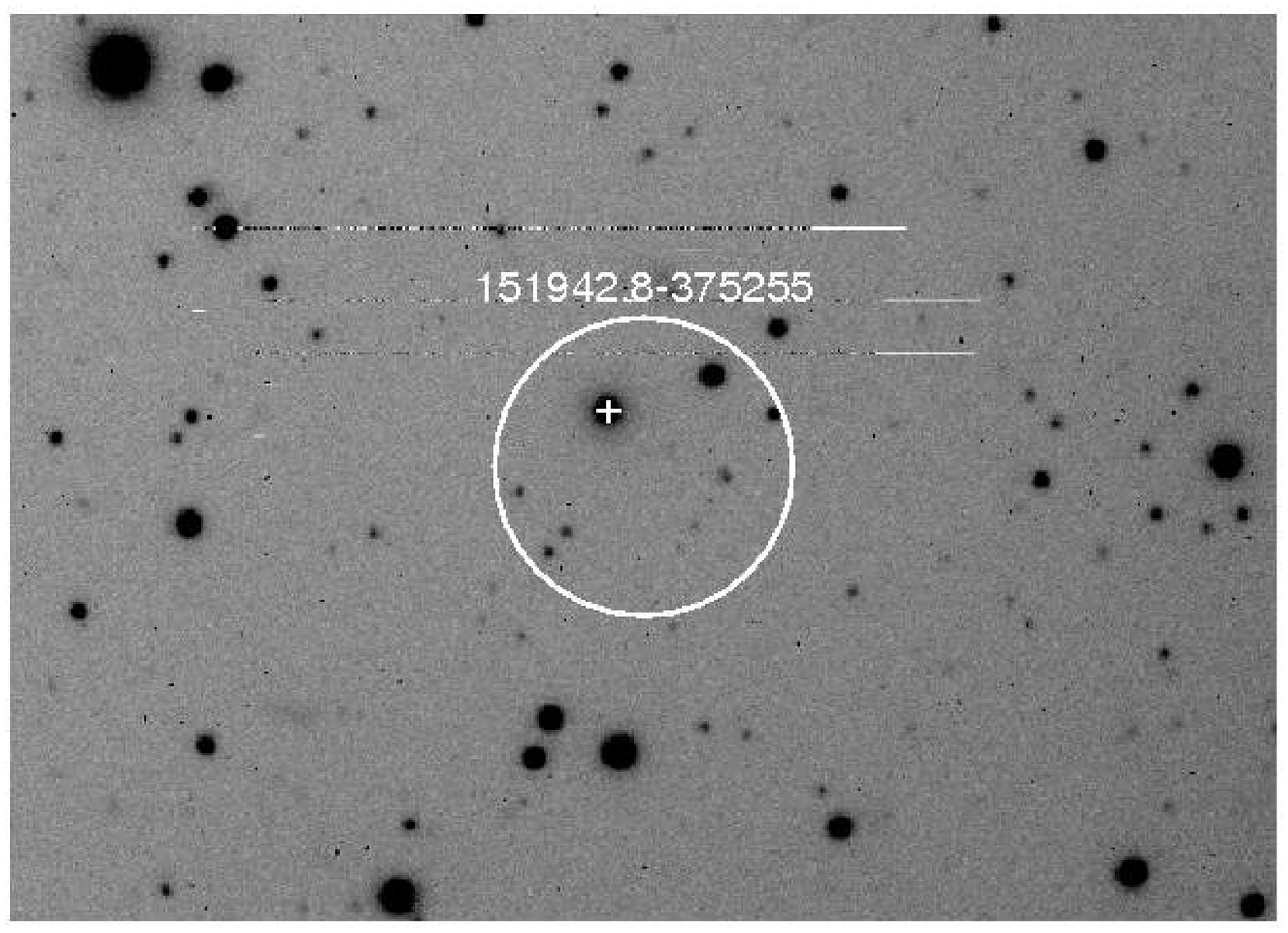}{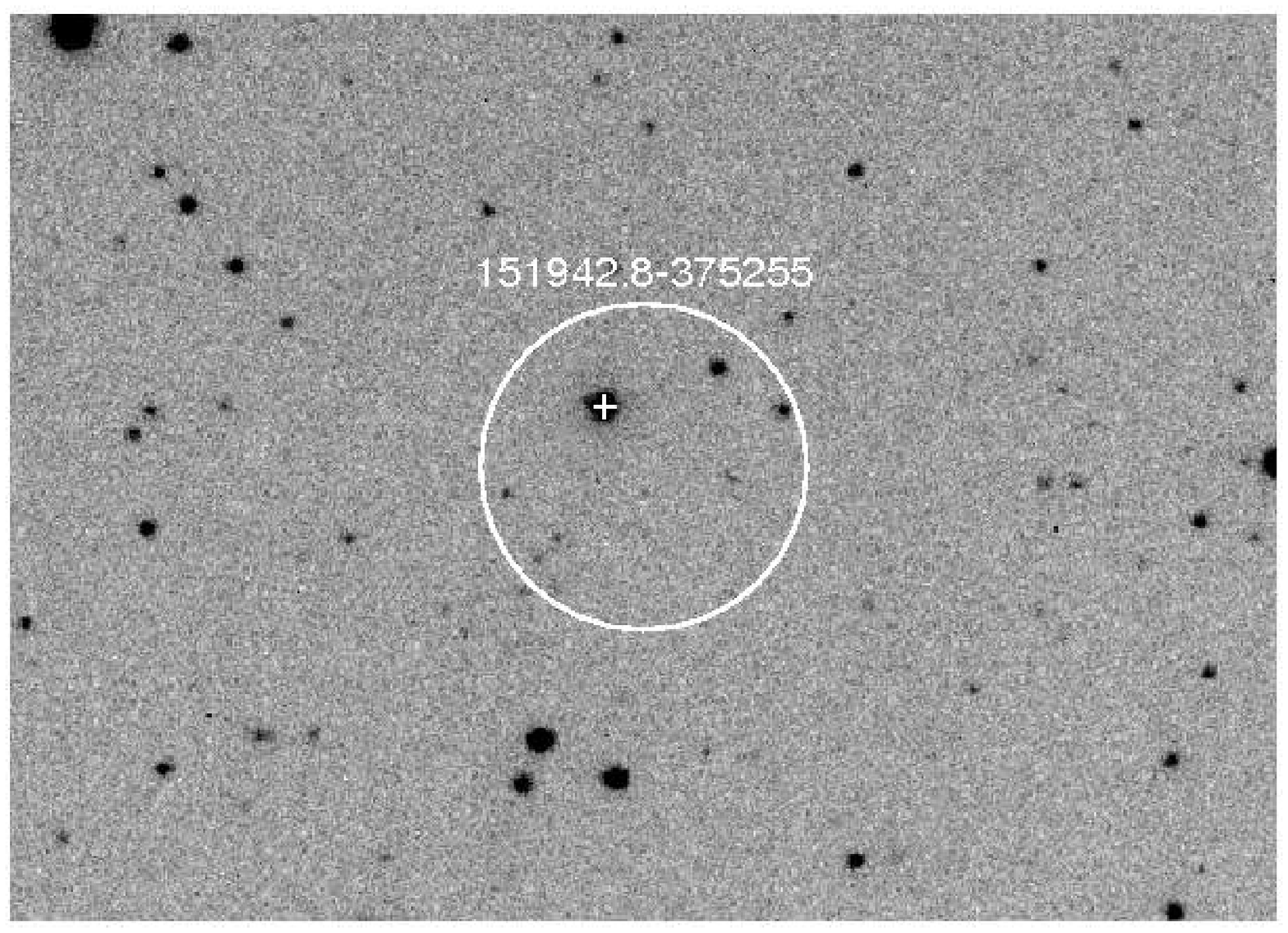}
\caption{Magellan images of 1RXS~J151942.8$-$375255.  Left: MagIC
  $R$-band; right: PANIC $K_s$-band.  The images are
  $\approx45\arcsec\times30\arcsec$, with North up and East to the left.
  The BSC X-ray position uncertainties are indicated by the circles,
  and the \chandra\ positions and optical counterparts are shown by
  the crosses.
\label{fig:mag}}
\end{figure*}

\begin{description}
\item[1RXS~J193458.1+335301] The \chandra\ source is extended.  See \S~\ref{sec:rxsa}.
\item[1RXS~J193228.6+345318] The \chandra\ source is point-like, and
  is at $19\hr32\mn27\fs25$, $+34\degr53\arcmin14\fs8$ ($17\arcsec$
  away from the BSC position).  It is coincident with the
  $K_s=14.2$~mag source 2MASS~J19322722+3453148, the $B=16.8$~mag
  source USNO~1248-0333432, and with the NVSS
  source identified in Table~\ref{tab:id}.  It was identified as a
  flare star by \citet{fs03} after our initial source selection.
\item[1RXS~J205042.9+284643] There is no \chandra\ source in the followup
  observation.  See \S~\ref{sec:rxsb}.
\item[AX~J2049.6+2939] This \textit{ASCA} source was identified as a
  possible neutron star by \citet{mtt+98,mot+01}. The
  \chandra\ source is point-like, and is at $20\hr49\mn35\fs41$,
  $+29\degr38\arcmin50\farcs9$.  It is coincident with the
  $K_s=10.0$~mag source 2MASS~J20493540+2938509 (also
  USNO~1196$-$0518650), which is presumably the $V=12.6$~mag G star
  discussed by \citet{mtt+98}. Our observed count-rate for this source
  ($\approx 0.03\mbox{ s}^{-1}$ in the 0.3--2.0~keV band) is roughly
  comparable with that predicted from the latest \textit{ASCA}
  spectroscopy, although as noted in \citet{mot+01} the source appears
  to be variable.  Given the variability of this source and the
  extremely tight coincidence with a G star (in this region, there are
  $\expnt{(1.97 \pm 0.07)}{-5}\mbox{ arcsec}^{-2}$ stars with $K\leq10.0$~mag
  in 2MASS, and to find one $<0\farcs2$ away from the X-ray source has
  a chance rate of $\approx \expnt{2}{-6}$; see Fig.~\ref{fig:stars}),
  the X-ray emission is very likely from an active star.
  \citet{mtt+98} had dismissed this possibility because of the high
  X-ray intensity, but it is in fact consistent with the majority of
  the stars that we detect here (Fig.~\ref{fig:fxk}; the X-ray flux is
  $\sim 10^{-13}\mbox{ ergs s}^{-1}\mbox{ cm}^{-2}$).
\item[1RXS~J045707.4+452751] The \chandra\ source is point-like, and
  is at $04\hr57\mn08\fs31$, $+45\degr27\arcmin49\farcs8$ ($10\arcsec$
  away from the BSC position).  It is coincident with the
  $K_s=14.5$~mag source 2MASS~J04570832+4527499.
\item[1RXS~J050339.8+451715] The \chandra\ source is point-like, and
  is at $05\hr03\mn39\fs59$, $+45\degr16\arcmin59\farcs5$ ($15\arcsec$
  away from the BSC position).  It is coincident with the
  $K_s=15.0$~mag source 2MASS~J05033958+4516594 and with the NVSS
  source identified in Table~\ref{tab:id}.  There is no USNO counterpart.
\item[1RXS~J062740.3+073103] The \chandra\ source is point-like, and
  is at $06\hr27\mn40\fs12$, $+07\degr31\arcmin00\farcs3$ ($4\arcsec$
  from the BSC position).  It is coincident with the $K_s=10.1$~mag
  source\\2MASS~J06274012+0731006.
\item[\lnsa]The \chandra\ source is point-like, and
  is at $06\hr30\mn05\fs29$, $+05\degr45\arcmin40\farcs8$.  It is coincident with the
  $K_s=10.0$~mag source 2MASS~J06300529+0545407.
\item[\lnsb]The \chandra\ source is point-like, and
  is at $06\hr33\mn33\fs22$, $+06\degr08\arcmin39\farcs5$.  It is coincident with the
  $K_s=13.4$~mag source 2MASS~J06333322+0608396.
\item[\lnsc]The \chandra\ source is point-like, and
  is at $06\hr39\mn25\fs67$, $+05\degr14\arcmin30\farcs1$.  It is coincident with the
  $K_s=11.6$~mag source 2MASS~J06392566+0514301.
\item[1RXS~J150818.8$-$401730]The \chandra\ source is point-like, and
  is at $15\hr08\mn18\fs17$, $-40\degr17\arcmin26\farcs0$ ($8\arcsec$
  away from the BSC position).  It is coincident with the
  $K_s=9.3$~mag source 2MASS~J15081819$-$4017261.
\item[1RXS~J150139.6$-$403815] The \chandra\ source is extended.  See \S~\ref{sec:rxsc}.
\item[1RXS~J151942.8$-$375255]The \chandra\ source is point-like, and
  is at $15\hr19\mn42\fs98$, $-37\degr52\arcmin51\farcs4$ ($4\arcsec$
  away from the BSC position).  There is no 2MASS counterpart, but we
  do identify a counterpart on our Magellan MagIC optical and PANIC infrared
  observations (\S~\ref{sec:oir} and Fig.~\ref{fig:mag}).  The source has $R=19.1$~mag and $K_s=15.79$~mag.
\end{description}

\begin{figure*}
\centering
\includegraphics[width=0.5\textwidth]{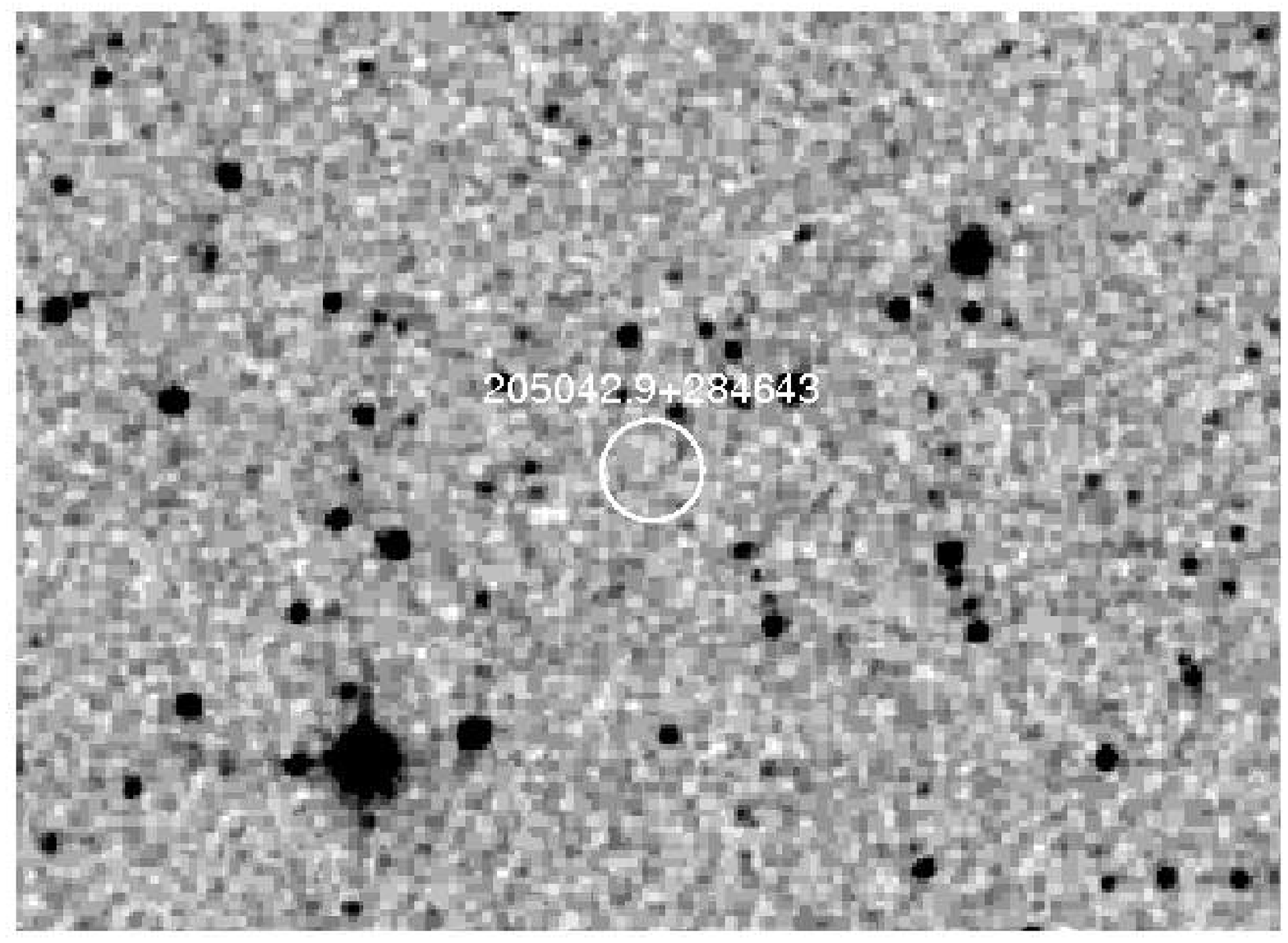}\includegraphics[width=0.5\textwidth]{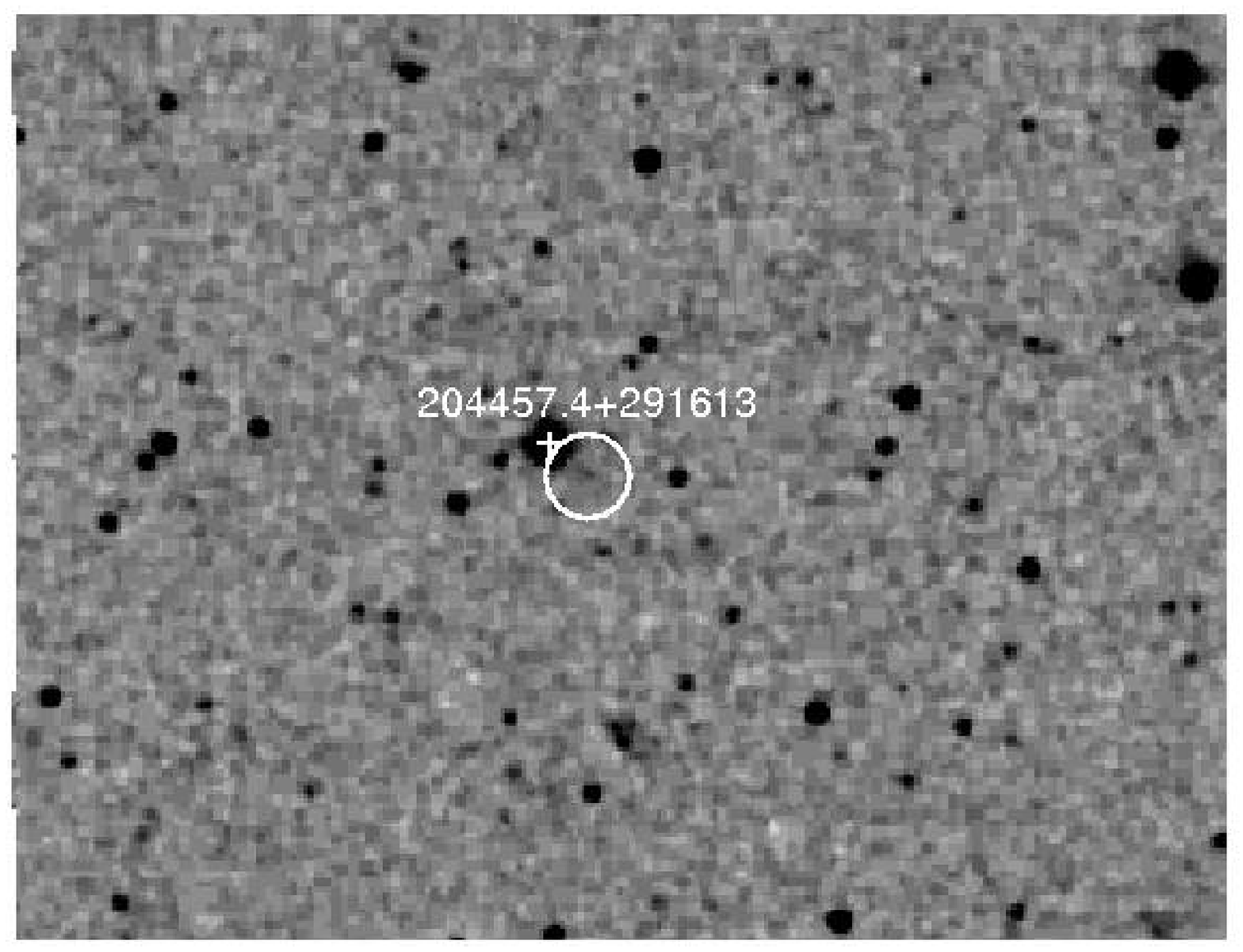}
\includegraphics[width=0.5\textwidth]{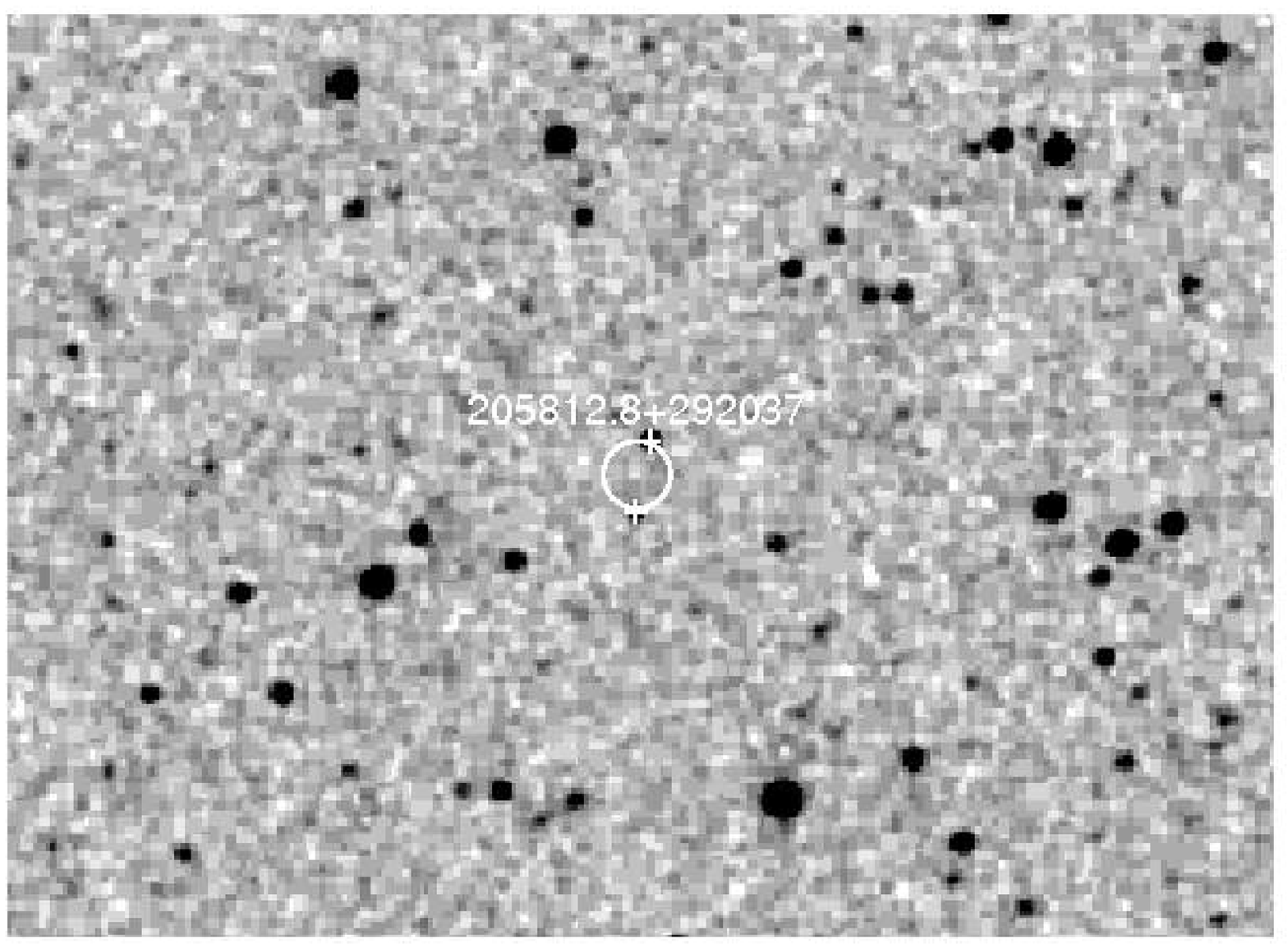}\includegraphics[width=0.5\textwidth]{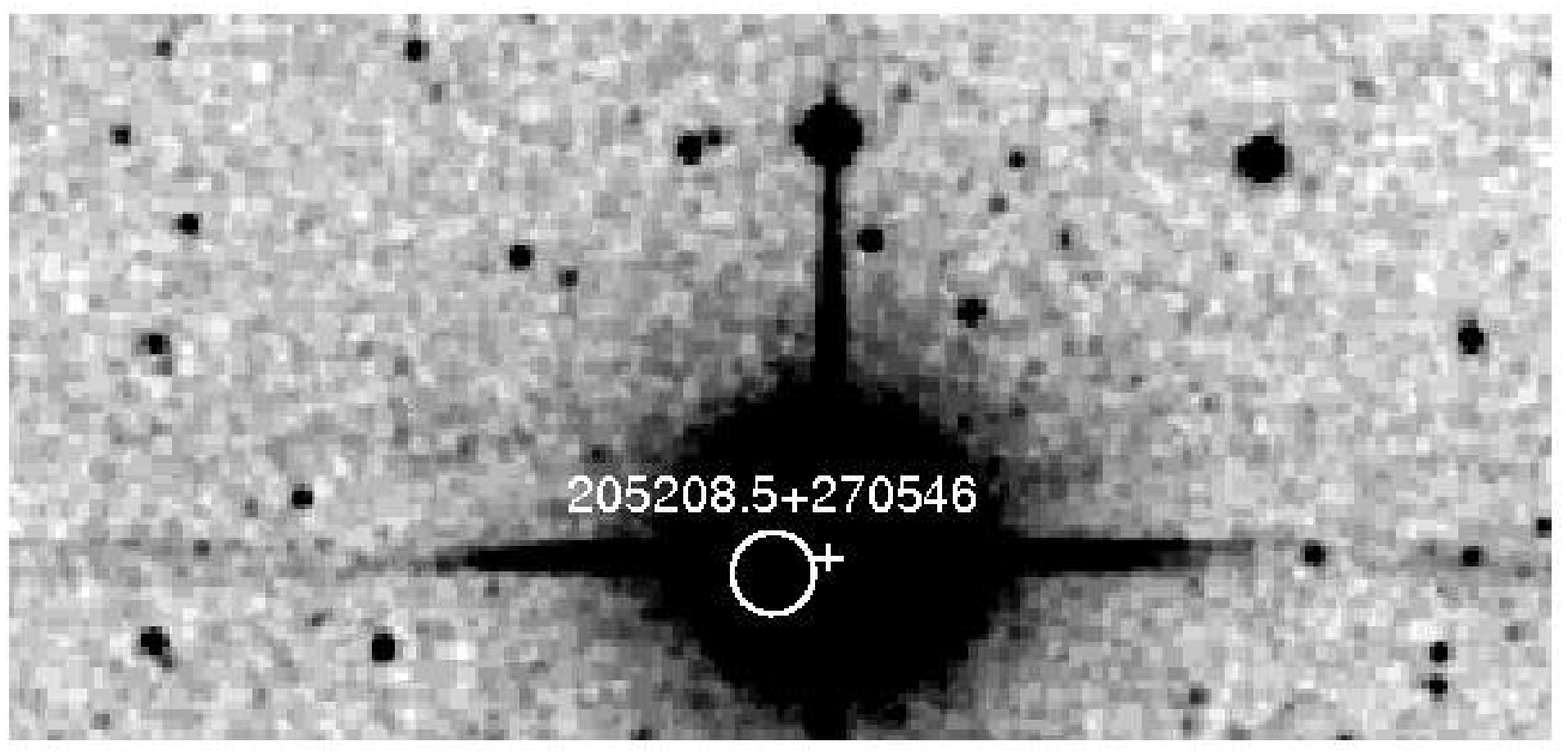}
\caption{2MASS $K_s$-band images of the sources in \snrb\ from
  Table~\ref{tab:srcs}.  The images are $5\arcmin\times3.5\arcmin$, with
  North up and East to the left.  The X-ray position uncertainties are
  indicated by the circles, and the proposed optical counterparts are
  shown by the crosses.\label{fig:opt4}}
\end{figure*}

\subsection{Optical and Infrared Followup}
\label{sec:oir}
We obtained optical and infrared followup observations of three of the
four sources where the \chandra\ followup did not immediately identify
a counterpart, namely \rxsa, 1RXS~J151942.8$-$375255, and \rxsc.  For
\rxsa, we got a 300-s $g^{\prime}$-band exposure with the Large Format
Camera (LFC) on the Palomar 200-inch telescope and 1200-s exposures in
H$\alpha$ and a narrowband filter located away from major emission
lines with the CCD camera (P60CCD) on the Palomar 60-inch telescope.
For 1RXS~J151942.8$-$375255, we obtained a 930-s $R$-band exposure
with the Raymond and Beverly Sackler Magellan Instant Camera (MagIC)
on the 6.5-m Clay (Magellan II) telescope, and a 200-s $K_s$-band
exposure with Persson's Auxiliary Nasmyth Infrared Camera (PANIC;
\citealt{mpm+04}), also on the 6.5-m Clay (Magellan II) telescope. For
\rxsc, we got a 160-s $B$-band exposure and a 60-s $R$-band exposure
with MagIC.  The log of the observations is given in
Table~\ref{tab:optobs}.  Reduction and calibration followed standard
procedures.  The Magellan data were taken during the same observing
runs as data presented in \citetalias{kfg+04}, and details can be
found there.

For the reduction of the remaining data, we used standard
\texttt{IRAF} routines to subtract the bias, flat-field, and then
combine separate exposures.  Significant focal-plane distortion
prevented simple addition of the LFC data, so we used the
\texttt{IRAF} \texttt{MSCRED} package to flatten each image with
custom distortion maps prior to addition.  We then performed absolute
astrometry, solving for plate-scale, rotation, and central position
relative to stars the 2MASS catalog, and getting residuals in each
coordinate of $0\farcs13$ (2100 stars) and $0\farcs17$ (4000 stars)
for P60CCD and LFC, respectively.

\subsection{Remaining Sources}
\label{sec:remain}
We find that after investigating 50 BSC sources plus four sources from
the literature and obtaining \chandra\ followup of 13 of these
sources, there remain four X-ray sources that do not have very
likely optical counterparts and are therefore worthy of extended
discussion.  As noted in \S\S~\ref{sec:cpt} and \ref{sec:notescxo},
these sources are: \rxsa, \rxsb, \rxsd, and \rxsc. The first and
fourth have extended X-ray emission, while the second has no apparent
\chandra\ counterpart.  The third has a probable but not definite
association with optical/IR sources.  We now discuss all of these
sources in more detail.

\subsubsection{\rxsa}
\label{sec:rxsa}
\begin{figure}
\plotone{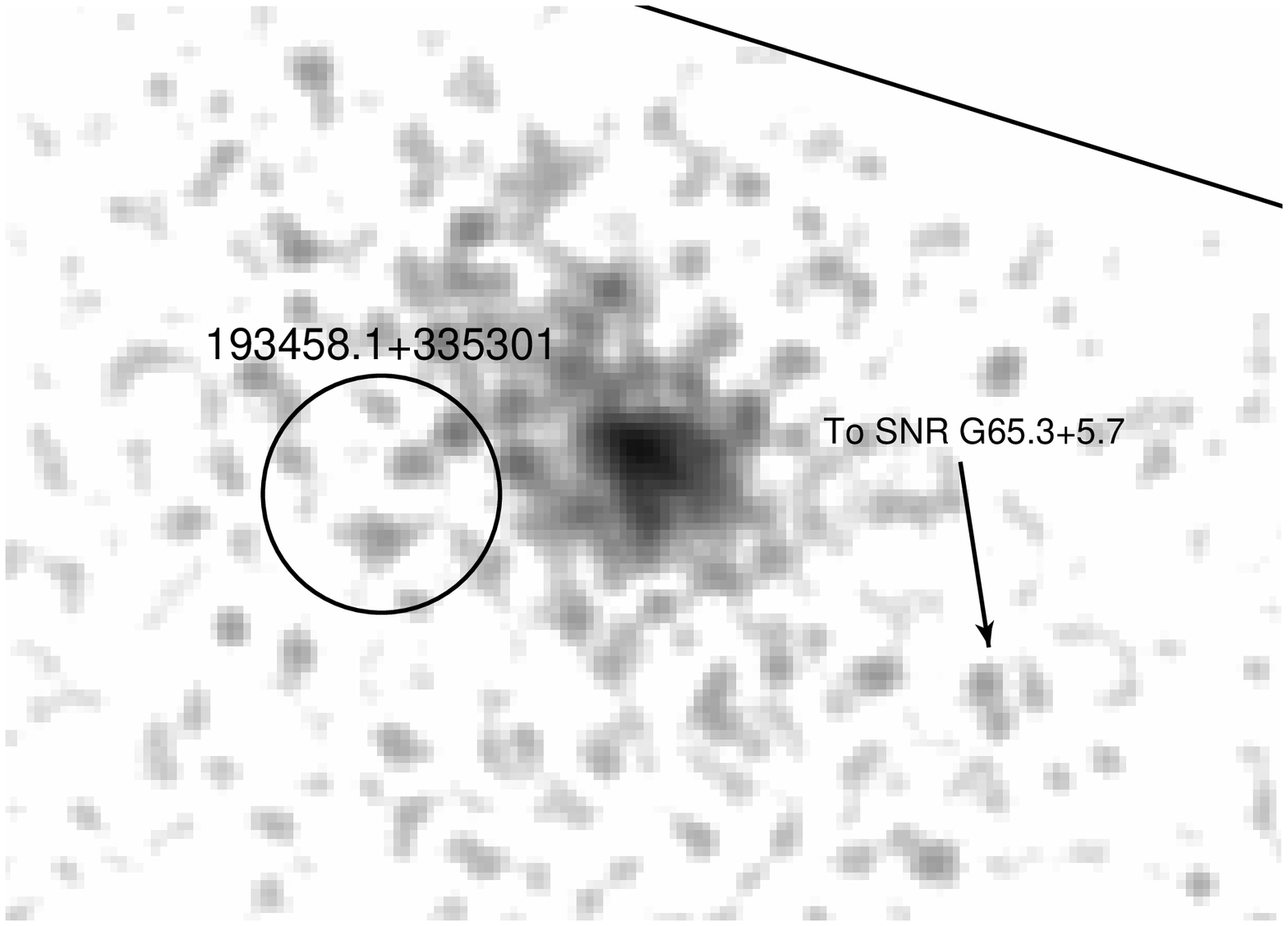}
\caption{\chandra\ ACIS-S3 image of \rxsa.  The BSC source and
  uncertainty are shown by the circle with radius $14\arcsec$.  The
  greyscale is proportional to the logarithm of the counts in 2-pixel
  bins, and the image has been smoothed with a Gaussian filter with a
  radius of 3 pixels.  The contours are in steps from 0.2--2.2~counts
  per bin, with spacing proportional to the square root of the counts.
  The arrow indicates the direction to the center of \snra.  The box
  shows the approximate extent of the ACIS subarray.  The image is
  $\approx 150\arcsec\times110\arcsec$, and has North up and East to
  the left.
\label{fig:1934}}
\end{figure}

The BSC lists \rxsa\ as having $0.051(12)\mbox{ count s}^{-1}$ in the
PSPC, with hardness ratios of ${\rm HR1}=1.00(17)$ and ${\rm
HR2}=0.10(23)$ (see \citealt{rbs2} for definitions of bands and
hardness ratios).  The corresponding \chandra\ source is clearly
extended, as shown in Figure~\ref{fig:1934}.  There are no other X-ray
sources nearby, indicating that the \chandra\ source is very likely
\rxsa\ despite the $\approx 32\arcsec$ distance between the two (this
is somewhat larger than the separations between the optical and X-ray
sources in Fig.~\ref{fig:offset}, but given the extended nature of the
X-ray source that is not that surprising).  The peak of the \chandra\
emission is at $19\hr34\mn55\fs61$, $+33\degr53\arcmin06\farcs0$, and
has an extent of $\approx 5\arcsec$.  The overall source is larger and
asymmetric, with a maximum visible distance of $\approx 40\arcsec$
from the peak to the North-East and a minimum distance of $\approx
15\arcsec$ from the peak to the South-West, although there is some
diffuse emission that extends further.  Averaged over azimuth, the
half-power radius is $11\arcsec$, and 95\% of the power is within
$42\arcsec$.  Fitting the spatial profile to a $\beta$-model (surface
brightness $\propto (1+(r/r_{c})^2)^{-3\beta+0.5}$, typical for galaxy
clusters) in \texttt{Sherpa} was successful, with core radius
$r_{c}=3.5(1)\arcsec$, $\beta=0.451(5)$, an amplitude of
$0.64(3)\mbox{ count pixel}^{-2}$, and $\chi^{2}=9.6$ for 12 degrees
of freedom.


We extracted photon events from a $45\arcsec\times 22\arcsec$ region
and created source and background response files using the
\texttt{CIAO} task \texttt{acisspec}.  We used versions of 3.0.2 of
\texttt{CIAO} and 2.26 \texttt{CALDB} that compensate for low-energy
degradation of the ACIS
detector\footnote{\url{http://cxc.harvard.edu/ciao/threads/apply\_acisabs/}.}.
We then fit the data in \texttt{sherpa}.  The events were binned so
that each bin had $\geq 25$~counts.

\begin{figure}
\plotone{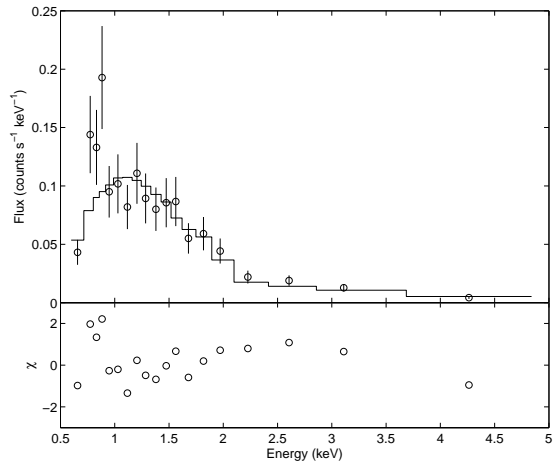}
\caption{\chandra\ ACIS-S3 spectrum of \rxsa, with the best-fit
  power-law model ($N_{\rm H}=\expnt{3.0(6)}{21}\mbox{ cm}^{-2}$,
  photon index $\Gamma=2.4(2)$, and an amplitude of
  $\expnt{4.6(8)}{-4}\mbox{ photons cm}^{-2}\mbox{ s}^{-1}\mbox{
  keV}^{-1}$ at 1~keV).  The residuals are plotted in the bottom
  panel.\label{fig:1934spec} }
\end{figure}

While there are not very many counts (601 source counts before
background subtraction, with 77.2 background counts), the data are
well fit (Fig.~\ref{fig:1934spec}) by an absorbed power-law model,
with $N_{\rm H}=\expnt{3.0(6)}{21}\mbox{ cm}^{-2}$, $\Gamma=2.4(2)$,
and an amplitude of $\expnt{4.6(8)}{-4}\mbox{ photons cm}^{-2}\mbox{
s}^{-1}\mbox{ keV}^{-1}$ at 1~keV (giving $\chi^{2}=18.7$ for 16
degrees of freedom; all uncertainties are 1-$\sigma$).  The observed
flux from this model is $\expnt{1.0}{-12}\mbox{ ergs cm}^{-2}\mbox{
s}^{-1}$ (0.3--8.0~keV), while the unabsorbed flux is
$\expnt{2.2}{-12}\mbox{ ergs cm}^{-2}\mbox{ s}^{-1}$ (0.3--8.0~keV).
\citet{rs77} plasma models do not fit, giving $\chi^2_{\nu}=70$/16.

\begin{figure}
\plotone{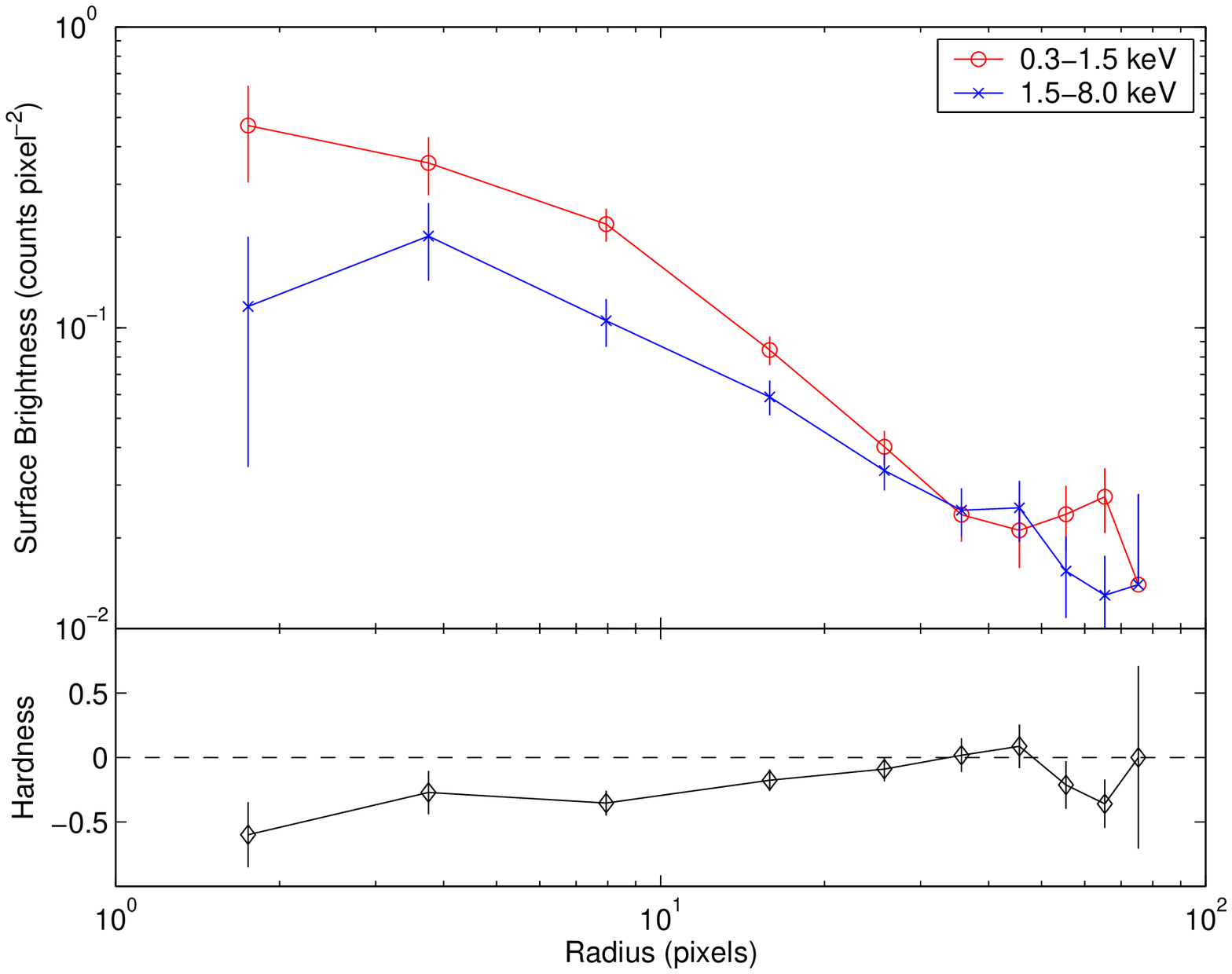}
\caption{Top: radial profiles of the events for \rxsa\ in two different
energy bands.  The background-subtracted surface brightness is plotted
against radius for soft (0.3--1.5~keV) and hard (1.5--8.0~keV) bands.
Bottom: hardness of the radial profile, defined as $(B_{\rm H}-B_{\rm
  S})/(B_{\rm H}+B_{\rm S})$, where $B_{\rm H}$ ($B_{\rm S}$) is the
surface brightness in the 1.5--8.0~keV (0.3--1.5~keV) band.
There appears to be a slight excess of soft photons toward the center.
\label{fig:radprof}}
\end{figure}

The extended morphology and the hard spectrum of the source suggest
several models: (1) a very hot nebula  of Galactic origin, (2) a
very hot nebula but of extra-galactic origin (gas from a cluster or an
early-type galaxy; \citealt{fabbiano89,bb98}), (3) the superposition of
many bright LMXBs (with power-law spectra) in an early-type
galaxy \citep{mma+94}, or (4) a pulsar wind nebula radiating via
synchrotron emission.  Here we discuss each interpretation in some detail.


\begin{description}
\item[Galactic Nebula] This would require a very bright central source
(i.e.\ an OB star) to heat the nebula, which would also be visible as
an extended optical/IR source and should show H$\alpha$ emission.
While the northern 2MASS source (2MASS~J19345557+3353136) in
Figure~\ref{fig:1934ha} is moderately bright (with about a 5\% chance
of a star this bright occurring randomly within $7\arcsec$ of a
position), it does not appear to be an OB star.  The colors are fairly
common for the field, they are more consistent with a star of type
$\approx$G0, and the extinction that would be required of an OB star
is higher than expected for this line of sight ($A_V \gsim 2$~mag,
compared to a maximum of $A_V \approx 1$ as determined by
\citealt*{dcllc03} and
W3COLDEN\footnote{\url{http://asc.harvard.edu/toolkit/colden.jsp}.}).
The southern 2MASS source (2MASS~J19345569+3353063) is fainter and
redder---consistent with being a late-type star---but its position
near the peak of the X-ray emission (chance probability of $<1$\%)
suggests that there might be an actual association between it and the
X-ray emission.  Since there is no diffuse broadband optical,
broadband IR, or H$\alpha$ emission, we do not believe that the
extended X-ray emission is powered by any of the stars, although it
may be related to one or both of the 2MASS sources.

\item[Extragalactic Nebula]The X-ray emission is much more
compact than is typical for galaxy groups or clusters (even clusters
at $z\sim 0.5$--1 have $r_{c} \gg 10\arcsec$; \citealt{aml+02,cbd+02}),
and the spectrum is wrong (thermal plasma models do not fit).

\item[Early-type Galaxy] The size is similar to what is often seen for
early-type (E and S0) galaxies.  In those galaxies the X-ray emission
comes from a combination of hot gas (plasma with $kT\approx
0.5$--1.0~keV; \citealt{fabbiano89,bb98}) and the superposition of many
hard X-ray point sources (whose spectra are power-laws with
$\Gamma\approx 1.7$) --- reasonably compatible with the observed
spectrum of \rxsa.

For these galaxies, the hard X-ray luminosity scales reasonably well
with the integrated $B$ or $K$ band luminosities as ${L_{\rm
X}}/{L_{B}} \approx {10^{30}\mbox{ ergs s}^{-1}}{L_{\odot,B}^{-1}}$ or
${L_{\rm X}}/{L_{K}} \approx {10^{29}\mbox{ ergs
s}^{-1}}{L_{\odot,K}^{-1}}$, where the X-ray luminosity is in the
0.3--8.0~keV band and all luminosities are corrected for extinction
\citep{kf04}; the scatter in
this (from a sample of 14 galaxies) is a factor of 2--3 (there is more
scatter in \citealt{bb98}, but their work concerns the soft emission
more than the hard emission, and the scatter is still only a factor of
$\sim 10$).  We can convert the relations from \citet{kf04} into
relations for fluxes and magnitudes (i.e.\ observables), giving
\begin{eqnarray}
\frac{L_{\rm X}}{L_{B}} & = & \expnt{4 \pi}{39} F_{\rm X}
10^{(m_B-M_{\odot,B}-N_{\rm H,21}/{1.3})/2.5} \frac{\mbox{ ergs
    s}^{-1}}{L_{\odot,B}}\nonumber \\
\frac{L_{\rm X}}{L_{K}} & = & \expnt{4 \pi}{39} F_{\rm X}
10^{(m_K-M_{\odot,K}-N_{\rm H,21}/{16})/2.5} \frac{\mbox{ ergs
    s}^{-1}}{L_{\odot,K}},
\label{eqn:flux}
\end{eqnarray}
where $F_{\rm X}$ is the absorption-corrected X-ray flux in the
0.3--8.0~keV band in units of $\mbox{ ergs s}^{-1}\mbox{ cm}^{-2}$,
$m_B$ ($m_K$) is the observed $B$-band ($K$-band) magnitude, $N_{\rm
  H,21}$ is the Galactic column density along the line of sight in
units of $10^{21}\mbox{ cm}^{-2}$, and $M_{\odot,B}=5.47$~mag
($M_{\odot,K}=3.33$~mag) is the $B$-band ($K$-band) absolute magnitude
of the Sun (we have assumed $N_{\rm H,21}=1.79A_V$ following
\citealt{ps95}).

The relation in Equation~\ref{eqn:flux} implies that for $F_{\rm
X}=10^{-12}\mbox{ ergs s}^{-1}\mbox{ cm}^{-2}$ and $N_{\rm H,21}=1$,
$m_{B}\approx 12$~mag and $m_{K}\approx 5$~mag (this indeed matches
what is seen for sources in \citealt{kf04}).  Therefore, such sources
should be readily visible on even shallow images.  Examining 2MASS and
Palomar (Fig.~\ref{fig:1934ha}) images of \rxsa\ we see that there are
two optical/IR sources near the peak of the X-ray emission: the
northern source appears stellar (FWHM$\approx 1\farcs4$), while the
southern source may have some extended emission\footnote{There are no
data from the 2MASS Extended Source Catalog (XSC) in this region owing
to the presence of a bright ($V=6.7$~mag) M star $90\arcsec$ to the
South.} to the North-East (although this could be a superposition of
point sources).  However, neither of these sources is a great
candidate for the origin of the X-ray emission, as they are too faint
by several orders of magnitude ($K_{s}=14.4$~mag and $K_s=12.6$~mag
for the northern and southern sources, respectively) and not extended
enough.  Therefore, while it is not impossible that \rxsa\ is an
early-type galaxy, we consider it unlikely.  Deeper X-ray observations
should be conclusive: if \rxsa\ is a galaxy, it should resolve into
discrete point sources.  Optical spectroscopy would also be useful in
determining the natures of the optical/IR sources.

\begin{figure*}
\plotone{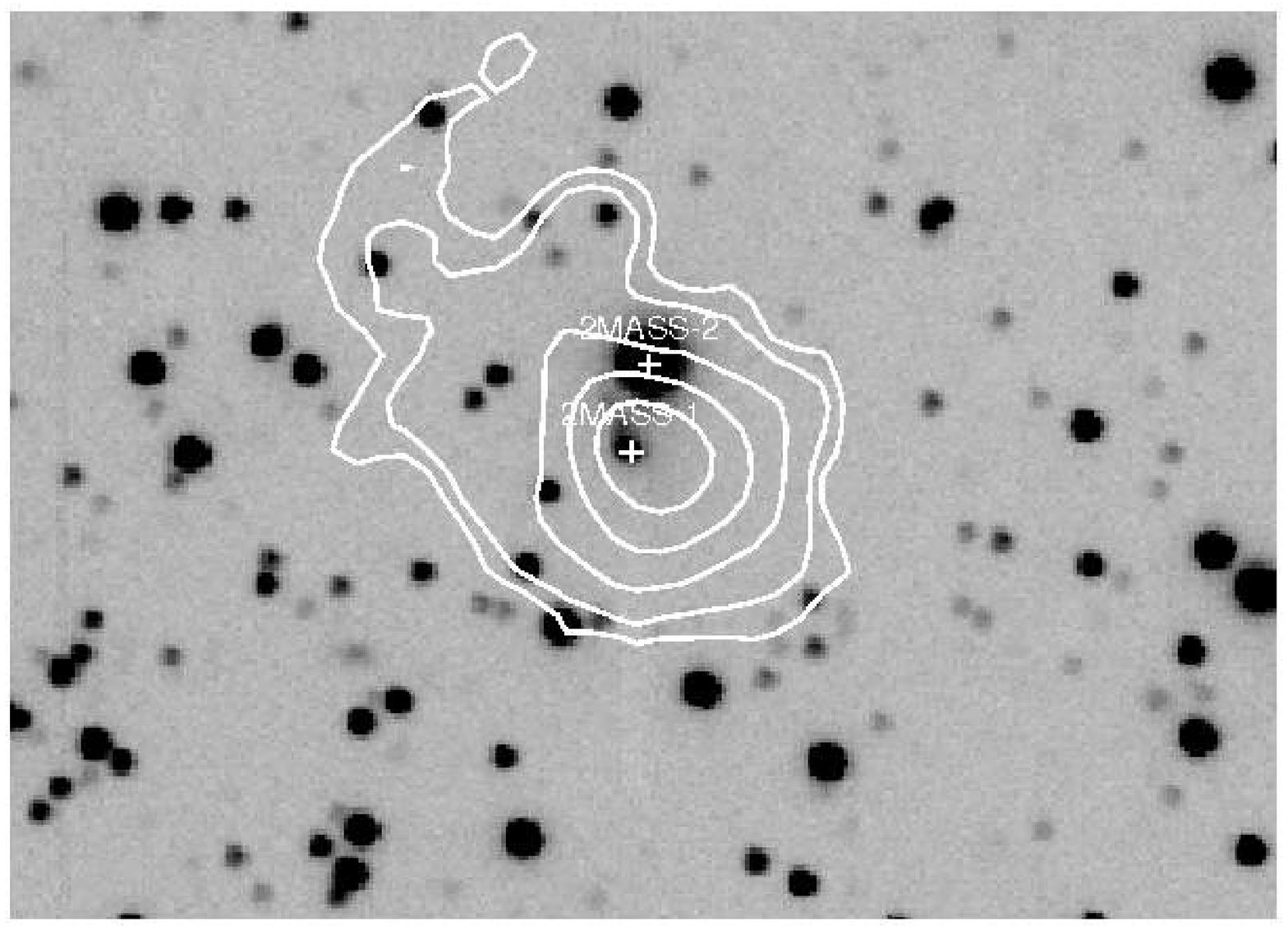}
\plottwo{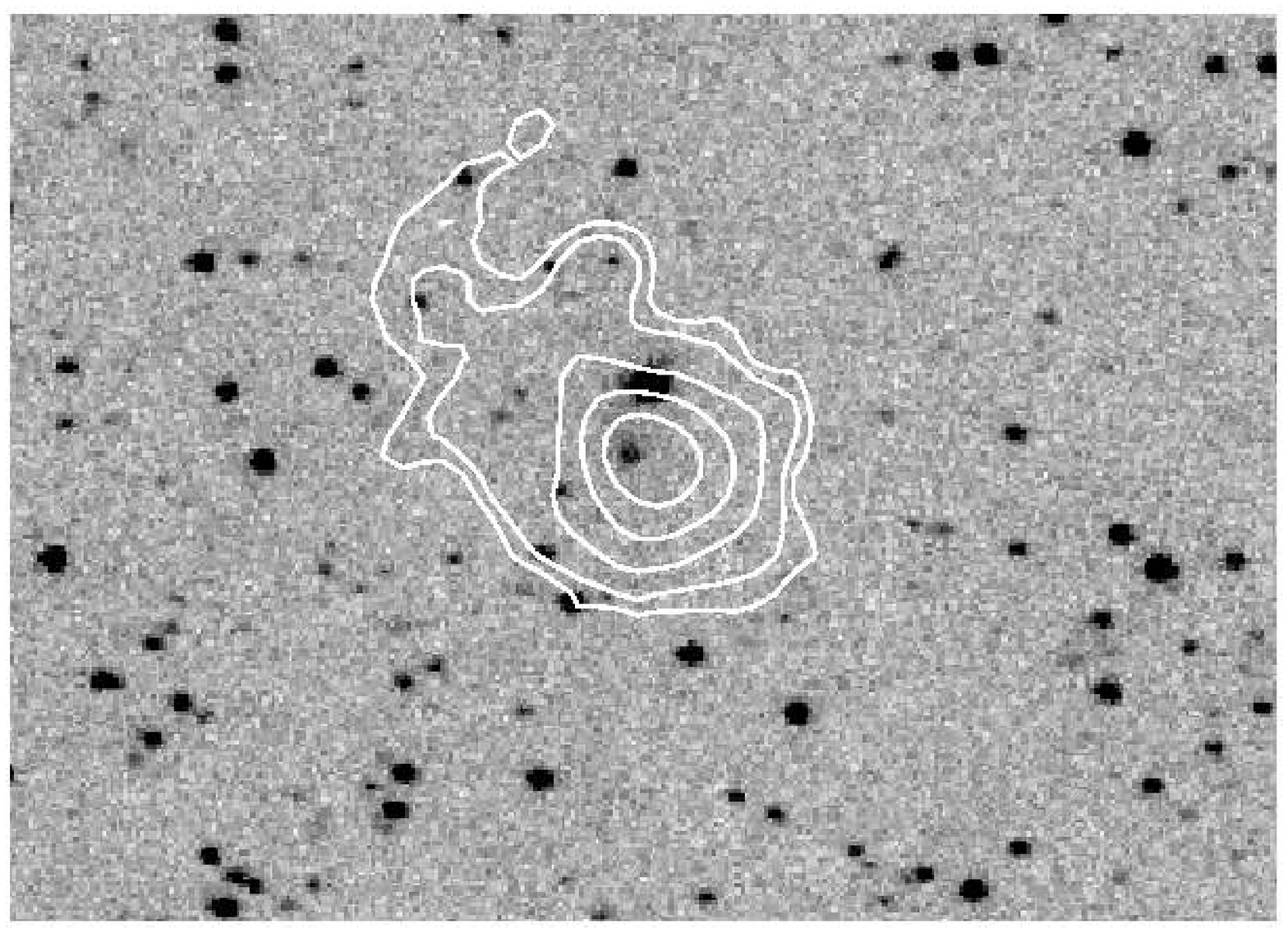}{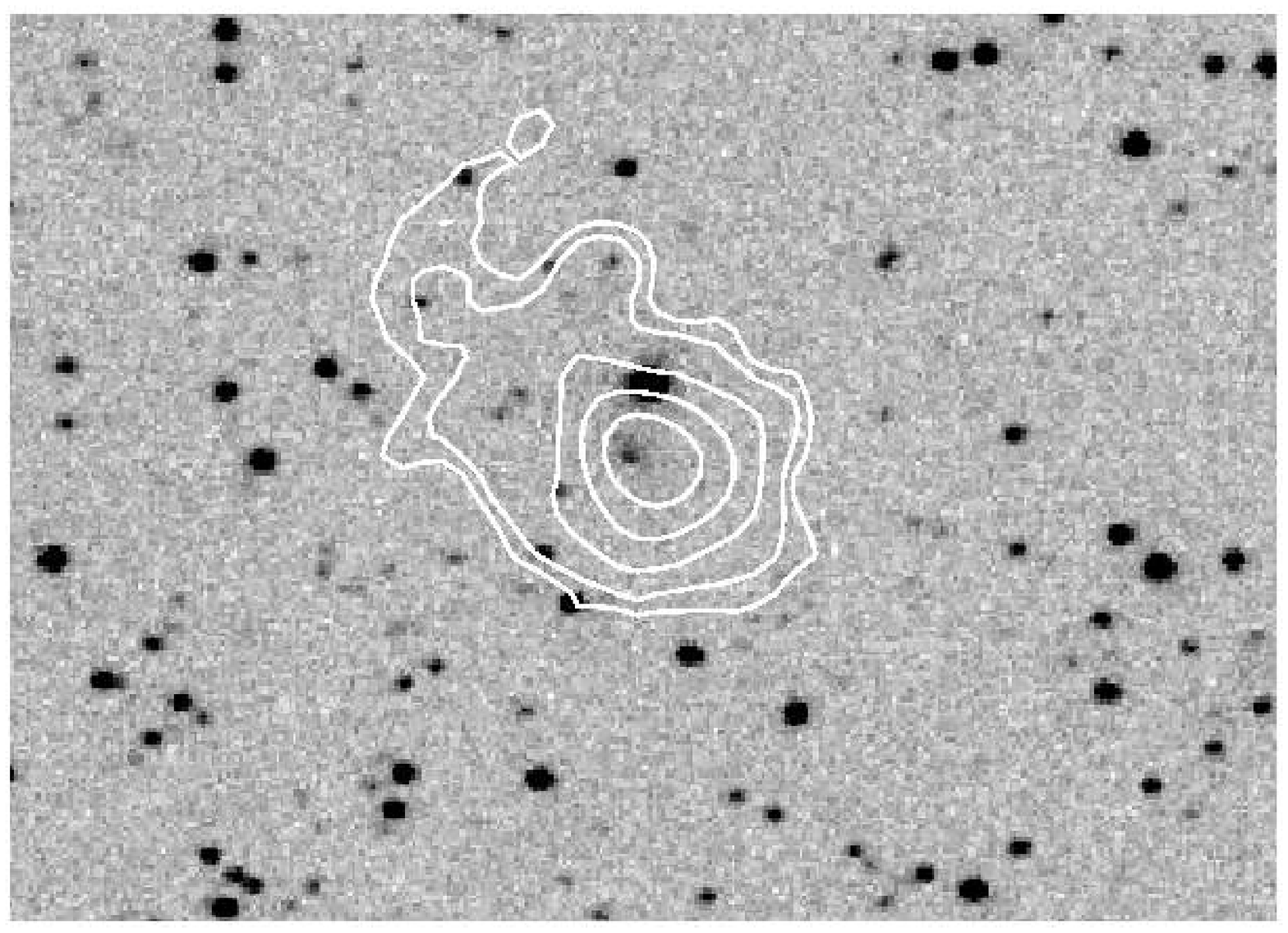}
\caption{Top: Palomar 200-inch $g^\prime$-band image of \rxsa.
  Bottom: Palomar 60-inch images of \rxsa, taken with H$\alpha$ (left)
  and off-band (right, 6584~\AA) filters.  The images are $\approx
  110\arcsec\times80\arcsec$, and have North up and East to the left.
  The contours are those from Figure~\ref{fig:1934} showing the extent
  of the X-ray emission.  The two 2MASS sources identified near the
  peak of the X-ray emission are indicated with the crosses in the top
  image; the southern source is 2MASS~J19345569+3353063 while the
  northern is 2MASS~J19345557+3353136.  We do not detect any diffuse
  H$\alpha$ emission associated with \rxsa.
\label{fig:1934ha}
}
\end{figure*}

There are some early-type galaxies with significant excesses of X-ray
emission \citep{vmh+99}, largely due to increases in the amounts of
hot gas that give roughly the same X-ray-to-optical ratio as would be
necessary here.  However, the optical/IR sources in
Fig.~\ref{fig:1934ha} do not look like bright galaxies (unlike the
galaxies from \citealt{vmh+99} which are typically $>20\arcsec$ in the
optical) and the spectrum of \rxsa\ is wrong: again, thermal plasma
models do not fit.

\item[PWN]A pulsar wind nebula (i.e., a nebula excited by a pulsar or
PWN; for a review, see \citealt{gs06}) is consistent with the size and
spectrum of \rxsa, although the source is slightly softer toward the
center (Fig.~\ref{fig:radprof}), contrary to what is expected for
PWNe.  There is no obvious H$\alpha$ emission from \rxsa\ in
Figure~\ref{fig:1934ha} as there can be near PWNe \citep[associated
either with SN ejecta or with the passage of the pulsar through the
interstellar medium;][]{hgbg90,cc02}, but this could be because the
conditions are not favorable.

\rxsa\ is outside \snra.  If it were a bow-shock nebula that
originated in the interior of the SNR and then moved outside the shell
(and not a static PWN inflated by the wind of its central pulsar), one
would expect H$\alpha$ emission and for the X-ray nebula to trail
away from the direction of motion/toward the SNR center (e.g.,
\citealt{sgk+03}, although this is not always the case).  Since the
X-ray emission trails \textit{away} from the SNR center (suggesting
motion toward the SNR instead of out of it), we see no H$\alpha$, the
fitted value of $N_{\rm H}$ is just about at the maximum predicted for
this line of sight by W3COLDEN and is somewhat higher than the nominal
value for \snra\ (suggesting that the X-ray source may be more distant
and highly absorbed than \snra), we believe that an association
between the two is unlikely.  However, this is not entirely
unexpected, as there are a number of young, newly-discovered PWNe that
have no definitively associated SNRs (similar to the the Crab Nebula;
\citealt*{sgs05}).
\end{description}

\subsubsection{\rxsb}
\label{sec:rxsb}
\begin{figure*}
\plotone{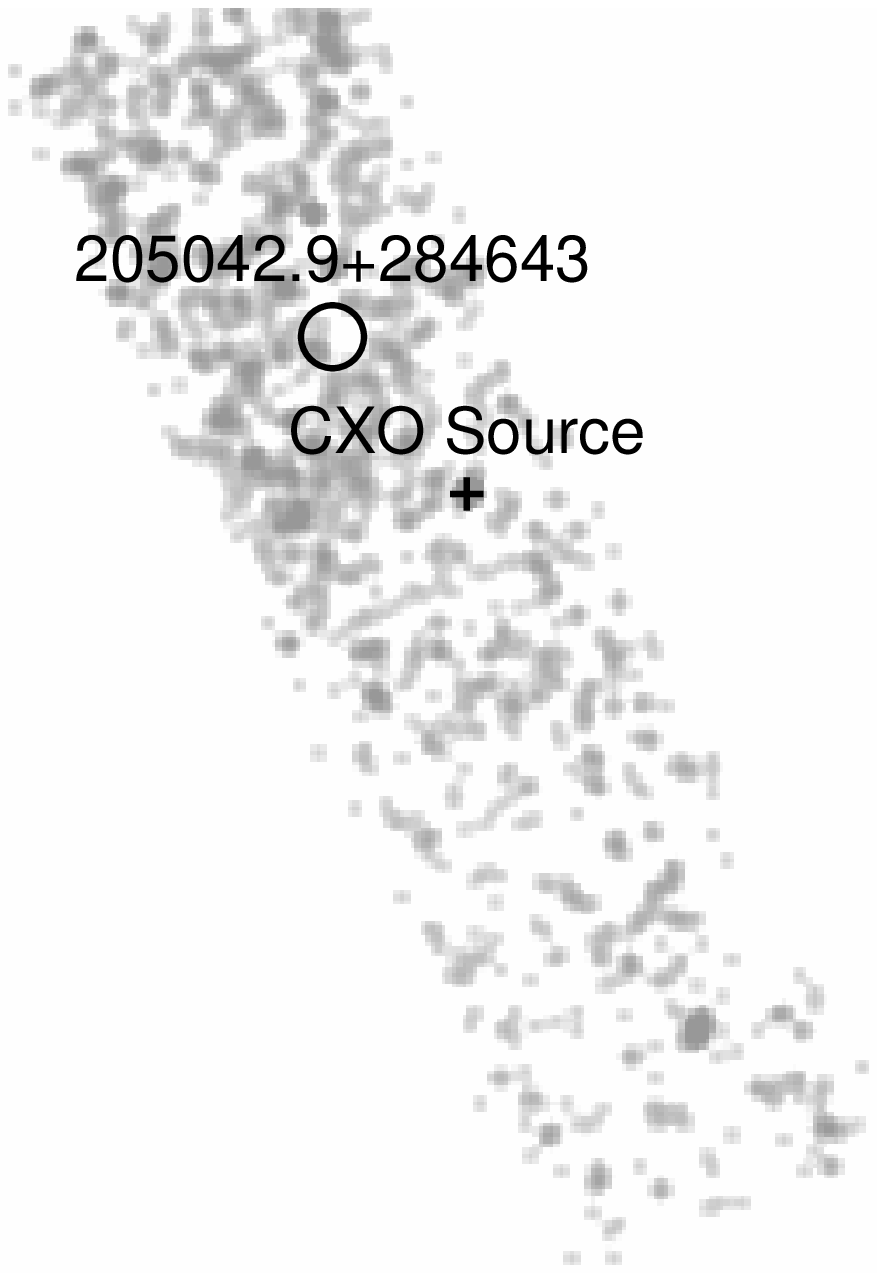}
\caption{\chandra\ ACIS-S image of the field around \rxsb.  The
  position of the \rosat\ source is shown by the circle with a radius
  of $12\arcsec$, which is the 1-$\sigma$ position uncertainty.  The
  only significant point source detected in the \chandra\ observation
  is shown by the cross, $79\arcsec$ away from the \rosat\ position:
  see \S~\ref{sec:rxsb}.  The greyscale is proportional to the
  logarithm of the 0.3--5.0~keV counts in 4-pixel bins, and the image
  has been smoothed with a Gaussian filter with a radius of 3 pixels.
\label{fig:rxsb}}
\end{figure*}

The BSC lists \rxsb\ as having $0.11(2)\mbox{ count s}^{-1}$ in the
PSPC, with hardness ratios of ${\rm HR1}=-0.03(17)$ and ${\rm
HR2}=-0.87(14)$.  The \chandra\ observation of \rxsb\ had a total
exposure time of 3.7~ksec, and which should give $\approx 1000$ ACIS-S
counts depending on the source spectrum.  However, as shown in
Figure~\ref{fig:rxsb} there are no point sources detected  anywhere
within three times the nominal position uncertainty (a conservative
limit, as seen in Fig.~\ref{fig:offset}): the only significant source
detected (using \texttt{wavdetect} on scales from 1--32~pixels; see
\citetalias{kfg+04} for the detection method) in the data-set is at
$20\hr50\mn39\fs01$, $+28\degr45\arcmin43\farcs6$ (with $12\pm3$
counts), which is $79\arcsec$ or 6-$\sigma$ away from \rxsb.  This
X-ray source is almost certainly not related to \rxsb.  We can then
set a limit of $\approx 3$~counts to any point source.  There are no
obvious extended sources, but such limits are more difficult to
quantify: overall, there are 1047 counts in the 0.3--5.0~keV energy
range over the whole $512\arcsec \times 128\arcsec$ image, so the
average background rate is $0.0160(5)\mbox{ arcsec}^{-2}$.  Then, in a
region $\theta\times \theta\mbox{ arcsec}^{2}$ in area, the 3-$\sigma$
limits will be $3\sqrt{0.016\theta^2+(\expnt{3}{-7})\theta^4}$~counts.
There are no regions in the event list with such concentrations, so no
extended sources are present.

One obvious explanation of the disappearance of \rxsb\ is 
variability.  This is not atypical among the most common class of soft
X-ray sources in the Galactic plane: active stars.  Flares and other
chromospheric/magnetospheric events often lead to dramatic changes in
the fluxes of these sources.  While other sources, such as X-ray
binaries, active galaxies, and some anomalous X-ray pulsars, do
exhibit variability, these sources have hard X-ray spectra generally
inconsistent with the BSC emission.  We therefore think it likely that
\rxsb\ is an active star, but of course this cannot be confirmed
without additional data.  It is also possible that the source is
extended, and therefore too diffuse to have been detected by \chandra.
The softness of the BSC emission make this unlikely, though, as most
known types of extended sources are relatively hard (e.g., \rxsa\ and
\rxsc).

It is possible, but unlikely, that \rxsb\ is a neutron star.  As
discussed above, most of
the neutron stars considered in \citetalias{kfg+04} have stable X-ray
emission: only some of the AXPs vary significantly.  However, the
spectrum of \rxsb\ is unlike those of AXPs (typically a power-law with
$\Gamma \sim 3$).

\subsubsection{\rxsd}
\label{sec:rxsd}
The BSC lists \rxsd\ as having $0.13(2)\mbox{ count s}^{-1}$ in the
PSPC, with hardness ratios of ${\rm HR1}=0.63(11)$ and ${\rm
HR2}=0.27(13)$.  This is moderately hard compared to the other sources
in Fig.~\ref{fig:hard}, but is not too extreme.

Unlike the rest of the sources without \chandra\ followup, the
counterpart(s) shown in Figure~\ref{fig:opt4} are not entirely secure.
Within $10\arcmin$ of \rxsd, there are 1627 2MASS sources, for an
average density of $\expnt{1.44(4)}{-3}\mbox{ arcsec}^{-2}$.  To find
a source within $9\arcsec$ (as in the case of \rxsd) has a chance
probability of 37\%, and the chance probability for two sources is
13\%.  These are not small enough for a definite association.  \rxsd\
is similar, in both hardness ratio and optical brightness, to other
sources like 1RXS~J193228.6+345318, \rxsb, 1RXS~J045707.4+452751, or
1RXS~J151942.8$-$375255.  These sources did not have certain
associations based on \rosat\ alone, but the \chandra\ data are
unambiguous.  These sources may represent a population of X-ray
sources that are somewhat fainter than the majority of the sources in
Table~\ref{tab:srcs}.  This faintness, together with the hardness of
the X-ray spectrum, likely reflects extragalactic origins of the
sources (i.e.\ they are active galaxies) : in Figure~\ref{fig:fxk}
these sources are largely those with the highest X-ray-to-IR flux
ratios most similar to the extragalactic sample.

\begin{figure}[b]
\plotone{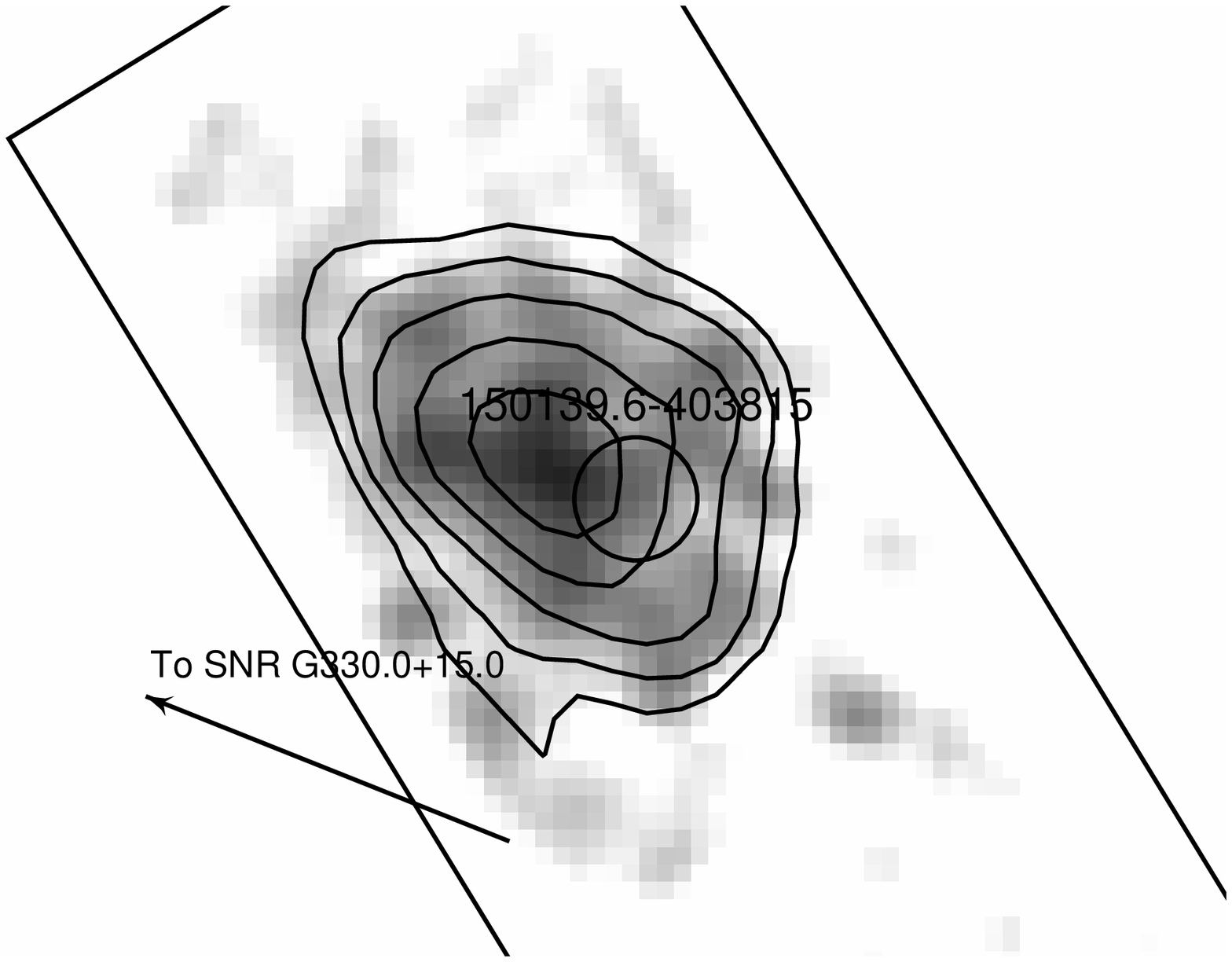}
\caption{\chandra\ ACIS-S3 image of \rxsc.  The BSC source and
  uncertainty are shown by the circle with radius $14\arcsec$.  The
  greyscale is proportional to the logarithm of the counts in 8-pixel
  bins, and the image has been smoothed with a Gaussian filter with a
  radius of 3 pixels.  The contours are in steps from 1.5--3.8~counts
  per bin, with spacing proportional to the square root of the counts.
  The box shows the approximate extent of the ACIS subarray.  The
  image is $\approx 300\arcsec\times 210\arcsec$, and has North up and
  East to the left.
\label{fig:1501}}
\end{figure}

Overall, \rxsd\ is consistent with having an association with one of
the identified 2MASS sources.  A \chandra\ followup observation would
have made the case secure, but it was not selected for \chandra\ due
to an oversight.  As with \rxsb, we do not believe that \rxsd\ is a
neutron star, but we cannot rule out this possibility.
\ \\
\subsubsection{\rxsc}
\label{sec:rxsc}

The BSC lists \rxsc\ as having $0.12(2)\mbox{ count s}^{-1}$ in the
PSPC, with hardness ratios of ${\rm HR1}=0.88(11)$ and ${\rm
HR2}=0.14(20)$.  The \chandra\ source is fainter than that of \rxsa,
but nonetheless it appears extended, as shown in
Figure~\ref{fig:1501}.  Since this source is more diffuse than \rxsa,
the spatial measurements are not as precise, but the center is at
approximately $15\hr01\mn41\fs1$, $-40\degr38\arcmin08\arcsec$.  The
total extent of the source is $\approx 1\arcmin$ in radius.  As with
\rxsa, while there is some offset between the \rosat\ and \chandra\
positions, this does not appear inconsistent with the position
uncertainties for such extended sources.  Again, we can be quite
confident that the \chandra\ source is \rxsc, since there are no other
sources nearby.

\begin{figure}[t]
\plotone{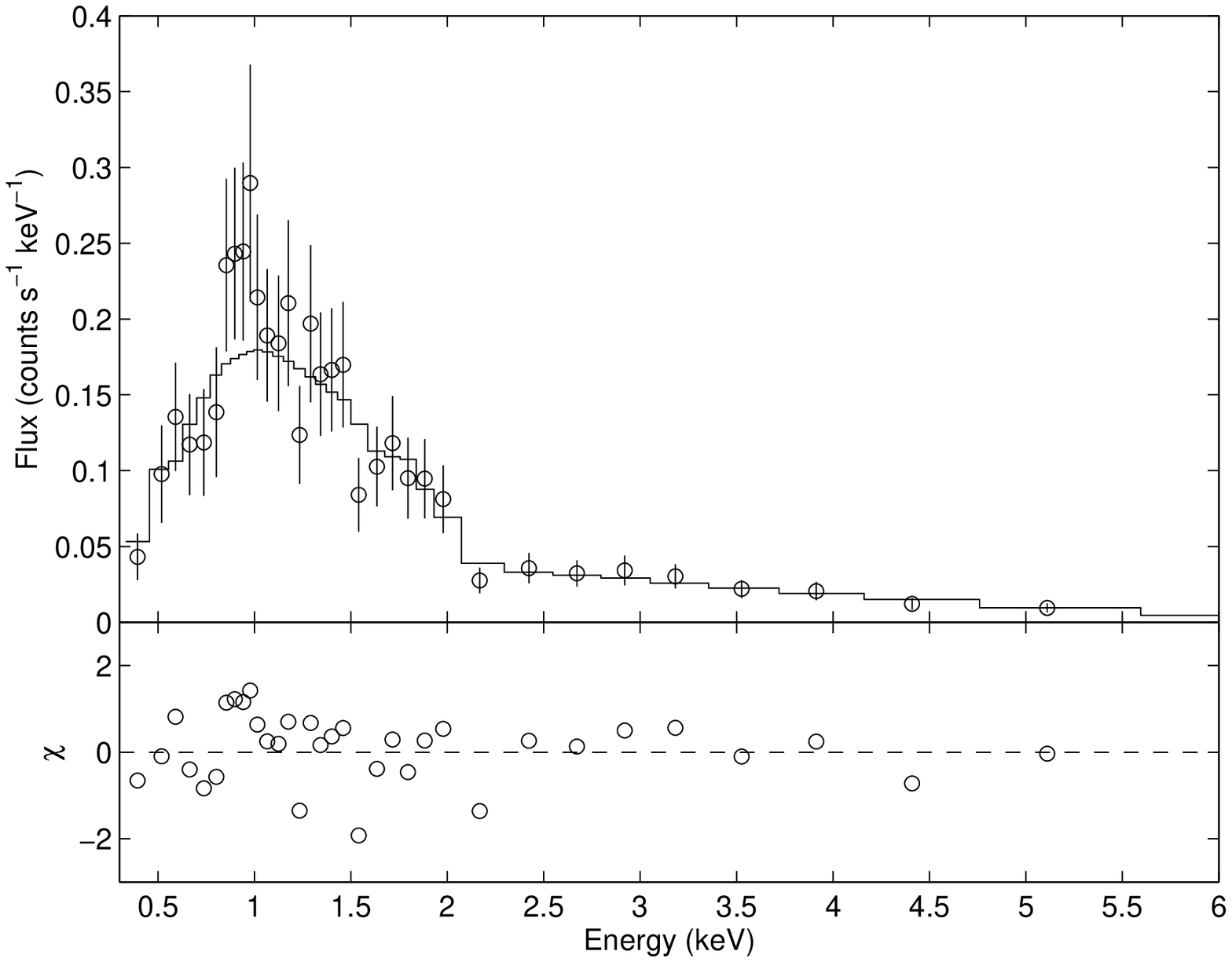}
\caption{\chandra\ ACIS-S3 spectrum of \rxsc, with the best-fit
  power-law model ($N_{\rm H}=\expnt{1.0(4)}{21}\mbox{ cm}^{-2}$,
  photon index $\Gamma=1.65(15)$, and an amplitude of
  $\expnt{4.7(6)}{-4}\mbox{ photons cm}^{-2}\mbox{ s}^{-1}\mbox{
  keV}^{-1}$ at 1~keV).  The residuals are plotted in the bottom
  panel.\label{fig:1501spec} }
\end{figure}

Similar to our analysis of \rxsa\ (\S~\ref{sec:rxsa}), we extracted
photon events from a $112\arcsec\times 90\arcsec$ region and created
source and background response files using the \texttt{CIAO} task
\texttt{acisspec}.  We then fit the data in \texttt{sherpa}, where the
events were binned so that each bin had $\geq 25$~counts.

There are 1305 source counts and 478.5 background counts. The data are
well fit (Fig.~\ref{fig:1501spec}) by an absorbed power-law model,
with $N_{\rm H}=\expnt{1.0(4)}{21}\mbox{ cm}^{-2}$, $\Gamma=1.65(15)$, and an
amplitude of $\expnt{4.7(6)}{-4}\mbox{ photons cm}^{-2}\mbox{
s}^{-1}\mbox{ keV}^{-1}$ at 1~keV (giving $\chi^{2}=21.6$ for 32
degrees of freedom; all uncertainties are 1-$\sigma$).  The observed
flux from this model is $\expnt{2.6}{-12}\mbox{ ergs cm}^{-2}\mbox{
s}^{-1}$ (0.3--8.0~keV), and the unabsorbed flux is
$\expnt{3.0}{-12}\mbox{ ergs cm}^{-2}\mbox{ s}^{-1}$ (0.3--8.0~keV).
The column density is higher than but consistent (given the
uncertainties) with both the column density of \snrf\ and the total
expected along this line of sight ($\expnt{6}{20}\mbox{ cm}^{-2}$).

\begin{figure*}
\plotone{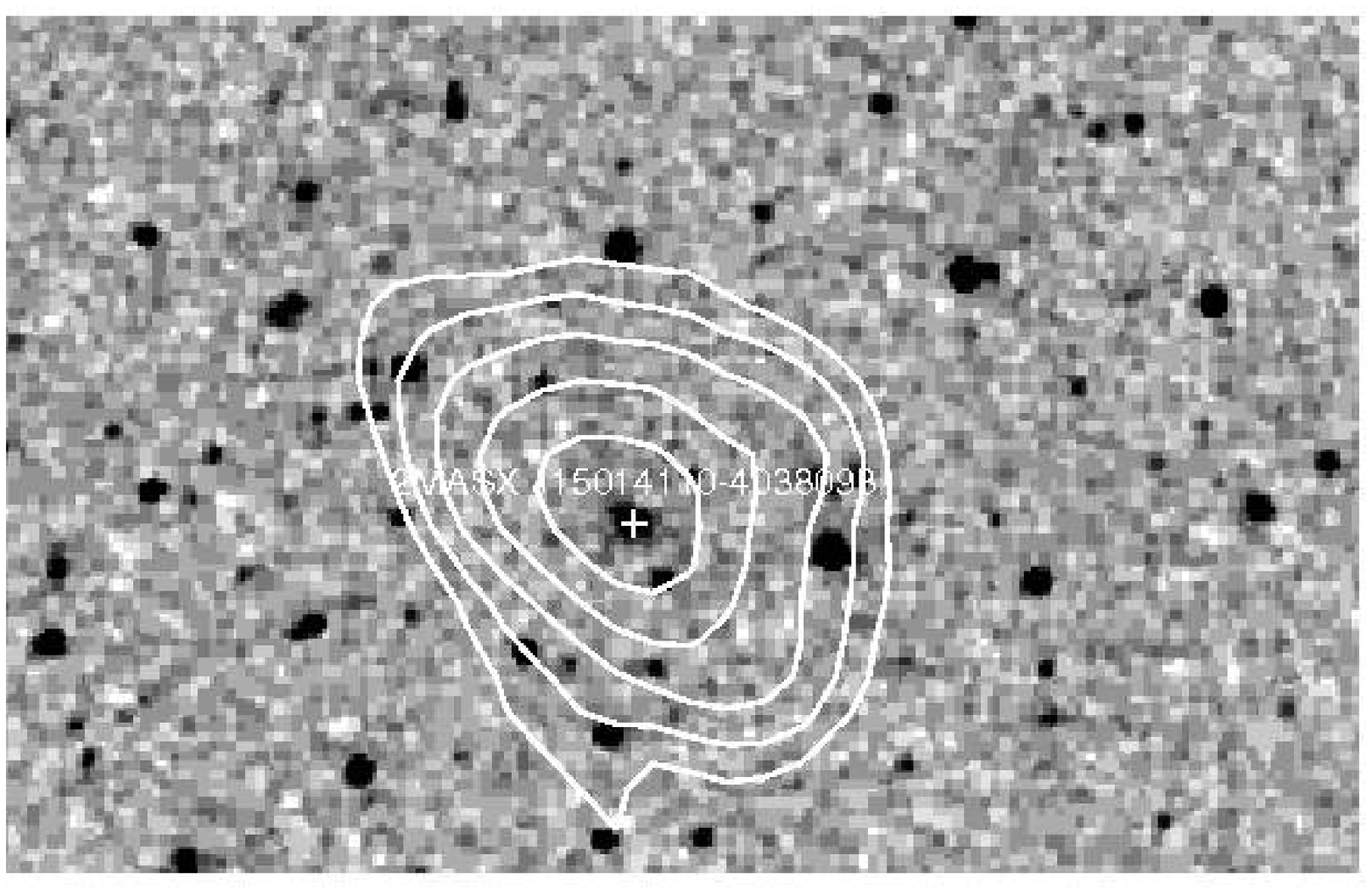}
\plottwo{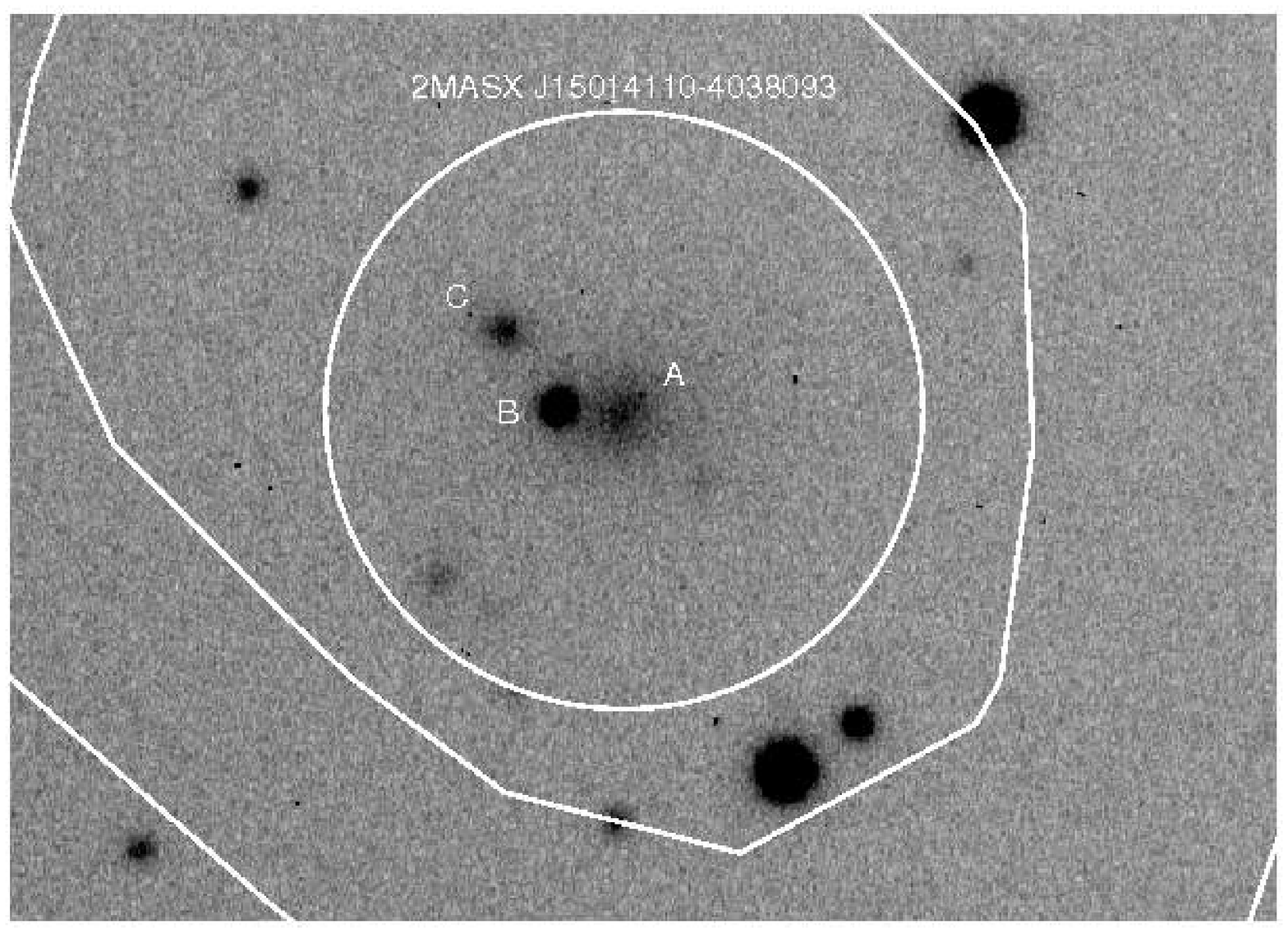}{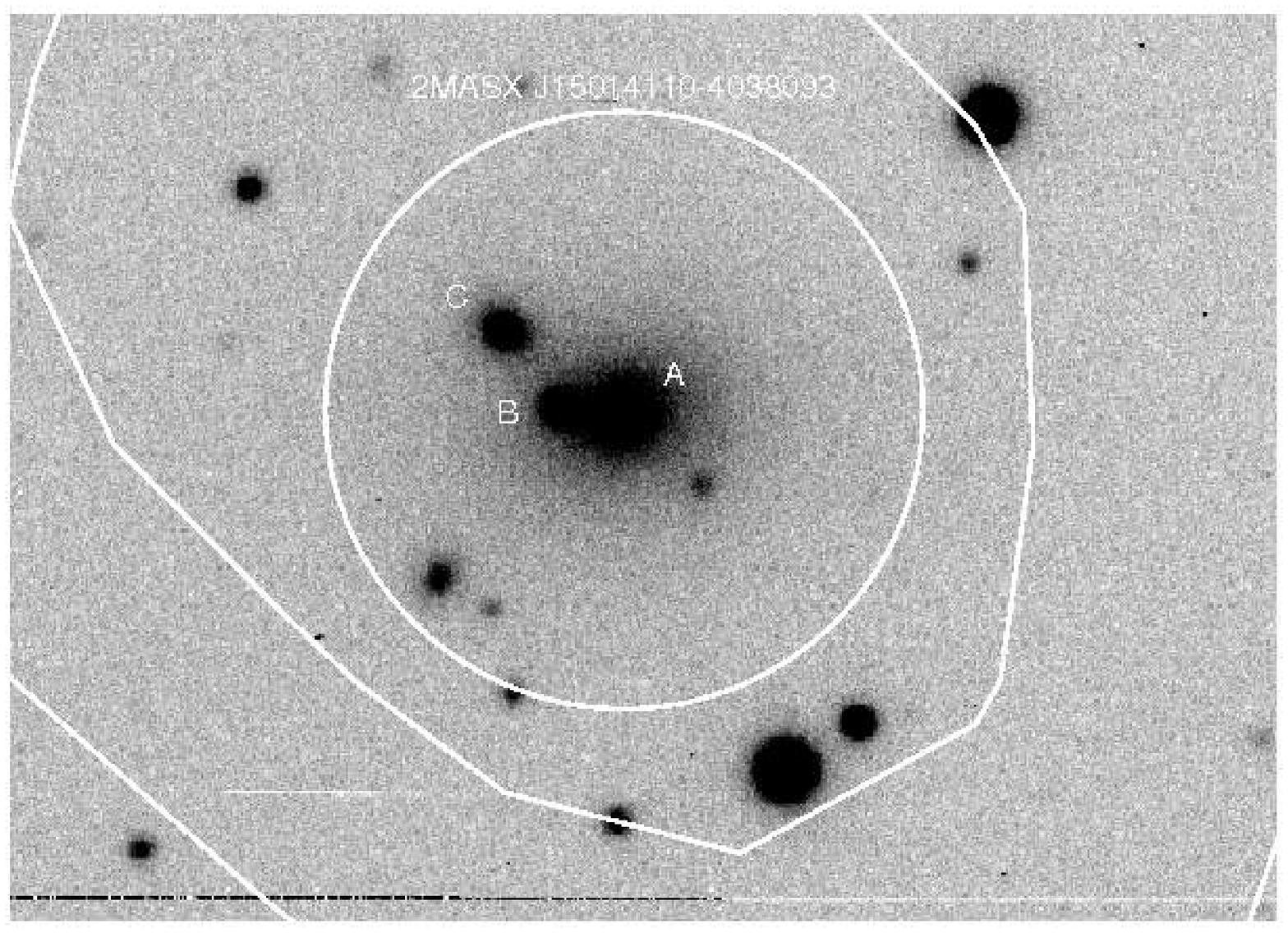}
\caption{Top: 2MASS $K_s$-band image of \rxsc.  The image is
  $5\arcmin\times3.5\arcmin$, with North up and East to the left.  The
  contours from Figure~\ref{fig:1501} are plotted, and the source
  2MASX~J15014110$-$4038093 is indicated by the cross.  Bottom:
  Magellan $B$ (left) and $R$ (right) images of \rxsc.  The images are
  $40\arcsec\times30\arcsec$, with North up and East to the left.
  Again, the contours from Figure~\ref{fig:1501} are plotted, and the
  source 2MASX~J15014110$-$4038093 is indicated by the circles
  ($10\arcsec$ radius).
\label{fig:15012mass}
}
\end{figure*}

As with \rxsa, we considered different models for \rxsc.
Figure~\ref{fig:15012mass} does not identify a single hot source, so a
thermal Galactic nebula is unlikely.  Our first idea
was that \rxsc\ is a PWN.  The size is about right and the spectrum is
typical for PWNe.  However, as with \rxsa\ there is a problem: \rxsc\
is outside of \snrc, and the largely symmetric morphology rules out an
association between \rxsc\ and \snrf\ (i.e.\ \rxsc\ cannot be a
bow-shock nebula).  \rxsc\ could instead be a pressure-confined bubble
PWN related to another supernova; We searched the Sydney University
Molongolo Sky Survey (SUMMS; \citealt*{bls99}) for evidence of radio
emission from or another supernova shell surrounding \rxsc, but there
is no extended or point-like emission present at the position of
\rxsc\ nor is there any sign of a new SNR around it.  SUMSS is
particularly sensitive to extended emission, and would almost
certainly have identified any SNR around \rxsc.  Like \rxsa, the lack
of a clear SNR shell does not mean that \rxsc\ is not a PWN.

We then examined possible extragalactic classifications for \rxsc.
This source is larger than \rxsa, and is more compatible with the
sizes of typical galaxy clusters ($\gsim 30\arcsec$): a fit to a
$\beta$ model has $r_{c}=32\arcsec$ and $\beta=0.4$. The spectral data
are reasonably well fit by a \citet{rs77} plasma model, having $N_{\rm
H}=\expnt{5(2)}{20}\mbox{ cm}^{-2}$, $kT=9^{+5}_{-2}$~keV, and a
normalization\footnote{The normalization follows the \texttt{xspec}
units of $10^{-14}\left(4\pi (D_{A}(1+z))^2\right)^{-1}\int
dV\,n_{e}n_{\rm H}$, where $D_A$ is the angular-size distance (in cm),
$n_e$ is the electron density (in ${\rm cm}^{-3}$), and $n_{\rm H}$ is
the hydrogen density (in ${\rm cm}^{-3}$).} of
$\expnt{5.6(5)}{-3}\mbox{ cm} $ (giving $\chi^{2}=22.0$ for 32 degrees
of freedom), such as what one would expect for a cluster
\citep*{wjf97}. With this model the observed flux is
$\expnt{2.7}{-12}\mbox{ ergs cm}^{-2}\mbox{ s}^{-1}$ (0.3--8.0~keV),
and the unabsorbed flux is $\expnt{2.9}{-12}\mbox{ ergs cm}^{-2}\mbox{
s}^{-1}$.

 Examining the 2MASS image  we see an extended elliptical source,
2MASX~J15014110$-$4038093, near the center of the X-ray emission
(Fig.~\ref{fig:15012mass}).  This source has a radius of $\approx
10\arcsec$ ($20\mbox{ mag arcsec}^{-2}$ isophotal radius), a $K_s$
magnitude of $12.7$~mag within that radius, and $J-K_s=1.2$~mag.
Higher-resolution optical images of \rxsc\ (Fig.~\ref{fig:15012mass})
show that\\2MASX~J15014110$-$4038093 is partially decomposed into two
sources: an extended source labeled A that is at the exact position of
2MASX~J15014110$-$4038093 (within uncertainties), and a source labeled
B $3\arcsec$ to the East.  There is also another extended source
labeled C $5\arcsec$ to the North-East, but this is a separate 2MASS
source (2MASS~J15014145$-$4038068).  We performed a rough photometric
calibration using 80 stars from the USNO-B1.0 catalog\footnote{This
calibration agreed with the nominal calibration at
\url{http://occult.mit.edu/instrumentation/magic/}.} and then ran
\texttt{sextractor} \citep{ba96} on the images: the results for
sources A--C are given in Table~\ref{tab:rxscopt}.  Source A is very
clearly extended, although it is not as large as
2MASX~J15014110$-$4038093 (the FWHM of the IR emission is $\approx
6\arcsec$).  Source B is very likely unresolved (within
uncertainties), and source C is extended.  Source A is very red
($B-R\approx 2.6$~mag), consistent with the 2MASS data.  We believe
that the 2MASS source is primarily due to source A, given the position
coincidence and the extreme redness of A compared to B or C.  If this
is the case, though, then A has the relatively blue color of
$R-K_s\approx -3.2$~mag, but this could be partly due to the
difficulties of measuring an extended source from images with
drastically different seeing (2MASS versus MagIC $R$-band).

The IR colors of 2MASX~J15014110$-$4038093 are similar to the brighter
galaxies in known clusters \citep[e.g.,][]{kb03}.  Therefore,
2MASX~J15014110$-$4038093 could be the central galaxy of an unknown
cluster.  The X-ray temperature is reasonably high, implying a high
luminosity ($\sim 10^{45}\mbox{ ergs s}^{-1}$; \citealt{mush04}), so
this source cannot be part of a nearby, low-$L$ group.  However, the
value of $\beta$ is lower than those of most known clusters
\citep*{vfj99}, and is more similar to those of low-$L$ systems
\citep{mdmb03}.

While \rxsc\ is compatible with the sizes and spectra of early-type
galaxies, and there is an extended optical/IR source near the peak of
the X-ray emission, the scenario is not entirely consistent.  The
optical/IR source is, like those in \rxsa, about 7 magnitudes fainter
than expected (the predicted magnitude following Eqn.~\ref{eqn:flux}
is $K_s\approx 5$~mag).  This is far greater than the variation seen
among galaxies.  We do not believe that the difference can be due to
an excess of soft emission in \rxsa\ or \rxsc\ (Eqn.~\ref{eqn:flux}
refers only to the contribution of hard point sources), as the spectra
of \rxsa\ and \rxsc\ are hard and similar to the prototypical sources
assumed in \citet{kf04} and when one fits primarily for the soft
emission (as in \citealt{bb98}) one finds a similar relation to that
of \citet{kf04}.  It is possible that \rxsc\ is an over-luminous
elliptical galaxy, such as those discussed in \citet{vmh+99}, as the
size, optical/X-ray flux ratio, and luminosity are similar to these
sources ($L_{\rm X}/L_{\rm opt}\sim 10^{32}\mbox{ ergs s}^{-1}
L_{\sun}^{-1}$; $L_{\rm opt}\sim 10^{11}L_{\sun}$ assuming $z\sim
0.1$), but again there are difficulties: the temperature of \rxsc\ is
considerably higher than those of \citet{vmh+99}, and the value of
$\beta$ is too low.

We see that no scenario is entirely consistent for \rxsc.  PWNe,
isolated elliptical galaxies, and galaxy clusters all have problems.
We believe it likely that \rxsc\ does have an extragalactic origin, as
2MASX~J15014110$-$4038093 looks like an elliptical galaxy and it is
probably associated with the X-ray emission: there are
39 extended 2MASS sources within $20\arcmin$ of \rxsc, giving a
false-association rate of 0.005\% for a source within $1\farcs3$.
However, it is not clear exactly what \rxsc\ is.
As with \rxsa, deeper X-ray observations and optical spectroscopy
should be conclusive for \rxsc.

\section{Discussion \& Conclusions}
\label{sec:disc}
We have fully investigated the population of \rosat\ BSC point sources
in six large-diameter SNRs.  Our identifications of counterparts to 50
of the 54 sources are quite secure: in most cases the positional
coincidence between the X-ray and optical/IR sources has been
augmented by identification of an abnormal stellar type (variable, T
Tau, binary, etc.), by the extreme brightness (and hence rarity) of
the optical source, or by a previous classification in the literature.
This conclusion echoes that of \citet{rfbm03}, who searched for older
neutron stars using \rosat\ and found only previously identified
neutron stars, along with 17 sources that are definitely not neutron
stars and 13 that are probably not.

The remaining sources, as discussed in Section~\ref{sec:remain}, are
more intriguing.  However, none of them is likely to be a neutron star
associated with one of the SNRs in Table~\ref{tab:snrs}.  To begin
with, all are outside their SNRs.  While this is not impossible for
older sources and high velocity neutron stars \citep[e.g.,][]{gj95}, it
lessens the chance of association.

For \rxsa\ and \rxsc\ the X-ray morphologies rule out associations,
since any PWNe outside the SNRs would likely have elongated bow-shock
appearances, in contrast to what we see (of course, \rxsa\ and/or
\rxsc\ could be extragalactic).  \rxsb\ and \rxsd, neither of which
has a \chandra\ detection, are more uncertain.  \rxsb\ is likely a
flare star.  \rxsd\ does not have a provisional classification but 
is probably extragalactic in origin.

\begin{figure*}
\epsscale{.9}\plotone{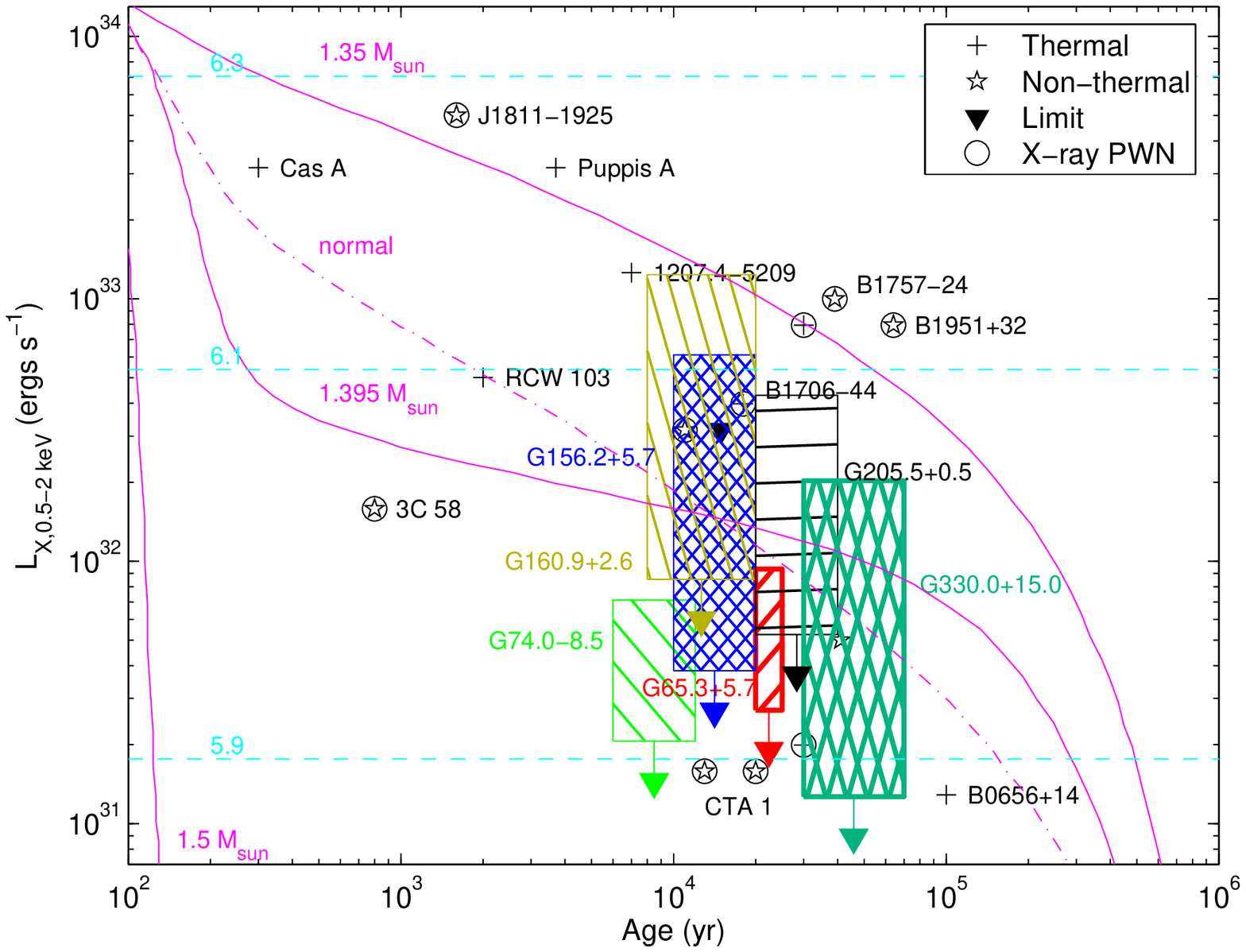}
\caption{ X-ray luminosities (0.5--2~keV) as a function of age for
young neutron stars.  Sources whose emission is primarily thermal are
indicated with plus symbols, those whose emission is primarily
non-thermal are indicated with stars, and those with only limits are
indicated with triangles; see \citetalias{kfg+04} for source data and
additional labels.  The sources that have X-ray PWNe, which are
typically $>10$ times the X-ray luminosity of the neutron stars
themselves, are circled.  We also plot the limits to blackbody
emission from sources in SNRs \Ga\ (red hatched region), \Gb\ (green
hatched region), \Gc\ (blue cross-hatched region), \Gd\ (gold hatched
region), \Ge\ (black hatched region), and \Gf\ (dark green
cross-hatched region).  An uncertainty of 30\% (for SNRs with
kinematic distances) or 60\% (for SNRs with distances from X-ray fits)
in the distance has been added to the luminosities given in
Table~\ref{tab:snrs} and \S~\ref{sec:snrs} and the likely range of
ages is also shown (for \snrb, the luminosity has been increased to
account for uncertain associations with \rxsb\ and \rxsd).  The
cooling curves are the 1p proton superfluid models from \citet{ygk+03}
(solid lines, with mass as labeled) and the normal (i.e.,
non-superfluid) $M=1.35\,M_{\odot}$ model (dot-dashed line), assuming
blackbody spectra and $R_{\infty}=10$~km.  These curves are meant to
be illustrative of general cooling trends, and should not be
interpreted as detailed predictions.  The horizontal lines show the
luminosity produced by blackbodies with $R_{\infty}=10$~km and $\log
T_{\infty}$ (K) as indicated.  Compare to Fig.~37 of
\citetalias{kfg+04}.
\label{fig:cool}
}
\end{figure*}

Since we have ruled out (to some degree of certainty) neutron stars in
all six SNRs considered here, we can then follow \citetalias{kfg+04} and
draw the X-ray luminosity limits on a cooling diagram.  This is shown
in Figure~\ref{fig:cool}.  To account for the uncertainties of \rxsb\
and \rxsd, both in \snrb, we have adjusted the luminosity of that SNR
from Table~\ref{tab:snrs} to correspond to $0.15\mbox{ count s}^{-1}$
--- above the count-rates of both of the uncertain sources and
therefore a more secure limit.  Further X-ray observations of 
\rxsd\ would very likely detect counterparts (for \rxsb, it might have
only been included in the BSC due to a flare, and therefore
significantly deeper X-ray observations may be necessary).  With
secure counterparts, the limit for \snrb\ would decrease by a factor
of 3.

The limits in Figure~\ref{fig:cool} are not as uniform or as
constraining as those from \citetalias{kfg+04}.  The lack of
uniformity is due to the sample construction: the different distances
and column densities of the SNRs make the BSC limit of $0.05\mbox{
counts s}^{-1}$ translate into different luminosities.  So, SNRs \Gb,
\Gf, and \Ga\ all have reasonably tight limits (and those of \snrb\
could get better).  SNRs \Gd\ and \Ge\ have loose limits primarily due
to uncertain distances: we have used the upper limit of 4~kpc for
\snrd\ and the full range of 0.8--1.6~kpc for \snre\ in
Figure~\ref{fig:cool}.  Finally, \snrc\ is more highly absorbed than
the other SNRs.

While all of the limits are below the luminosities of central sources
in Cas~A, \object[SNR 260.4-03.4]{Puppis~A}, and \object[SNR
G296.5+10.0]{SNR~G296.5+10.0} (and are therefore in concordance with
our original survey design from \citetalias{kfg+04}), some are further
below than others.  The utility of these limits is somewhat lessened,
though, as the SNRs are all reasonably large and are older
(10--30~kyr) than the sources in \citetalias{kfg+04} (3--10~kyr).
Therefore the cooling curves have descended, and there are other SNRs
that have similar or even lower neutron star luminosities (\object[SNR
119.5+10.2]{CTA~1}, \object[SNR 189.1+03.0]{IC~443}, \object[SNR
034.7-00.4]{W44} for the SNRs with tighter limits, and \object[SNR
263.4-03.0]{Vela}, SNRs~\object[SNR 114.3+00.3]{G114.3+0.3},
\object[SNR 343.1-02.3]{G343.1$-$2.3}, and \object[SNR
354.1+0.1]{G354.1+00.1} for the remaining SNRs), although 5/7 of these
sources have X-ray PWNe that increase their luminosities by a factor
of $\sim 10$.

In one sense, though, the limits here are tighter than those of
\citetalias{kfg+04}.  By using the BSC to go to twice the SNR radii, we
have virtually eliminated the possibility that there are high-velocity
neutron stars in these SNRs (as discussed in \S~\ref{sec:ext}
confusion is most likely not a limiting factor in detecting X-ray
sources), while in \citetalias{kfg+04} we only searched a portion of
the SNR interiors.  It is of course possible that the SN explosions were
type~Ia or produced black holes, but as discussed in \citetalias{kfg+04}
these alternate scenarios are not very likely for an ensemble.

Therefore, while not as tight as those of \citetalias{kfg+04} (or
e.g., \citealt*{shm02}; \citealt{hgc+04}), our limits are still
useful.  They are not below all detected neutron stars, so do not
require appeals to exotic physics or cooling processes, but they
conclusively demonstrate that there is a significant range in the
observed luminosities of neutron stars, even including experimental
uncertainties.  It is clear that the neutron stars of a single age
must be able to produce a luminosities differing by a factor of $>10$.
Whether the unknown parameter that controls the luminosity is one of
the usual culprits (mass, rotation, composition, magnetic field) or
something entirely different is not known.  It is also clear that
there is a significant number of objects that do not show non-thermal
emission and would therefore not go on to evolve as traditional radio
pulsars.

\acknowledgements

We thank an anonymous referee for helpful comments.  D.~L.~K.\ was
partially supported by a fellowship from the Fannie and John Hertz
Foundation.  B.~M.~G.\ and P.~O.~S.\ acknowledge support from NASA
Contract NAS8-39073 and Grant G02-3090.  B.~M.~G.\ is supported by
NASA LTSA grant NAG5-13032. S.~R.~K.\ is supported by grants from NSF
and NASA.  Support for this work was provided by the National
Aeronautics and Space Administration through Chandra award GO3-4088X
issued by the \textit{Chandra X-Ray Observatory} Center, which is
operated by the Smithsonian Astrophysical Observatory for and on
behalf of NASA under contract NAS8-39073.  The NRAO is a facility of
the National Science Foundation operated under cooperative agreement
by Associated Universities, Inc.  The Digitized Sky Surveys were
produced at the Space Telescope Science Institute under
U.S. Government grant NAG W-2166.  We have made extensive use of the
SIMBAD database, and we are grateful to the astronomers at the Centre
de Donn\'{e}es Astronomiques de Strasbourg for maintaining this
database.  We would like to thank E.~Persson and C.~Rakowski for
assistance with the PANIC observing, and D.~Fox for assistance with
the P60 observing.

{\it Facilities:} \facility{CXO (ACIS-S3)}, \facility{PO:1.5m (P60CCD)}, 
\facility{Magellan:Clay (MagIC,PANIC)}, \facility{Hale (LFC)}


\begin{thebibliography}{}

\bibitem[{Arnaud} {et~al.}(2002){Arnaud}, {Majerowicz}, {Lumb}, {Neumann},  {Aghanim}, {Blanchard}, {Boer}, {Burke}, {Collins}, {Giard}, {Nevalainen},  {Nichol}, {Romer}, \& {Sadat}]{aml+02}
{Arnaud}, M., {et al.} 2002, \aap,  390, 27

\bibitem[{Baade} \& {Zwicky}(1934){Baade} \& {Zwicky}]{bz34}
{Baade}, W. \& {Zwicky}, F. 1934, Proceedings of the National Academy of  Science, 20, 254

\bibitem[{Bertin} \& {Arnouts}(1996){Bertin} \& {Arnouts}]{ba96}
{Bertin}, E. \& {Arnouts}, S. 1996, \aaps, 117, 393

\bibitem[{Biggs} \& {Lyne}(1996){Biggs} \& {Lyne}]{bl96}
{Biggs}, J.~D. \& {Lyne}, A.~G. 1996, \mnras, 282, 691

\bibitem[{Blair} {et~al.}(1999){Blair}, {Sankrit}, {Raymond}, \&  {Long}]{bsrl99}
{Blair}, W.~P., {Sankrit}, R., {Raymond}, J.~C., \& {Long}, K.~S. 1999, \aj,  118, 942

\bibitem[{Bock} {et~al.}(1999){Bock}, {Large}, \& {Sadler}]{bls99}
{Bock}, D.~C.-J., {Large}, M.~I., \& {Sadler}, E.~M. 1999, \aj, 117, 1578

\bibitem[{Brown} \& {Bregman}(1998){Brown} \& {Bregman}]{bb98}
{Brown}, B.~A. \& {Bregman}, J.~N. 1998, \apjl, 495, L75

\bibitem[{Burrows} {et~al.}(2004){Burrows}, {Ott}, \& {Meakin}]{bom04}
{Burrows}, A., {Ott}, C.~D., \& {Meakin}, C. 2004, in Cosmic
explosions in  three dimensions, ed. P.~Hoflich, P.~Kumar, \&
J.~C. Wheeler (Cambridge: Cambridge  University Press), 209 (astro-ph/0309684)

\bibitem[{Camilo}(2003){Camilo}]{camilo03}
{Camilo}, F. 2003, in ASP Conf. Ser. 302: Radio Pulsars, ed. M.~Bailes, D.~J.  Nice, \& S.~E. Thorsett (San Fransisco: ASP), 145 (astro-ph/0210620)

\bibitem[{Camilo} {et~al.}(2002){Camilo}, {Stairs}, {Lorimer}, {Backer},  {Ransom}, {Klein}, {Wielebinski}, {Kramer}, {McLaughlin}, {Arzoumanian}, \&  {M{\" u}ller}]{csl+02}
{Camilo}, F., {et al.} 2002, \apjl, 571, L41

\bibitem[{Case} \& {Bhattacharya}(1998){Case} \& {Bhattacharya}]{cb98}
{Case}, G.~L. \& {Bhattacharya}, D. 1998, \apj, 504, 761

\bibitem[{Chatterjee} \& {Cordes}(2002){Chatterjee} \& {Cordes}]{cc02}
{Chatterjee}, S. \& {Cordes}, J.~M. 2002, \apj, 575, 407

\bibitem[{Chevalier}(2005){Chevalier}]{che05}
{Chevalier}, R.~A. 2005, \apj, 619, 839

\bibitem[{Condon} {et~al.}(1998){Condon}, {Cotton}, {Greisen}, {Yin},  {Perley}, {Taylor}, \& {Broderick}]{ccg+98}
{Condon}, J.~J., {Cotton}, W.~D., {Greisen}, E.~W., {Yin}, Q.~F., {Perley},  R.~A., {Taylor}, G.~B., \& {Broderick}, J.~J. 1998, \aj, 115, 1693

\bibitem[{Condon} \& {Kaplan}(1998){Condon} \& {Kaplan}]{ck98}
{Condon}, J.~J. \& {Kaplan}, D.~L. 1998, \apjs, 117, 361

\bibitem[{Cotter} {et~al.}(2002){Cotter}, {Buttery}, {Das}, {Jones},  {Grainge}, {Pooley}, \& {Saunders}]{cbd+02}
{Cotter}, G., {Buttery}, H.~J., {Das}, R., {Jones}, M.~E., {Grainge}, K.,  {Pooley}, G.~G., \& {Saunders}, R. 2002, \mnras, 334, 323

\bibitem[{Damashek} {et~al.}(1978){Damashek}, {Taylor}, \& {Hulse}]{dth78}
{Damashek}, M., {Taylor}, J.~H., \& {Hulse}, R.~A. 1978, \apjl, 225, L31

\bibitem[{Drimmel} {et~al.}(2003){Drimmel}, {Cabrera-Lavers}, \& {L{\'  o}pez-Corredoira}]{dcllc03}
{Drimmel}, R., {Cabrera-Lavers}, A., \& {L{\' o}pez-Corredoira}, M. 2003, \aap,  409, 205

\bibitem[{Fabbiano}(1989){Fabbiano}]{fabbiano89}
{Fabbiano}, G. 1989, \araa, 27, 87

\bibitem[{Fuhrmeister} \& {Schmitt}(2003){Fuhrmeister} \& {Schmitt}]{fs03}
{Fuhrmeister}, B. \& {Schmitt}, J.~H.~M.~M. 2003, \aap, 403, 247

\bibitem[{Gaensler} \& {Johnston}(1995){Gaensler} \& {Johnston}]{gj95}
{Gaensler}, B.~M. \& {Johnston}, S. 1995, \mnras, 275, L73

\bibitem[Gaensler \& Slane(2006)Gaensler \& Slane]{gs06}
Gaensler, B.~M. \& Slane, P.~O. 2006, \araa, 44, in press, (astro-ph/0601081)

\bibitem[{Garmire} {et~al.}(2003){Garmire}, {Bautz}, {Ford}, {Nousek}, \&  {Ricker}]{gbf+03}
{Garmire}, G.~P., {Bautz}, M.~W., {Ford}, P.~G., {Nousek}, J.~A., \& {Ricker},  G.~R. 2003, \procspie, 4851, 28

\bibitem[{Gorham} {et~al.}(1996){Gorham}, {Ray}, {Anderson}, {Kulkarni}, \&  {Prince}]{gra+96}
{Gorham}, P.~W., {Ray}, P.~S., {Anderson}, S.~B., {Kulkarni}, S.~R., \&  {Prince}, T.~A. 1996, \apj, 458, 257

\bibitem[Green(2000)Green]{g00}
Green, D.~A. 2000, A Catalogue of Galactic Supernova Remnants (2000 August  version), Mullard Radio Astronomy Observatory, Cavendish Laboratory,  Cambridge, UK (available on the World-Wide-Web at  http://www.mrao.cam.ac.uk/surveys/snrs/)

\bibitem[{Gull} {et~al.}(1977){Gull}, {Kirshner}, \& {Parker}]{gkp77}
{Gull}, T.~R., {Kirshner}, R.~P., \& {Parker}, R.~A.~R. 1977, \apjl, 215, L69

\bibitem[{Halpern} {et~al.}(2004){Halpern}, {Gotthelf}, {Camilo}, {Helfand},  \& {Ransom}]{hgc+04}
{Halpern}, J.~P., {Gotthelf}, E.~V., {Camilo}, F., {Helfand}, D.~J., \&  {Ransom}, S.~M. 2004, \apj, 612, 398

\bibitem[{Hester} {et~al.}(1990){Hester}, {Graham}, {Beichman}, \&  {Gautier}]{hgbg90}
{Hester}, J.~J., {Graham}, J.~R., {Beichman}, C.~A., \& {Gautier}, T.~N. 1990,  \apj, 357, 539

\bibitem[{Huang} \& {Thaddeus}(1985){Huang} \& {Thaddeus}]{ht85}
{Huang}, Y.-L. \& {Thaddeus}, P. 1985, \apjl, 295, L13

\bibitem[{Kaplan} {et~al.}(2004){Kaplan}, {Frail}, {Gaensler}, {Gotthelf},  {Kulkarni}, {Slane}, \& {Nechita}]{kfg+04}
{Kaplan}, D.~L., {Frail}, D.~A., {Gaensler}, B.~M., {Gotthelf}, E.~V.,  {Kulkarni}, S.~R., {Slane}, P.~O., \& {Nechita}, A. 2004, \apjs, 153, 269

\bibitem[{Kaspi} \& {Helfand}(2002){Kaspi} \& {Helfand}]{kh02}
{Kaspi}, V.~M. \& {Helfand}, D.~J. 2002, in ASP Conf. Ser. 271: Neutron Stars  in Supernova Remnants, ed. P.~O. Slane \& B.~M. Gaensler (San Fransisco:  ASP), 3 (astro-ph/0201183)

\bibitem[{Kaspi} {et~al.}(1996){Kaspi}, {Manchester}, {Johnston}, {Lyne}, \&  {D'Amico}]{kmj+96}
{Kaspi}, V.~M., {Manchester}, R.~N., {Johnston}, S., {Lyne}, A.~G., \&  {D'Amico}, N. 1996, \aj, 111, 2028

\bibitem[{Kassim} {et~al.}(1994){Kassim}, {Hertz}, {van Dyk}, \&  {Weiler}]{khvdw94}
{Kassim}, N.~E., {Hertz}, P., {van Dyk}, S.~D., \& {Weiler}, K.~W. 1994, \apjl,  427, L95

\bibitem[{Kim} \& {Fabbiano}(2004){Kim} \& {Fabbiano}]{kf04}
{Kim}, D. \& {Fabbiano}, G. 2004, \apj, 611, 846

\bibitem[{Kodama} \& {Bower}(2003){Kodama} \& {Bower}]{kb03}
{Kodama}, T. \& {Bower}, R. 2003, \mnras, 346, 1

\bibitem[Large {et~al.}(1968)Large, Vaughan, \& Mills]{lvm68}
Large, M.~I., Vaughan, A.~E., \& Mills, B.~Y. 1968, \nat, 220, 340

\bibitem[{Leahy} \& {Aschenbach}(1995){Leahy} \& {Aschenbach}]{la95}
{Leahy}, D.~A. \& {Aschenbach}, B. 1995, \aap, 293, 853

\bibitem[{Leahy} {et~al.}(1986){Leahy}, {Naranan}, \& {Singh}]{lns86}
{Leahy}, D.~A., {Naranan}, S., \& {Singh}, K.~P. 1986, \mnras, 220, 501

\bibitem[{Leahy} {et~al.}(1991){Leahy}, {Nousek}, \& {Hamilton}]{lnh91}
{Leahy}, D.~A., {Nousek}, J., \& {Hamilton}, A.~J.~S. 1991, \apj, 374, 218

\bibitem[{Leahy} \& {Roger}(1991){Leahy} \& {Roger}]{lr91}
{Leahy}, D.~A. \& {Roger}, R.~S. 1991, \aj, 101, 1033

\bibitem[{Levenson} {et~al.}(2002){Levenson}, {Graham}, \& {Walters}]{lgw+02}
{Levenson}, N.~A., {Graham}, J.~R., \& {Walters}, J.~L. 2002, \apj, 576, 798

\bibitem[{Lorimer} {et~al.}(1998){Lorimer}, {Lyne}, \& {Camilo}]{llc98}
{Lorimer}, D.~R., {Lyne}, A.~G., \& {Camilo}, F. 1998, \aap, 331, 1002

\bibitem[{Lozinskaya}(1981){Lozinskaya}]{loz81}
{Lozinskaya}, T.~A. 1981, Soviet Astronomy Letters, 7, 17

\bibitem[Martini {et~al.}(2004)Martini, Persson, Murphy, Birk, Shectman,  Gunnels, \& Koch]{mpm+04}
Martini, P., Persson, S.~E., Murphy, D.~C., Birk, C., Shectman, S.~A., Gunnels,  S.~M., \& Koch, E. 2004, \procspie, 5492 (astro-ph/0406666)

\bibitem[{Matsushita} {et~al.}(1994){Matsushita}, {Makishima}, {Awaki},  {Canizares}, {Fabian}, {Fukazawa}, {Loewenstein}, {Matsumoto}, {Mihara},  {Mushotzky}, {Ohashi}, {Ricker}, {Serlemitsos}, {Tsuru}, {Tsusaka}, \&  {Yamazaki}]{mma+94}
{Matsushita}, K., {et al.} 1994, \apjl, 436, L41

\bibitem[{Mavromatakis} {et~al.}(2002){Mavromatakis}, {Boumis},  {Papamastorakis}, \& {Ventura}]{mbpv02}
{Mavromatakis}, F., {Boumis}, P., {Papamastorakis}, J., \& {Ventura}, J. 2002,  \aap, 388, 355

\bibitem[{Miyata} {et~al.}(2001){Miyata}, {Ohta}, {Torii}, {Takeshima},  {Tsunemi}, {Hasegawa}, \& {Hashimoto}]{mot+01}
{Miyata}, E., {Ohta}, K., {Torii}, K., {Takeshima}, T., {Tsunemi}, H.,  {Hasegawa}, T., \& {Hashimoto}, Y. 2001, \apj, 550, 1023

\bibitem[{Miyata} {et~al.}(1998a){Miyata}, {Tsunemi}, {Kohmura},  {Suzuki}, \& {Kumagai}]{mtk+98}
{Miyata}, E., {Tsunemi}, H., {Kohmura}, T., {Suzuki}, S., \& {Kumagai}, S.  1998a, \pasj, 50, 257

\bibitem[{Miyata} {et~al.}(1998b){Miyata} {et~al.}]{mtt+98}
{Miyata}, E. {et~al.} 1998b, \pasj, 50, 475

\bibitem[{Mulchaey} {et~al.}(2003){Mulchaey}, {Davis}, {Mushotzky}, \&  {Burstein}]{mdmb03}
{Mulchaey}, J.~S., {Davis}, D.~S., {Mushotzky}, R.~F., \& {Burstein}, D. 2003,  \apjs, 145, 39


\bibitem[{{Mushotzky}(2004)}]{mush04}
{Mushotzky}, R.~F. 2004, in Clusters of Galaxies: Probes of Cosmological
  Structure and Galaxy Evolution, ed. J.~S. Mulchaey, A.~Dressler, \& A.~Oemler
  (Cambridge: Cambridge University Press), 123, (astro-ph/0311105)


\bibitem[{Odegard}(1986){Odegard}]{ode86}
{Odegard}, N. 1986, \apj, 301, 813

\bibitem[{Pfeffermann} {et~al.}(1991){Pfeffermann}, {Aschenbach}, \&  {Predehl}]{pap91}
{Pfeffermann}, E., {Aschenbach}, B., \& {Predehl}, P. 1991, \aap, 246, L28

\bibitem[{Predehl} \& {Schmitt}(1995){Predehl} \& {Schmitt}]{ps95}
{Predehl}, P. \& {Schmitt}, J.~H.~M.~M. 1995, \aap, 293, 889

\bibitem[{Raymond} \& {Smith}(1977){Raymond} \& {Smith}]{rs77}
{Raymond}, J.~C. \& {Smith}, B.~W. 1977, \apjs, 35, 419

\bibitem[{Richman}(1996){Richman}]{r96}
{Richman}, H.~R. 1996, \apj, 462, 404

\bibitem[{Rutledge} {et~al.}(2003){Rutledge}, {Fox}, {Bogosavljevic}, \&  {Mahabal}]{rfbm03}
{Rutledge}, R.~E., {Fox}, D.~W., {Bogosavljevic}, M., \& {Mahabal}, A. 2003,  \apj, 598, 458

\bibitem[{Schaudel} {et~al.}(2002){Schaudel}, {Becker}, {Lu}, \&  {Aschenbach}]{sbla02}
{Schaudel}, D., {Becker}, W., {Lu}, F., \& {Aschenbach}, B. 2002, Advances in  Space Research, 34, in press

\bibitem[{Seward} {et~al.}(2005){Seward}, {Gorenstein}, \& {Smith}]{sgs05}
{Seward}, F.~D., {Gorenstein}, P., \& {Smith}, R.~K. 2005, \apj, in press  (astro-ph/0509636)

\bibitem[{Shelton} {et~al.}(2004){Shelton}, {Kuntz}, \& {Petre}]{skp04}
{Shelton}, R.~L., {Kuntz}, K.~D., \& {Petre}, R. 2004, \apj, 615, 275

\bibitem[{Skrutskie} {et~al.}(1997){Skrutskie}, {Schneider}, {Stiening},  {Strom}, {Weinberg}, {Beichman}, {Chester}, {Cutri}, {Lonsdale}, {Elias},  {Elston}, {Capps}, {Carpenter}, {Huchra}, {Liebert}, {Monet}, {Price}, \&  {Seitzer}]{2mass}
{Skrutskie}, M.~F., {et al.} 1997, in ASSL Vol.  210: The Impact of Large Scale Near-IR Sky Surveys, 25

\bibitem[{Slane} {et~al.}(2002){Slane}, {Helfand}, \& {Murray}]{shm02}
{Slane}, P.~O., {Helfand}, D.~J., \& {Murray}, S.~S. 2002, \apjl, 571, L45

\bibitem[Staelin \& Reifenstein(1968)Staelin \& Reifenstein]{sr68}
Staelin, D.~H. \& Reifenstein, E.~C. 1968, Science, 162, 1481

\bibitem[{Stappers} {et~al.}(2003){Stappers}, {Gaensler}, {Kaspi}, {van der  Klis}, \& {Lewin}]{sgk+03}
{Stappers}, B.~W., {Gaensler}, B.~M., {Kaspi}, V.~M., {van der Klis}, M., \&  {Lewin}, W.~H.~G. 2003, Science, 299, 1372

\bibitem[{Stelzer} {et~al.}(2003){Stelzer}, {Hu{\' e}lamo}, {Hubrig},  {Zinnecker}, \& {Micela}]{shh+03}
{Stelzer}, B., {Hu{\' e}lamo}, N., {Hubrig}, S., {Zinnecker}, H., \& {Micela},  G. 2003, \aap, 407, 1067

\bibitem[{Tananbaum}(1999){Tananbaum}]{t99}
{Tananbaum}, H. 1999, \iaucirc, 7246, 1

\bibitem[{Ueda} {et~al.}(2001){Ueda}, {Ishisaki}, {Takahashi}, {Makishima}, \&  {Ohashi}]{uit+01}
{Ueda}, Y., {Ishisaki}, Y., {Takahashi}, T., {Makishima}, K., \& {Ohashi}, T.  2001, \apjs, 133, 1

\bibitem[{van der Swaluw} {et~al.}(2004){van der Swaluw}, {Downes}, \&  {Keegan}]{vdsdk04}
{van der Swaluw}, E., {Downes}, T.~P., \& {Keegan}, R. 2004, \aap, 420, 937

\bibitem[{Verbunt} {et~al.}(1997){Verbunt}, {Bunk}, {Ritter}, \&  {Pfeffermann}]{vbrp97}
{Verbunt}, F., {Bunk}, W.~H., {Ritter}, H., \& {Pfeffermann}, E. 1997, \aap,  327, 602

\bibitem[{Vikhlinin} {et~al.}(1999a){Vikhlinin}, {Forman}, \&  {Jones}]{vfj99}
{Vikhlinin}, A., {Forman}, W., \& {Jones}, C. 1999a, \apj, 525, 47

\bibitem[{Vikhlinin} {et~al.}(1999b){Vikhlinin}, {McNamara},  {Hornstrup}, {Quintana}, {Forman}, {Jones}, \& {Way}]{vmh+99}
{Vikhlinin}, A., {McNamara}, B.~R., {Hornstrup}, A., {Quintana}, H., {Forman},  W., {Jones}, C., \& {Way}, M. 1999b, \apjl, 520, L1

\bibitem[{Voges} {et~al.}(1999){Voges}, {Aschenbach}, {Boller},  {Br{\"a}uninger}, {Briel}, {Burkert}, {Dennerl}, {Englhauser}, {Gruber},  {Haberl}, {Hartner}, {Hasinger}, {K{\"u}rster}, {Pfeffermann}, {Pietsch},  {Predehl}, {Rosso}, {Schmitt}, {Tr{\"u}mper}, \& {Zimmermann}]{rbs2}
{Voges}, W., {et al.} 1999, \aap, 349, 389

\bibitem[{White} {et~al.}(1997){White}, {Jones}, \& {Forman}]{wjf97}
{White}, D.~A., {Jones}, C., \& {Forman}, W. 1997, \mnras, 292, 419

\bibitem[{Yakovlev} {et~al.}(2004){Yakovlev}, {Gnedin}, {Kaminker},  {Levenfish}, \& {Potekhin}]{ygk+03}
{Yakovlev}, D.~G., {Gnedin}, O.~Y., {Kaminker}, A.~D., {Levenfish}, K.~P., \&  {Potekhin}, A.~Y. 2004, Advances in Space Research, 33, 523  (astro-ph/0306143)

\bibitem[{Yamauchi} {et~al.}(1999){Yamauchi}, {Koyama}, {Tomida}, {Yokogawa},  \& {Tamura}]{ykt+99}
{Yamauchi}, S., {Koyama}, K., {Tomida}, H., {Yokogawa}, J., \& {Tamura}, K.  1999, \pasj, 51, 13

\end{thebibliography}


\clearpage

\begin{landscape}
\LongTables
\begin{deluxetable}{r c c r r c c l c r c r}
\tabletypesize{\scriptsize}
\setlength{\tabcolsep}{0.05in}
\tablewidth{0pt}
\tablecaption{\rosat\ Point Sources  and Stellar Counterparts\label{tab:srcs}}
\tablehead{
\colhead{N} & \colhead{1RXS J} & \colhead{PSPC} & \colhead{$\Delta R$\tablenotemark{b}}
& \colhead{$\sigma r$\tablenotemark{c}} & 
 \colhead{RA\tablenotemark{a}} & \colhead{DEC\tablenotemark{a}}  & \colhead{Star} & \colhead{2MASS~J} &
\colhead{$\delta r$\tablenotemark{d}} & \colhead{$V$} & \colhead{$K_s$} \\ \cline{6-7} \cline{11-12}
& & \colhead{(count/s)} & \colhead{(arcmin)} &
\colhead{(asec)} & \mc{2}{c}{(J2000)}  & & &\colhead{(asec)} & \mc{2}{c}{(mag)}\\}
\startdata
\mc{2}{c}{{\bf SNR G65.3+5.7:}}\\
1  & \object[1RXS J193445.6+303100]{193445.6+303100} & 0.066& 45.1 & 11 & $19\hr34\mn45\fs23$ & $+30\degr30\arcmin58\fs9$ &  \object{HD 184738}              & 19344524+3030590        & 5.0 &  10.41    &   8.108\\
2  & \object[1RXS J193840.0+303035]{193840.0+303035} & 0.083& 82.7 & 9 & $19\hr38\mn40\fs10$ & $+30\degr30\arcmin28\fs0$ &  \object{V* EM Cyg}	        & 19384012+3030284        & 7.1 &  12.6     &  11.150\\
3  & \object[1RXS J193922.4+300921]{193922.4+300921} & 0.054& 101.8 & 12 & $19\hr39\mn22\fs61$ & $+30\degr09\arcmin12\fs0$ &  \object{HD 185734}	        & 19392261+3009119        & 9.9 &  4.685    &   2.499\\
4  & \object[1RXS J194337.2+322523]{194337.2+322523} & 0.256& 155.8 & 8 & $19\hr43\mn36\fs80$ & $+32\degr25\arcmin20\fs7$ &  \object{BDS 9566 B}        & 19433674+3225206        & 5.6 &  9.9	        &   8.082\\
 '' &  '' & ''& '' & '' & $19\hr43\mn37\fs90$ & $+32\degr25\arcmin12\fs7$ &  \object{HD 331149}	        & 19433790+3225124        & 13.6 &  10.7      &   7.179\\[0.1in]
5  & \object[1RXS J193458.1+335301]{193458.1+335301} & 0.051& 165.0 & 14 & $19\hr34\mn58\fs10$ & $+33\degr53\arcmin01\fs5$ & \nodata & \nodata & \nodata & \nodata & \nodata\\
6  & \object[1RXS J192722.3+280934]{192722.3+280934} & 0.110& 194.4 & 15 & $19\hr27\mn21\fs91$ & $+28\degr09\arcmin42\fs8$ &  USNO 1181$-$0406270      & 19272197+2809452        & 9.8 &  11.620\tablenotemark{e} &   8.426\\
7  & \object[1RXS J194401.5+284456]{194401.5+284456} & 0.138& 202.7 & 8 & $19\hr44\mn01\fs37$ & $+28\degr45\arcmin09\fs9$ &  \object[GSC 02151-03394]{GSC 02151$-$03394}        & 19440138+2845099        & 14.0 & \nodata &   8.691\\
8  & \object[1RXS J194902.9+295258]{194902.9+295258} & 0.357& 219.9 & 7 & $19\hr49\mn02\fs99$ & $+29\degr52\arcmin58\fs3$ &  \object{HD 187460} 	        & 19490298+2952582       & 1.2 &  8.32      &   5.734\\
9  & \object[1RXS J193228.6+345318]{193228.6+345318} & 0.091& 223.4 & 9 & $19\hr32\mn28\fs59$ & $+34\degr53\arcmin18\fs5$ & \nodata & \nodata & \nodata & \nodata & \nodata\\[0.1in]
10 & \object[1RXS J191449.0+315131]{191449.0+315131} & 0.057& 237.0 & 14 & $19\hr14\mn50\fs21$ & $+31\degr51\arcmin37\fs3$ &  \object{HD 180314} 	        & 19145022+3151371        & 16.5 &  6.618    &   4.312\\
11 & \object[1RXS J193856.2+351407]{193856.2+351407} & 0.290& 255.7 & 8 & $19\hr38\mn55\fs77$ & $+35\degr14\arcmin13\fs0$ &  \object{HD 185696} 	        & 19385576+3514132        & 8.0 &  8.29      &   6.858\\
12 & \object[1RXS J193936.8+263718]{193936.8+263718} & 0.074& 285.6 & 8 & $19\hr39\mn36\fs67$ & $+26\degr37\arcmin16\fs1$ &  \object{AG+26 2090}	        & 19393666+2637169        & 3.0 &  11.1      &   7.345\\
13 & \object[1RXS J193113.0+360730]{193113.0+360730} & 0.153& 298.4 & 7 & $19\hr31\mn12\fs57$ & $+36\degr07\arcmin30\fs0$ &  \object[G 125-15]{G 125$-$15}	        & 19311257+3607300        & 13.1 & \nodata &   8.839\\
14 & \object[1RXS J191151.1+285012]{191151.1+285012} & 0.052& 305.3 & 11 & $19\hr11\mn50\fs81$ & $+28\degr50\arcmin07\fs6$ &  USNO 1188$-$0330651      & 19115080+2850075        & 5.7 &  12.510\tablenotemark{e} &   8.898\\[0.1in]
\mc{2}{c}{{\bf SNR G74.0$-$8.5:}}\\
1 & \object[1RXS J205042.9+284643]{205042.9+284643} & 0.111& 113.3 & 12 & $20\hr50\mn42\fs90$ & $+28\degr46\arcmin43\fs5$ & \nodata & \nodata & \nodata & \nodata & \nodata\\
2 & \object[1RXS J204457.4+291613]{204457.4+291613} & 0.104& 114.4 & 10 & $20\hr44\mn58\fs09$ & $+29\degr16\arcmin21\fs3$ &  \object{HD 335070}	        & 20445809+2916211        & 11.9 &  10.8      &   8.746\\
3 & \object[1RXS J205812.8+292037]{205812.8+292037} & 0.129& 122.3 & 8 & $20\hr58\mn12\fs82$ & $+29\degr20\arcmin28\fs6$ &  USNO 1193$-$0519643      & 20581282+2920282        & 8.9 &  15.980\tablenotemark{e} &  14.258\\
 '' &  '' & ''& '' & '' & $20\hr58\mn12\fs80$ & $+29\degr20\arcmin37\fs5$ & \nodata & 20581257+2920454        & 8.5 & \nodata &  13.405\\
4 & \object[1RXS J205208.5+270546]{205208.5+270546} & 0.225& 214.7 & 8 & $20\hr52\mn07\fs68$ & $+27\degr05\arcmin49\fs1$ &  \object{HD 198809}	        & 20520768+2705491        & 11.4 &  4.576    &   2.722\\[0.1in]
\mc{2}{c}{{\bf SNR G156.2+5.7:}}\\
1 & \object[1RXS J050639.1+513607]{050639.1+513607} & 0.051& 75.3 & 20 & $05\hr06\mn40\fs63$ & $+51\degr35\arcmin51\fs8$ &  \object{HD 32537} 	        & 05064067+5135519        & 20.8 &  4.980    &   4.124\\
\mc{2}{c}{{\bf SNR G160.9+2.6:}}\\
1 & \object[1RXS J045707.4+452751]{045707.4+452751} & 0.061& 82.5 & 9 & $04\hr57\mn07\fs40$ & $+45\degr27\arcmin51\fs0$ & \nodata & \nodata & \nodata & \nodata & \nodata\\
2 & \object[1RXS J050339.8+451715]{050339.8+451715} & 0.061& 87.2 & 11 & $05\hr03\mn39\fs80$ & $+45\degr17\arcmin15\fs0$ & \nodata & \nodata & \nodata & \nodata & \nodata\\
3 & \object[1RXS J045222.2+455619]{045222.2+455619} & 0.052& 99.0 & 10 & $04\hr52\mn21\fs51$ & $+45\degr56\arcmin23\fs7$ &  \object{HD 30736}	        & 04522151+4556236        & 8.6 &  6.695    &   5.407\\
\mc{2}{c}{{\bf SNR G205.5+0.5:}}\\
1 & \object[1RXS J064108.3+052250]{064108.3+052250} & 0.086& 74.3 & 11 & $06\hr41\mn08\fs07$ & $+05\degr22\arcmin52\fs1$ &  USNO 0155$-$01104$-$1      & 06410807+0522522        & 4.0 &  10.780\tablenotemark{e} &   8.905\\[0.1in]
 '' &  '' & ''& '' & '' & $06\hr41\mn07\fs96$ & $+05\degr22\arcmin43\fs8$ & \nodata & 06410796+0522438        & 8.0 & \nodata &  12.121\\
2 & \object[1RXS J064136.2+080218]{064136.2+080218} & 0.054& 100.1 & 11 & $06\hr41\mn35\fs94$ & $+08\degr02\arcmin05\fs6$ &  \object{HD 262113} 	        & 06413601+0802055        & 13.0 &  10.3      &   8.712\\
3 & \object[1RXS J064109.3+044733]{064109.3+044733} & 0.076& 107.4 & 17 & $06\hr41\mn09\fs61$ & $+04\degr47\arcmin35\fs8$ &  USNO 0947$-$0100759      & 06410953+0447354        & 5.4 &  17.650\tablenotemark{e} &  15.084\\
 '' &  '' & ''& '' & '' & $06\hr41\mn09\fs82$ & $+04\degr47\arcmin35\fs1$ &  USNO 0947$-$0100763      & 06410988+0447350        & 8.1 &  15.840\tablenotemark{e} &  14.636\\
 '' &  '' & ''& '' & '' & $06\hr41\mn09\fs00$ & $+04\degr47\arcmin23\fs9$ & \nodata & 06410900+0447239        & 10.1 & \nodata &  12.182\\[0.1in]
 '' &  '' & ''& '' & '' & $06\hr41\mn09\fs27$ & $+04\degr47\arcmin18\fs7$ &  USNO 0155$-$02167$-$1      & 06410924+0447187        & 14.3 &  11.100\tablenotemark{e} &   8.688\\
4 & \object[1RXS J064641.1+082152]{064641.1+082152} & 0.092& 160.1 & 10 & $06\hr46\mn40\fs73$ & $+08\degr21\arcmin47\fs3$ &  \object{HD 49015}	        & 06464073+0821471        & 7.6 &  7.5        &   6.080\\
5 & \object[1RXS J062740.3+073103]{062740.3+073103} & 0.082& 179.5 & 8 & $06\hr27\mn40\fs30$ & $+07\degr31\arcmin03\fs0$ & \nodata & \nodata & \nodata & \nodata & \nodata\\
6 & \object[1RXS J062937.2+082930]{062937.2+082930} & 0.088& 183.9 & 9 & $06\hr29\mn36\fs89$ & $+08\degr29\arcmin32\fs8$ &  \object{HD 45759}	        & 06293689+0829327        & 5.1 &  7.62      &   6.306\\
7 & \object[1RXS J063715.7+032005]{063715.7+032005} & 0.109& 191.7 & 10 & $06\hr37\mn15\fs22$ & $+03\degr20\arcmin08\fs0$ &  USNO 0150$-$00332$-$1      & 06371522+0320081        & 7.6 &  10.640\tablenotemark{e} &   9.323\\[0.1in]
 '' &  '' & ''& '' & '' & $06\hr37\mn15\fs85$ & $+03\degr20\arcmin04\fs3$ & \nodata & 06371585+0320043        & 2.5 & \nodata &  14.508\\
 '' &  '' & ''& '' & '' & $06\hr37\mn15\fs55$ & $+03\degr20\arcmin02\fs6$ & \nodata & 06371555+0320026        & 3.7 & \nodata &  11.813\\
8 & \object[1RXS J062554.8+065543]{062554.8+065543} & 0.106& 196.7 & 11 & $06\hr25\mn55\fs24$ & $+06\degr55\arcmin38\fs8$ &  USNO 0145$-$01717$-$1      & 06255524+0655386        & 7.8 &   9.660\tablenotemark{e} &   8.019\\
\mc{2}{c}{{\bf SNR G330.0+15.0:}}\\
1 & \object[1RXS J150818.8-401730]{150818.8$-$401730} & 0.077& 26.1 & 10 & $15\hr08\mn18\fs80$ & $-40\degr17\arcmin30\fs0$ & \nodata & \nodata & \nodata & \nodata & \nodata\\
2 & \object[1RXS J151059.6-392655]{151059.6$-$392655} & 0.082& 35.0 & 12 & $15\hr10\mn59\fs06$ & $-39\degr26\arcmin58\fs5$ &  USNO 0505$-$0350285      & 15105908$-$3926590        & 7.2 &   \nodata &  15.066\\[0.1in]
 '' &  '' & ''& '' & '' & $15\hr10\mn58\fs25$ & $-39\degr26\arcmin50\fs2$ &  USNO 7826$-$00179$-$1      & 15105821$-$3926499        & 16.4 &  10.600\tablenotemark{e} &   8.374\\
 '' &  '' & ''& '' & '' & $15\hr10\mn59\fs72$ & $-39\degr26\arcmin56\fs9$ &  USNO 0505$-$0350290      & \nodata & 2.4 &   \nodata & \nodata\\
3  & \object[1RXS J150814.0-403445]{150814.0$-$403445} & 0.069& 40.3 & 16 & $15\hr08\mn12\fs12$ & $-40\degr35\arcmin02\fs1$ &  \object{HD 133880}	        & 15081213$-$4035022        & 27.1 &  5.762    &   5.934\\
4  & \object[1RXS J150428.9-392423]{150428.9$-$392423} & 0.051& 72.7 & 11 & $15\hr04\mn28\fs65$ & $-39\degr24\arcmin26\fs1$ &  \object{CD-38 9913}	        & 15042865$-$3924261        & 3.9 &  10.7      &   8.293\\
5  & \object[1RXS J150526.4-385709]{150526.4$-$385709} & 0.073& 81.8 & 8 & $15\hr05\mn26\fs01$ & $-38\degr57\arcmin000\fs8$ &  \object[RX J1505.4-3857]{RX J1505.4$-$3857}        & 15052586$-$3857031        & 9.4 &  12.55	        &   9.124\\[0.1in]
6  & \object[1RXS J150139.6-403815]{150139.6$-$403815} & 0.125& 103.2 & 14 & $15\hr01\mn39\fs60$ & $-40\degr38\arcmin15\fs5$ & \nodata & \nodata & \nodata & \nodata & \nodata\\
7  & \object[1RXS J151849.8-405108]{151849.8$-$405108} & 0.107& 113.6 & 17 & $15\hr18\mn52\fs82$ & $-40\degr50\arcmin52\fs8$ &  \object{V* LX Lup}	        & 15185282$-$4050528        & 38.4 &  11.01	        &   8.547\\
8  & \object[1RXS J145951.7-401158]{145951.7$-$401158} & 0.085& 117.1 & 11 & $14\hr59\mn52\fs44$ & $-40\degr11\arcmin59\fs5$ &  \object{HD 132349}	        & 14595244$-$4011594        & 8.6 &  9.90     &   8.401\\
9  & \object[1RXS J151659.2-382648]{151659.2$-$382648} & 0.065& 123.0 & 16 & $15\hr16\mn59\fs35$ & $-38\degr26\arcmin51\fs4$ &  \object{HD 135549}	        & 15165935$-$3826514        & 3.4 &  6.876    &   5.789\\
10 & \object[1RXS J152046.2-405405]{152046.2$-$405405} & 0.074& 135.1 & 10 & $15\hr20\mn46\fs97$ & $-40\degr53\arcmin52\fs7$ &  \object{HD 136206}	        & 15204697$-$4053526        & 15.1 &  7.83      &   6.518\\[0.1in]
11 & \object[1RXS J145837.5-391507]{145837.5$-$391507} & 0.084& 138.2 & 10 & $14\hr58\mn37\fs56$ & $-39\degr15\arcmin02\fs7$ &  USNO 0507$-$0344267      & 14583744$-$3915033        & 4.9 &  11.850\tablenotemark{e} &   8.648\\
12 & \object[1RXS J152211.8-395958]{152211.8$-$395958} & 0.080& 140.1 & 11 & $15\hr22\mn11\fs75$ & $-39\degr59\arcmin49\fs6$ &  \object{V* LZ Lup}	        & 15221162$-$3959509        & 8.4 &  12.02	        &   9.100\\
13 & \object[1RXS J145721.8-401401]{145721.8$-$401401} & 0.099& 145.9 & 9 & $14\hr57\mn22\fs07$ & $-40\degr13\arcmin58\fs6$ & \nodata & 14572207$-$4013586        & 4.2 & \nodata &  13.333\\
14 & \object[1RXS J151806.7-380423]{151806.7$-$380423} & 0.115& 148.5 & 12 & $15\hr18\mn07\fs15$ & $-38\degr04\arcmin23\fs8$ & \nodata & 15180715$-$3804238        & 5.3 & \nodata &  10.600\\
 '' &  '' & ''& '' & '' & $15\hr18\mn07\fs61$ & $-38\degr04\arcmin23\fs6$ &  USNO 7822$-$00433$-$1      & 15180762$-$3804237        & 10.7 &  10.650\tablenotemark{e} &   7.999\\[0.1in]
15 & \object[1RXS J151446.3-422020]{151446.3$-$422020} & 0.054& 150.7 & 13 & $15\hr14\mn47\fs48$ & $-42\degr20\arcmin14\fs9$ &  \object[RX J1514.8-4220]{RX J1514.8$-$4220}        & 15144748$-$4220149        & 14.2 & \nodata &   9.011\\
16 & \object[1RXS J152012.2-382159]{152012.2$-$382159} & 0.153& 152.8 & 8 & $15\hr20\mn12\fs53$ & $-38\degr21\arcmin57\fs9$ &  \object{CD-37 10147C}	        & 15201253$-$3821579        & 4.2 & \nodata &   8.454\\
17 & \object[1RXS J151942.8-375255]{151942.8$-$375255} & 0.061& 169.1 & 10 & $15\hr19\mn42\fs80$ & $-37\degr52\arcmin55\fs0$ & \nodata & \nodata & \nodata & \nodata & \nodata\\
18 & \object[1RXS J152445.4-394238]{152445.4$-$394238} & 0.105& 170.4 & 11 & $15\hr24\mn45\fs01$ & $-39\degr42\arcmin37\fs0$ &  \object{HD 136933}	        & 15244501$-$3942367        & 4.7 &  5.367    &   5.495\\
19 & \object[1RXS J145613.6-385121]{145613.6$-$385121} & 0.176& 172.5 & 8 & $14\hr56\mn14\fs04$ & $-38\degr51\arcmin20\fs1$ &  \object{HD 131675}	        & 14561404$-$3851200        & 5.2 &  9.15      &   7.323\\[0.1in]
20 & \object[1RXS J145744.3-414140]{145744.3$-$414140} & 0.051& 173.7 & 15 & $14\hr57\mn44\fs90$ & $-41\degr41\arcmin38\fs8$ &  USNO 0483$-$0366208      & 14574495$-$4141394        & 6.9 &  11.790\tablenotemark{e} &   9.351\\
 '' &  '' & ''& '' & '' & $14\hr57\mn44\fs14$ & $-41\degr41\arcmin40\fs8$ & \nodata & 14574414$-$4141408        & 1.8 & \nodata &  14.698\\
\enddata
\tablenotetext{a}{This is the position of the optical counterpart if known,
 otherwise it is the X-ray position.}
\tablenotetext{b}{Separation between the X-ray source and the nominal SNR
  center given by \citet{g00}.}
\tablenotetext{c}{X-ray position uncertainty.}
\tablenotetext{d}{Separation between the X-ray and optical sources.}
\tablenotetext{e}{No $V$ magnitude was available from SIMBAD, so this is the $R2$ magnitude from USNO-B1.0.}
\tablecomments{Stellar identifications were made only on the basis of
  the \rosat\ data and SIMBAD.  See also Table~\ref{tab:id} and
  \S~\ref{sec:cpt}.} 
\end{deluxetable}
\clearpage
\end{landscape}

\LongTables
\begin{deluxetable}{l c p{1.25in} c c p{2.5in}}
\tabletypesize{\scriptsize}
\tablewidth{0pt}
\tablecaption{Comments on Identification of X-ray Sources\label{tab:id}}
\tablehead{
\colhead{ID\tablenotemark{a}} & \colhead{N\tablenotemark{b}} & \colhead{Star(s)\tablenotemark{c}} & \colhead{Optical} &
\colhead{\cxo?\tablenotemark{e}} & \colhead{Additional Notes\tablenotemark{f}} \\
 & & &\colhead{Figure(s)\tablenotemark{d}} & \\
}
\startdata
\mc{1}{l}{\bf \snra:}\\
\object[1RXS J193445.6+303100]{193445.6+303100} &1& \object{HD 184738} & \ref{fig:opt1} & N & Planetary nebula, associated with  235.2~mJy NVSS source at $19\hr34\mn45\fs20$ $+30\degr30\arcmin58\farcs8$ \citep{ck98}\\
\object[1RXS J193840.0+303035]{193840.0+303035} &2& \object{V* EM Cyg} & \ref{fig:opt1} &N &  Dwarf nova; X-ray emission  reported by \citet{r96} and \citet{vbrp97}\\
\object[1RXS J193922.4+300921]{193922.4+300921} &3& \object{HD 185734} &\ref{fig:opt1} &N &  Spectroscopic binary; type G8III\\
\object[1RXS J194337.2+322523]{194337.2+322523} &4& \object[BDS 9566 B]{BDS 9566~B}/ \object[HD 331149]{HD~331149} &\ref{fig:opt1} &N &  Binary system, late-type\\
\object[1RXS J193458.1+335301]{193458.1+335301} &5&\nodata &\ref{fig:opt1},\ref{fig:1934ha} &Y & Extended \chandra\ source; see \S~\ref{sec:rxsa} \\
\object[1RXS J192722.3+280934]{192722.3+280934} &6& USNO 1181$-$0406270 & \ref{fig:opt1}& N & \\
\object[1RXS J194401.5+284456]{194401.5+284456} &7& \object[GSC 02151-03394]{GSC 02151$-$03394} & \ref{fig:opt2}& N & Late-type\\
\object[1RXS J194902.9+295258]{194902.9+295258} &8& \object{HD 187460}& \ref{fig:opt2}& N & Pulsating variable star of type K2II-III\\
\object[1RXS J193228.6+345318]{193228.6+345318} &9&2MASS~J19322722+3453148&\ref{fig:opt2},\ref{fig:cxoopt1} & Y & 35.4~mJy NVSS counterpart  at $19\hr32\mn27\fs20$, $+34\degr53\arcmin14\farcs8$; flare star \citep{fs03}\\
\object[1RXS J191449.0+315131]{191449.0+315131} &10& \object{HD 180314}& \ref{fig:opt2}& N & Late-type\\
\object[1RXS J193856.2+351407]{193856.2+351407} &11& \object{HD 185696}& \ref{fig:opt2}& N & Double star system; late-type\\
\object[1RXS J193936.8+263718]{193936.8+263718} &12& \object{AG+26 2090}& \ref{fig:opt2}& N & Late-type\\
\object[1RXS J193113.0+360730]{193113.0+360730} &13& \object[G 125-15]{G 125$-$15}& \ref{fig:opt3}& N\\
\object[1RXS J191151.1+285012]{191151.1+285012} &14& USNO~1188$-$0330651& \ref{fig:opt3}& N\\[0.1in]
\mc{1}{l}{\bf \snrb:}\\
\object[1RXS J205042.9+284643]{205042.9+284643} &1& \nodata & \ref{fig:opt4} & Y & Flare star? See \S~\ref{sec:rxsb}\\
\object[1RXS J204457.4+291613]{204457.4+291613} &2& \object{HD 335070}& \ref{fig:opt4} & N & Late-type\\
\object[1RXS J205812.8+292037]{205812.8+292037} &3& USNO~1193$-$0519643/ 2MASS~J20581257+2920454&\ref{fig:opt4} &N &Association may be somewhat  questionable; see \S~\ref{sec:rxsd}\\
\object[1RXS J205208.5+270546]{205208.5+270546} &4& \object{HD 198809} &\ref{fig:opt4} & N& Variable type G7III star\\
\object{AX J2049.6+2939} &\nodata& 2MASS~J20493540+2938509 &\ref{fig:cxoopt1} & Y & Late-type\\[0.1in]
\mc{1}{l}{\bf \snrc:}\\
\object[1RXS J050639.1+513607]{050639.1+513607} &1& \object{HD 32537} & \ref{fig:opt5} & N & Variable star of the $\gamma$-Dor type\\[0.1in]
\mc{1}{l}{\bf \snrd:}\\
\object[1RXS J045707.4+452751]{045707.4+452751} &1& 2MASS~J04570832+4527499& \ref{fig:opt6},\ref{fig:cxoopt1} & Y \\
\object[1RXS J050339.8+451715]{050339.8+451715} &2& 2MASS~J05033958+4516594& \ref{fig:opt6},\ref{fig:cxoopt1} & Y & Associated with the 34.3~mJy NVSS source at $05\hr03\mn39\fs59$, $+45\degr16\arcmin58\farcs9$; flare star?\\
\object[1RXS J045222.2+455619]{045222.2+455619} &3& \object{HD 30736} & \ref{fig:opt6} & N & Late-type\\[0.1in]
\mc{1}{l}{\bf \snre:}\\
\object[1RXS J064108.3+052250]{064108.3+052250} &1& USNO 0155$-$01104$-$1/ 2MASS~J06410796+0522438 & \ref{fig:opt7} & N & Binary system?\\
\object[1RXS J064136.2+080218]{064136.2+080218} &2& \object{HD 262113} & \ref{fig:opt7} & N & Late-type\\
\object[1RXS J064109.3+044733]{064109.3+044733} &3& USNO~0947$-$0100759/ USNO~0947$-$0100763/ 2MASS~J06410900+0447239/ USNO~0155$-$02167$-$1 & \ref{fig:opt7} & N & Multiple-star system\\
\object[1RXS J064641.1+082152]{064641.1+082152} &4& \object{HD 49015} &\ref{fig:opt7} & N & Variable star of the $\gamma$-Dor type\\
\object[1RXS J062740.3+073103]{062740.3+073103} &5& 2MASS~J06274012+0731006&\ref{fig:opt7},\ref{fig:cxoopt1} & Y &  \\
\object[1RXS J062937.2+082930]{062937.2+082930} &6& \object{HD 45759}&\ref{fig:opt7} & N & Late-type\\
\object[1RXS J063715.7+032005]{063715.7+032005} &7& USNO~0150$-$00332$-$1/ 2MASS~J06371585+0320043/ 2MASS~J06371555+0320026&\ref{fig:opt8} & N & Multiple-star system?\\
\object[1RXS J062554.8+065543]{062554.8+065543} &8& USNO~0145$-$01717$-$1&\ref{fig:opt8} & N\\
\lnsa &\nodata& 2MASS~J06300529+0545407 & \ref{fig:cxoopt1} & Y & \\
\lnsb &\nodata& 2MASS~J06333322+0608396& \ref{fig:cxoopt2} & Y &\\
\lnsc &\nodata& 2MASS~J06392566+0514301& \ref{fig:cxoopt2} & Y & \\[0.1in]
\mc{1}{l}{\bf \snrf:}\\
\object[1RXS J150818.8-401730]{150818.8$-$401730} &1& 2MASS~J15081819$-$4017261 & \ref{fig:opt9},\ref{fig:cxoopt2} & Y& Also 1AXG~J150818$-$4016  \citep{uit+01}\\
\object[1RXS J151059.6-392655]{151059.6$-$392655} &2& USNO~0505$-$0350285/ USNO~7826$-$00179$-$1/ USNO~0505$-$0350290& \ref{fig:opt9} & N& Multiple-star system?\\
\object[1RXS J150814.0-403445]{150814.0$-$403445} &3& \object{HD 133880}& \ref{fig:opt9} & N & Variable star of the $\alpha^2$-CVn type; has late-type companion \citep{shh+03}\\
\object[1RXS J150428.9-392423]{150428.9$-$392423} &4& \object[CD-38 9913]{CD$-$38 9913}& \ref{fig:opt9} & N\\
\object[1RXS J150526.4-385709]{150526.4$-$385709} &5& \object[RX J1505.4-3857]{RX~J1505.4$-$3857}& \ref{fig:opt9} & N & T  Tauri star\\
\object[1RXS J150139.6-403815]{150139.6$-$403815} &6& \nodata& \ref{fig:opt9},\ref{fig:15012mass} & Y &  Extended \chandra\ source; see \S~\ref{sec:rxsc}\\
\object[1RXS J151849.8-405108]{151849.8$-$405108} &7& \object{V* LX Lup}& \ref{fig:opt10} & N & T  Tauri star\\
\object[1RXS J145951.7-401158]{145951.7$-$401158} &8& \object{HD 132349}& \ref{fig:opt10} & N & Late-type\\
\object[1RXS J151659.2-382648]{151659.2$-$382648} &9& \object{HD 135549}& \ref{fig:opt10} & N & Late-type\\
\object[1RXS J152046.2-405405]{152046.2$-$405405} &10& \object{HD 136206}& \ref{fig:opt10} & N & Late-type\\
\object[1RXS J145837.5-391507]{145837.5$-$391507} &11& USNO~0507$-$0344267& \ref{fig:opt10} & N\\
\object[1RXS J152211.8-395958]{152211.8$-$395958} &12& \object{V* LZ Lup}& \ref{fig:opt10} & N & T  Tauri star\\
\object[1RXS J145721.8-401401]{145721.8$-$401401} &13& 2MASS~J14572207$-$4013586& \ref{fig:opt11} & N & Extended on the DSS/2MASS images (2MASX~J14572207$-$4013588); likely a  galaxy\\
\object[1RXS J151806.7-380423]{151806.7$-$380423} &14& USNO~7822$-$00433$-$1/ 2MASS~J15180762$-$3804237&\ref{fig:opt11} & N & Multiple-star system?\\
\object[1RXS J151446.3-422020]{151446.3$-$422020} &15& \object[RX J1514.8-4220]{RX~J1514.8$-$4220}& \ref{fig:opt11} & N\\
\object[1RXS J152012.2-382159]{152012.2$-$382159} &16& \object[CD-37 10147C]{CD$-$37~10147C}& \ref{fig:opt11} & N& Multiple-star system?\\
\object[1RXS J151942.8-375255]{151942.8$-$375255} &17& \nodata& \ref{fig:mag} & Y & Star detected in Magellan data\\
\object[1RXS J152445.4-394238]{152445.4$-$394238} &18& \object{HD 136933}& \ref{fig:opt11} & N & Double star; type  A0sp$\ldots$\\
\object[1RXS J145613.6-385121]{145613.6$-$385121} &19& \object{HD 131675}& \ref{fig:opt12} & N & Late-type\\
\object[1RXS J145744.3-414140]{145744.3$-$414140} &20& USNO~0483$-$0366208/ 2MASS~J14574414$-$4141408& \ref{fig:opt12} & N& Multiple-star system?\\
\enddata
\tablenotetext{a}{ID of X-ray source, which is 1RXS~J unless otherwise indicated.}
\tablenotetext{b}{Number of X-ray source in the given SNR from
  Table~\ref{tab:srcs}.} 
\tablenotetext{c}{Name(s) of likely stellar companion(s).  In contrast
  to Table~\ref{tab:srcs}, this also includes identifications made
  from \chandra\ followup observations.}
\tablenotetext{d}{Figure(s) where optical/IR counterparts are
  identified.}
\tablenotetext{e}{Indicates if source was selected for \chandra\
  followup; see \S~\ref{sec:notescxo}.}
\tablenotetext{f}{Classifications are from SIMBAD unless otherwise
  noted.  ``Late-type'' means that the star is of type mid-F or later,
  and hence is likely to have intrinsic X-ray emission \citep[e.g.,][]{shh+03}.}
\end{deluxetable}



\begin{deluxetable}{l c c c}
\tablewidth{0pt}
\tablecaption{Log of \chandra/ACIS-S Observations\label{tab:cxo}}
\tablehead{
\colhead{Source} & \colhead{Date} & \colhead{Exp.} & \colhead{Subarray} \\
 & & \colhead{(ksec)} & \colhead{Mode}\\
}
\startdata
\mc{1}{l}{\bf \snra:}\\
\object{1RXS J193228.6+345318} &  \dataset[ADS/Sa.CXO#obs/3887]{2002-Dec-08} &	3.7 & 1/4\\	
\object{1RXS J193458.1+335301} &  \dataset[ADS/Sa.CXO#obs/3888]{2003-Jan-26} &	3.5 & 1/4\\[0.1in]	
\mc{1}{l}{\bf \snrb:}\\
\object{AX J2049.6+2939}      & \dataset[ADS/Sa.CXO#obs/3889]{2003-Mar-19}   &	3.2 & 1/2\\	
\object{1RXS J205042.9+284643} & \dataset[ADS/Sa.CXO#obs/3890]{2003-Mar-19}  &	3.9 & 1/4\\[0.1in]	
\mc{1}{l}{\bf \snrd:}\\
\object{1RXS J045707.4+452751} & \dataset[ADS/Sa.CXO#obs/3878]{2003-Jan-04} & 5.2 & 1/4\\[0.1in]
\mc{1}{l}{\bf \snre:}\\
\object{1RXS J050339.8+451715} & \dataset[ADS/Sa.CXO#obs/3879]{2003-Jan-08} & 5.4 & 1/4\\
\object{1RXS J062740.3+073103} & \dataset[ADS/Sa.CXO#obs/3880]{2003-Mar-11} & 3.4 & 1/4\\
\lnsa\ & \dataset[ADS/Sa.CXO#obs/3881]{2002-Dec-07} & 3.5 & 1/2\\
\lnsb\ & \dataset[ADS/Sa.CXO#obs/3882]{2003-Apr-22} & 3.6 & 1/2\\
\lnsc\ & \dataset[ADS/Sa.CXO#obs/3883]{2003-Mar-11} & 3.8 & 1/2\\[0.1in]
\mc{1}{l}{\bf \snrf:}\\
\object[1RXS J150139.6-403815]{1RXS J150139.6$-$403815} & \dataset[ADS/Sa.CXO#obs/3884]{2003-Mar-18} & 2.9 & 1/4\\
\object[1RXS J150818.8-401730]{1RXS J150818.8$-$401730} & \dataset[ADS/Sa.CXO#obs/3885]{2003-Mar-18} &	3.7 & 1/4\\
\object[1RXS J151942.8-375255]{1RXS J151942.8$-$375255} & \dataset[ADS/Sa.CXO#obs/3886]{2003-Mar-10} &	2.9 & 1/4\\	
\enddata
\end{deluxetable}

\begin{deluxetable}{l l l l c r}
\tablewidth{0pt}
\tablecaption{Log of Optical/IR Observations\label{tab:optobs}}
\tablehead{
\colhead{Source} & \colhead{Date} & \colhead{Telescope} &
\colhead{Instrument} & \colhead{Band(s)} & \colhead{Exp.} \\
 & & & & & \colhead{(sec)} \\
}
\startdata
1RXS J193458.1+335301 & 2003-Jul-03 & P200 & LFC & $g^\prime$ & 300 \\
1RXS J193458.1+335301  & 2003-Jul-24 & P60 & P60CCD & H$\alpha$/Offband & 1200\\
1RXS J151942.8$-$375255  & 2003-Apr-04 & Magellan II/Clay & MagIC & $R$ & 930 \\
& 2003-Apr-18 & Magellan II/Clay & PANIC &$K_s$ & 200 \\
1RXS J150139.6$-$403815\tablenotemark{a} & 2004-Feb-16 & Magellan II/Clay & MagIC & $R$& 60 \\
 & & & & $B$ & 160 \\
\enddata
\tablecomments{The telescopes/instruments used were LFC: the Large
  Format Camera on the  Palomar 200-inch; P60CCD: the CCD 
  camera on the Palomar 60-inch; MagIC: Raymond and Beverly
  Sackler Magellan Instant Camera on the 6.5-m Clay (Magellan II)
  telescope; and PANIC: Persson's Auxiliary Nasmyth Infrared Camera on
  the 6.5-m Clay (Magellan II) telescope \citep{mpm+04}.}
\tablenotetext{a}{Observed by C.~Rakowski.}
\end{deluxetable}

\begin{deluxetable}{c c c c c c c c c }
\tablewidth{0pt}
\tablecaption{Properties of Optical Sources in Figure~\ref{fig:15012mass}\label{tab:rxscopt}}
\tablehead{
\colhead{Source} & \colhead{$\alpha-15\hr01\mn$} &
\colhead{$-\delta-40\deg38\arcmin$} && \mc{2}{c}{$B$} && \mc{2}{c}{$R$}
\\ \cline{5-6} \cline{8-9}
 & \colhead{(sec)} & \colhead{(arcsec)} &&  \colhead{FWHM\tablenotemark{a}} & \colhead{Mag} &&\colhead{FWHM\tablenotemark{b}} & \colhead{Mag} \\\cline{2-3} 
 & \mc{2}{c}{(J2000)} & & \colhead{(arcsec)} & & & \colhead{(arcsec)}\\
}
\startdata
A & 41.12 & 09.4 && 3.2 & 12.1 &&3.6 & 9.5 \\ 
B & 41.30 & 09.3 && 0.8 & 12.1 && 0.8 & 11.1\\
C & 41.46 & 06.7 && 1.3 & 12.9 && 3.6 & 11.2\\
\enddata
\tablecomments{The astrometry has absolute uncertainties of $\approx
  0\farcs2$ in each coordinate owing to uncertainties in 2MASS.  The
  photometry has systematic uncertainties of $\approx 0.5$~mag owing
  to uncertain zero-point calibration.}
\tablenotetext{a}{The seeing was $\approx 0\farcs77$.}
\tablenotetext{b}{The seeing was $\approx 0\farcs66$.}
\end{deluxetable}

\end{document}